\documentclass[natbib]{svjour3}

\usepackage{graphicx,mathptmx,jab}
\usepackage[dvips]{color}

\usepackage{lineno}
\begin{document}
\title{Solar Wind Turbulence and the Role of Ion Instabilities}
\author{{O. Alexandrova} \and {C. H. K. Chen}  \and {L. Sorriso-Valvo}  \and {T. S. Horbury}   \and {S. D. Bale} }
\institute{
O. Alexandrova \at
LESIA, Observatoire de Paris
5, place Jules Janssen
92190 Meudon, France\\
\email{olga.alexandrova@obspm.fr}
\and
C. H. K. Chen and S. D. Bale \at
Space Sciences Laboratory, University of California, Berkeley, CA 94720, USA
\and
L. Sorriso-Valvo \at
IPCF/CNR - UOS di Cosenza, 87036 Rende (CS), Italy.
\email{sorriso@fis.unical.it}\\
\& Space Sciences Laboratory, University of California, Berkeley, CA 94720, USA
\and
T. S. Horbury \at
The Blackett Laboratory, Imperial College London, London SW7 2AZ, United Kingdom
}
\date{Received: date / Accepted: date}
\date{\today}
\maketitle

\abstract{ 

Solar wind is probably the best laboratory to study turbulence in astrophysical plasmas. In addition to the presence of magnetic field, the differences with neutral fluid isotropic turbulence are: (i) weakness of collisional dissipation and (ii) presence of several characteristic space and time scales. In this paper we discuss observational properties of solar wind turbulence in a large range from the MHD to the electron scales.  At MHD scales, within the inertial range, turbulence cascade of magnetic fluctuations develops mostly in the plane perpendicular to the mean field, with the Kolmogorov scaling $k_{\perp}^{-5/3}$ for the perpendicular cascade and  $k_{\|}^{-2}$ for the parallel one.   Solar wind turbulence is compressible in nature: density fluctuations at MHD scales  have the Kolmogorov spectrum. Velocity fluctuations do not follow magnetic field ones: their spectrum is a power-law with  a $-3/2$ spectral index. Probability distribution functions of different plasma parameters are not Gaussian, indicating  presence of  intermittency.  At the moment there is no  global model taking into account all these observed properties of the inertial range.  At ion scales, turbulent spectra have a break, compressibility increases and the density fluctuation spectrum has a local flattening. 
Around ion scales, magnetic spectra are variable and ion instabilities occur as a function of the local plasma parameters. Between ion and electron scales, a small scale turbulent cascade seems to be established. It is characterised by a well defined power-law spectrum in magnetic and density fluctuations  with a spectral index close to $-2.8$. Approaching electron scales, the fluctuations are no more self-similar: an exponential cut-off is    usually observed (for time intervals without quasi-parallel whistlers)  indicating an onset of dissipation. The small scale inertial range  between ion and electron scales and the  electron dissipation range can be  together  described by $\sim k_{\perp}^{-\alpha}\exp(-k_{\perp}\ell_d)$, with $\alpha\simeq 8/3$ and the dissipation scale $\ell_d$ close to the electron Larmor radius $\ell_d \simeq \rho_e$. The nature of this small scale cascade and a possible dissipation mechanism are still under debate. 
}

\section{Introduction}

Natural plasmas are frequently in a turbulent state characterized by large, irregular fluctuations of the physical parameters. The spatial and temporal scales of these fluctuations cover a large range, usually extending down to the smallest scales resolved by the observations. Well known examples are provided by the solar wind, the magnetosheath of planetary magnetospheres, the interstellar medium, etc... 
 
Is there a certain degree of generality  in the physics of the various astrophysical situations where turbulent states are observed? If this is the case, is it of the same nature as what happens in incompressible neutral (or magnetized) fluid turbulence, which is a non-linear process, non-reproducible locally but with some  ``universal" statistical properties? These ``universal" statistical properties are thought to result from the combination of (1) an infinite number of degrees of freedom, each characterized by its spatial and temporal scale; 
(2) the absence of characteristic spatial and temporal scales, which implies some sort of equivalence between all of the degrees of freedom;  (3) a nonlinear transfer of energy between these degrees of freedom, often called a cascade of energy. 

To be more specific, the incompressible fluid turbulence occurs at large Reynolds numbers $R_e = L V_L / \eta \gg 1$ (where $L$ is the scale at which the energy is injected in the system,  that is of the order of the correlation length of the largest turbulent eddy, $V_L$ the typical value of velocity fluctuations at scale $L$ and $\eta$ the kinematic viscosity).  This is verified when the energy injection scale is sufficiently  far from the dissipation scale $\ell_d$ ($L \gg \ell_d$). Thanks to a number of observations, numerical simulations and theoretical works,  the following universal properties of a turbulent system have been firmly established:  
\begin{itemize}
    \item In Fourier space, at intermediate scales $L^{-1} \ll k\ll \ell_d^{-1}$ ($k$ being a wave-number), within the so called {\it inertial range}, the power spectrum of the velocity fluctuations is observed to follow a $k^{-5/3}$ law, independently of how the energy is injected in the system, and of how it is dissipated at small scales. This suggests scale invariance, i.e., at each scale the same physical description is valid (the Navier-Stokes equation for fluids and the magnetohydrodymanic equations for magnetized plasmas are scale invariant and describe well self-similar turbulent fluctuations).
    \item Intermittency, due to spatial nonuniformity of the energy transfer across scales,  manifests itself as a  scale dependent departure from Gaussian distributions of the probability distribution functions of the turbulent fluctuations. 
\end{itemize}

To date, 3D fluid turbulence is far from being understood, and there is no satisfactory  theory, based on first principles, that fully describes it in a sufficiently general frame.  Therefore one has to rely on ``phenomenologies" which attempt to provide a framework for the interpretation of experimental results; for example the  empirical  $k^{-5/3}$ law is well described  by the Kolomogorov's phenomenology (hereafter K41) \citep{k41,frisch95}.  In this simple model of turbulence, kinetic energy $E_c$ is supposed to cascade from large scales to small scales and the {\it cascade rate} (an energy per unit time) is constant over the inertial range $\epsilon= \partial E_c/\partial t = const$.  Since the only timescale that appears in the system is  the time of the energy exchange between the  fluctuations  (the {\it eddies}), also called the {\it non-linear} or {\it eddy turnover time}  $\tau_{nl} = \ell/\delta v$, the cascade rate can be approximated by $\epsilon \approx (\delta v)^2 /\tau_{nl}=const$. It follows that the velocity field fluctuations $\delta v \approx (\epsilon \ell)^{1/3}$  so that the power spectrum $(\delta v)^2/k$ goes like $\ell^{5/3}$ or $k^{-5/3}$.

Intermittency is beyond the Kolmogorov phenomenology but it has been observed that in neutral fluids it appears in the form of coherent structures as filaments of vorticity.  Their characteristic length can be of the order of the energy injection scale $L$ but their cross-section is of the order of the dissipation scale $\ell_d$ (see the references of Section 8.9 in \citet{frisch95}).  Thus, in Fourier space, these filaments occupy all scales including the edges of the inertial range.

As we have said, in the phenomenological framework of turbulence, the majority of the results are based on the interpretation of experimental results. However, one important theoretical result   was obtained from   the Navier-Stokes equation, independently of K41 phenomenology: it  is known as Kolmogorov's $4/5$ law (hereafter K4/5). The K4/5 law prescribes that, for fully developed incompressible  turbulence  in a stationary state\footnote{In a stationary state, the energy injection rate $\epsilon_{inj}$ at large scales is equal to the energy transfer rate within the inertial range $\epsilon=(\delta v)^2/\tau_{nl}$ and to the energy dissipation rate  within the dissipation range of scales $\epsilon_{dis}=\eta \langle(\partial_x v(x))^2\rangle$, where $\eta$ is the kinematic viscosity: $\epsilon_{inj}=\epsilon_{dis}=\epsilon$.}, under conditions of isotropy, %\footnote{ {\bf More general law for anisotropic turbulence is known as Yaglom relation \citep{yaglom49}.}}, 
 local homogeneity, and vanishing dissipation (i.e., in the inertial range), the third order moment of the longitudinal (i.e. along the bulk flow) velocity fluctuations $\delta v$ scales linearly with the separation $\ell$  (or with the time scale $\tau=\ell/V$, with $V$ being a bulk flow speed):
\begin{equation}\label{eq:K4/5}
Y(\ell)=\langle \delta v^3 \rangle = -4/5 \epsilon \ell,
\end{equation}
the proportionality factor $\epsilon$ being the mean energy transfer  rate and dissipation rate  of the turbulent cascade (see \citep{frisch95}, Section 6.2, and references therein). This law has been  indeed observed in the neutral fluid turbulence, e.g. \citep{danaila01}.  
 Note that Kolmogorov 4/5 law can be obtained from the more general Yaglom law \citep{yaglom49} in case of Navier-Stokes isotropic turbulence.

When the energy cascade ``arrives" to the spatial (or time) scale of the order of the dissipation scale $\ell_d$, the spectrum  becomes curved \citep{grant62}, indicating a lack of self-similarity.  This spectrum is also universal  (see, e.g., Fig.~8.14 in \citep{frisch95}) and can be described by $\sim k^3\exp{(-ck\ell_d)}$ with $c\simeq 7$  \citep{chen93}. In  neutral fluids the dissipation sets in usually at scales of the order of the collisional mean free path.

We shall restrict ourselves here to  the solar wind turbulence, which is perhaps our best laboratory for studying astrophysical plasma turbulence  \citep{tu95,bruno05a,horbury05,matthaeus11}. Does the solar wind turbulence share  the above universal characteristics, such as power-law  spectra, intermittency and linear dependence between the third order moment of the fluctuations and the energy transfer rate? How does the dissipation set in? and is its spectrum universal?

The solar wind expands radially but not  with spherical symmetry. Fast, rather steady wind at around $700$~km/s flows from coronal holes, generally at high solar latitudes. More variable slow wind ($200-500$~km/s) is thought to have its source around coronal hole boundaries or in transiently open regions. In general, the properties of fluctuations within fast and slow wind  at 1~AU are rather different, with fast wind turbulence appearing less developed than that in slow wind, indicating different ``age of turbulence". Interactions between fast and slow wind, as well as transient events, produce compressions, rarefactions and shocks. When considering the innate properties of plasma turbulence, it is usually easier to treat steady, statistically homogeneous intervals of data from individual streams.

 {\it In situ} spacecraft  measurements in the solar wind  provide time series of local plasma parameters. Therefore, in Fourier space, we have a direct access to frequency spectra. When the flow speed of the solar wind $V_{sw}$  is much larger than the characteristic plasma speeds, one can invoke the Taylor's hypothesis \citep{taylor38,perri10a}  and convert a spacecraft-frame frequency $f$ to a flow-parallel wavenumber $k$ in the plasma frame $k=\frac{2 \pi f}{V_{sw}}$.
 At scales larger than the proton characteristic scales,  we can largely treat the solar wind fluctuations using magnetohydrodynamics (MHD) \citep{marsch87,biskamp93,schekochihin09}. The flow speed $V_{sw}$ is typically much larger than the Alfv\'en speed $V_A=B/\sqrt{4\pi\rho}\simeq 50$~km/s ($B$ being the magnetic field and $\rho$  the mass density) and far faster than spacecraft motions, so that one can use Taylor's hypothesis.  At plasma kinetic scales, the Taylor hypothesis can be used  in the absence of quasi-parallel propagating whistler waves, which have a phase speed higher than $V_{sw}$.

The solar wind is pervaded with fluctuations on all measured scales. These fluctuations form energy spectra following power laws as expected for developed turbulence.  For example, for magnetic fluctuations,  at very large scales (for the spacecraft-frame frequencies $f<10^{-4}$~Hz) the power spectrum goes as $\sim f^{-1}$.  This spectrum can be interpreted in terms of  uncorrelated large scale Alfv\'en waves \citep{matthaeus86,horbury05}. A recent work proposes that it originates due to the nonlinear coupling in the corona between outgoing and ingoing Alfv\'en waves with the help of multiple reflections on the non-homogeneous transition region \citep{verdini12}.  The corresponding frequency range is usually called the energy injection scales \citep{bruno05a}.   The maximal frequency $f_0$ of this range, or {\it outer scale} of the turbulent cascade,  is close to $10^{-4}$~Hz at 1~AU.  It was proposed by \citet{mangeney91,salem00,meyer07}, that at the outer scale there is a balance between the solar wind expansion time $\tau_{exp}=R/V_{sw}$ at a radial distance $R$ and the eddy-turnover time $\tau_{nl}$; and the turbulent cascade can develop at scales where $\tau_{nl}<\tau_{exp}$.  Estimations  at 1 AU for $V_{sw}= 600$~km/s give $\tau_{exp}\simeq 70$~h.  The characteristic non-linear time at $f_0$ is of the order of $\tau_{nl}\simeq70$~h as well\footnote{This is estimated using the Taylor hypothesis  $\ell=V_{sw}/f_0\simeq 6\cdot 10^6$~km and a typical value of $\delta v\simeq 25$~km/s/$\sqrt{Hz}$ at $f_0=10^{-4}$~Hz.}.  At smaller scales, i.e. at higher frequencies $f>10^{-4}$~Hz, the non-linear time becomes smaller than the expansion time and turbulent cascade develops. As $\tau_{exp}$ increases with $R$, the outer scale increases, i.e. $f_0$ shifts towards lower frequencies. This is indeed observed in the solar wind \citep{bruno05a}.  It will be interesting to verify the relationship between the outer-scale and $\tau_{exp}$  with solar wind observations for different turbulence levels and at different heliospheric distances.

Within the $\sim [10^{-4},10^{-1}]$~Hz range, magnetic spectrum is usually observed to follow the K41 scaling, interpreted as the inertial range (the details on the spectral slope of the inertial range will be discussed in Section 2).  The spectrum undergoes new changes at the proton characteristic scales (appearing in the measured  spectra at $\sim [0.1,1]$~Hz) and at the electron scales $\sim [50,100]$~Hz (see details in Section 3).

One of the important differences of the solar wind turbulence with the isotropic neutral fluid turbulence is the presence of the mean magnetic field ${\bf B}$, which introduces a privileged direction and so imposes an anisotropy of turbulent fluctuations.  In the inertial range, the observed  magnetic fluctuations $\delta B_{\|}$ along the  mean field  are  usually much smaller than the transverse Alfv\'enic fluctuations $\delta B_{\perp}$. The wave vector distributions are not isotropic either, $k_{\perp} > k_{\|}$.  In Section 2 we will discuss in more details  how this k-anisotropy has been detected  within the inertial range of the the solar wind turbulence and its possible interpretations.  We will discuss as well intermittency in the solar wind and show recent verification of the K4/5 law.

Another important difference between neutral fluid turbulence and solar wind turbulence is the weakness of collisional dissipation in the solar wind, as for most of the space plasmas. The dissipation process at work and the dissipation length are not known  precisely. There are observational indications and theoretical considerations that characteristic plasma scales may be good candidates to replace, in some sense, the dissipation scale of fluid turbulent cascade. The characteristic plasma scales are  the ion Larmor radius $\rho_{i}=\sqrt{2k_BT_{i\perp}/m_i}/(2\pi f_{ci})$ (with $T_{i\perp}$ being the ion temperature perpendicular to the magnetic field ${\bf B}$, $m_i$ being the ion mass),  the ion inertia length $\lambda_{i}=c/\omega_{pi}$ (with $c$ the speed of light and $\omega_{pi}$ the ion plasma frequency),  the corresponding electron scales $\rho_{e}, \lambda_{e}$, and the ion and electron cyclotron frequencies $f_{ci,e}=qB/(2\pi m_{i,e})$ (with $q$ being the charge of the  particle).  At these scales different kinetic effects may take place. 
However, the precise mechanism (or mechanisms)  which dissipates electromagnetic turbulent energy in the solar wind and the corresponding spatial and/or temporal scale(s) are still under debate. The details of the observations of solar wind turbulence around plasma kinetic scales will be discussed in Section 3.  In particular, in Section 3.2 we discuss the ion temperature anisotropy instabilities  which may control turbulent fluctuations around ion scales. Conclusions are found in Section~4.

\section{The MHD Scale Cascade}

An MHD theory of cascading turbulence similar to Kolmogorov, but carried by Alfv\'enic fluctuations propagating in the large-scale magnetic field ${\bf B}$ was proposed independently by  \citet{iroshnikov63} and \citet{kraichnan65} (IK hereafter). In this model, the fluctuations are still assumed to be isotropic but most of the energy transfer is due to interactions between Alfv\'enic fluctuations moving in opposite direction along ${\bf B}$ with the Alfv\'en speed $V_A$. This limits the time during which two eddies interact,  which is of the order of an Alfv\'en time $\tau_A\sim \ell/V_A$.  It is also assumed that  the interactions are weak such that   $\tau_A\ll \tau_{nl}$, and thus  a number of interactions proportional to  $\tau_{nl}/\tau_A$ is needed to transfer the energy
\citep{dobrowolny80}. Following the argument of Kolomogorov and under the assumption of equipartition between magnetic and kinetic energies, for incompressible fluctuations  and random interactions between the Alfv\'en wave packets, the velocity and magnetic turbulence spectra follow a $\sim k^{-3/2}$ scaling\footnote{For the detailed demonstration we refer to the problem 6.6.4 in the book of \citet{meyer07}.}.

However, the assumption of isotropy in IK model for the magnetized plasma  is quite strong. \citet{goldreich95} proposed an MHD model for anisotropic Alfv\'enic fluctuations. In that theory, the cascade energy is carried by perpendicular fluctuations $v_{\perp}$ with wavelength $\ell_{\perp} = 2 \pi/k_{\perp}$.  The  Alfv\'en time  is the time scale  along  ${\bf B}$, $\tau_A = \ell_{\|}/V_A$, and the  eddie turnover time $\tau_{nl} \approx \ell_{\perp}/v_{\perp}$ governs the energy exchange in the plane perpendicular to  ${\bf B}$. Goldreich and Sridhar proposed that the  turbulence is {\it strong}, so that these timescales are comparable, $\tau_{nl} \approx \tau_A$. This condition, called {\it critical balance}, implies that the nonlinear interaction occurs over  a single Alfv\'en wave period.  Using the argument of Kolomogorov, one can show that  the perpendicular energy transfer rate is $\epsilon(k_{\perp})\sim v_{\perp}^3/\ell_{\perp}$. Under the assumption of  $\epsilon(k_{\perp})=const$,
the  power spectral density of $k_{\perp}$--fluctuations goes  therefore like $\sim k_{\perp}^{-5/3}$.  For the parallel energy transfer rate $\epsilon(k_{\|})$ one gets $v^2_{\perp}V_A/\ell_{\|}$  and a  spectrum $v_{\perp}^2/k_{\|}\sim k_{\parallel}^{-2}$.  An interesting consequence  of the Goldreich-Sridhar model is the following:  since the cascade is carried by the perpendicular fluctuation spectrum (and indeed this property is reinforced as the energy arrives at  larger wavenumbers, where the k-anisotropy becomes important $k_{\perp} \gg k_{\|}$), the energy in the spectrum reaches dissipation scales (or characteristic plasma scales)  in the perpendicular spectrum long before it does so in the parallel spectrum.  This implies that relatively little energy of $k_{\|}$--fluctuations reaches the  characteristic plasma scales due to the nonlinear cascade. 

%%%%  CHRIS & LUCA  %%%

It should be pointed out that the model of \citet{goldreich95}  describes Alfv\'enic turbulence, i.e., the perpendicular magnetic $\delta B_{\perp}$ and velocity $\delta v_{\perp}$ fluctuations.  This model  has been extended to include the passive mixing of the compressive fluctuations by the Alfv\'enic turbulence 
\citep{goldreich95,goldreich97,lithwick01,schekochihin09}. However, the nature of compressible fluctuations observed in the solar wind, i.e. a passive scalar or an active turbulence ingredient,  remains under debate.

%%%%%%%%%   STUART  %%%%%%%%%%%%%%%
Some theoretical results and solar wind observations suggest that ion cyclotron wave-particle interactions are an important source  of heating for solar wind ions \citep{marsch01,isenberg01,kasper08,kasper13,bourouaine10,bourouaine11,marsch11,he11b}.  However, this interpretation requires substantial  turbulent energy at $k_{\parallel} \rho_i \approx 1$, that is in apparent contradiction to the Goldreich-Sridhar model and to the solar wind measurements described in the following section \citep{horbury08,podesta09a,luo10,wicks10,wicks11a,chen11a}. This is another puzzle that has important  ramifications for the coronal heating problem.

%%%%%%%%%%%%%%%%%%%%%%%

\subsection{Scaling and Anisotropy as observed in the solar wind}
\label{scalingandanisotropy}

\subsubsection*{Magnetic fluctuations}

\begin{figure}%[!htb] 
\centering
\includegraphics[width=8cm]{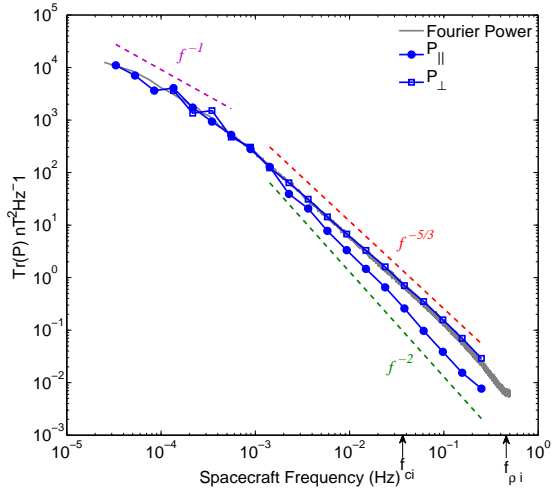}
\caption{Trace of the spectral matrix of magnetic field corresponding to the field being parallel ($\theta_{BV}\in [0,10]^{\circ}$) and perpendicular ($\theta_{BV}\in [80,90]^{\circ}$) to the plasma flow are shown by blue lines, the total Fourier spectrum is shown in gray. The field-perpendicular spectrum $P_{\perp}$ dominates turbulence within the inertial range, it follows a power-law with the spectral index $-5/3$. The field-parallel spectrum $P_{\|}$  has lower power, is steeper and has the spectral slope $-2$. At the energy injection scales $f<5\cdot 10^{-4}$~Hz ($k\rho_i< 2\cdot 10^{-3}$) the fluctuations are isotropic and their spectrum follows $\sim f^{-1}$.  Courtesy of R. Wicks. The same figure as a function of $k\rho_i$ can be found in \citep{wicks10}.}
\label{fig:inertial_range}
\end{figure}

It has long been known that in the inertial range the power spectrum of magnetic field fluctuations in the solar wind is $P(f) \propto f^{-5/3}$, i.e. the same spectrum as for the velocity fluctuations in hydrodynamics turbulence \citep{kolmogorov41a, frisch95}. One might conclude that the turbulence in the solar wind is similar to that in a neutral fluid, like air. However, turbulence in a magnetofluid is radically different to that in a neutral fluid, due to the presence of a magnetic field which breaks the isotropy of the turbulence \citep{shebalin83}, leading to a correlation length  parallel to the field longer than that across it, $\ell_{\|}>\ell_{\perp}$ \citep{matthaeus90} -- crudely, we can think of the turbulent eddies as being shorter perpendicular to the magnetic field than parallel to it, and more formally as having a dominance of turbulent power at wavevectors at large angles to the field, $k_{\perp}>k_{\|}$.

Measurements of the wave-vector anisotropy and of the corresponding spectra in the solar wind with one satellite  are not trivial. A satellite provides time series measurements along its orbit; therefore, applying the Fourier (or wavelet) transform  we obtain directly frequency spectra and not $k$--spectra. As we have discussed in the introduction, the Taylor hypothesis can be used, i.e. we can easily estimate $k$ along the bulk flow  through  $k=\frac{2 \pi f}{V_{sw}}$. Thus, if  ${\bf V_{sw}}$  is parallel to the mean field, the fluctuations with parallel wave vectors  $k_{\|}$ will be measured, and if ${\bf V_{sw}}$ is perpendicular to ${\bf B}$, the satellite resolves well fluctuations with $k_{\perp}$. We denote the local flow-to-field angle as $\theta_{BV}$. Fig.~\ref{fig:inertial_range} shows magnetic spectra in the fast  high latitude solar wind  measured by the {\it Ulysses} spacecraft  (at distance of 1.38 to 1.93 AU from the Sun). As the spacecraft only measures wave vectors ${\bf k}$ parallel to ${\bf V_{sw}}$,  for small flow-to-field angles $\theta_{BV}\in [0,10]^{\circ}$, $P_{\|}$~(nT$^2/$Hz)  represents an $E(k_{\|})$ spectrum, and for quasi-perpendicular  angles $\theta_{BV}\in [80,90]^{\circ}$, $P_{\perp}$~(nT$^2/$Hz),  is the proxy of $E(k_{\perp})$. The total Fourier power, without separation into different angles is also shown. Within the energy injection range, the fluctuations are found to be isotropic, $P_{\|}\simeq P_{\perp}$, and both spectra follow an $\sim f^{-1}$ power-law in agreement with previous observations \citep{bruno05a}. In the inertial range one observes a bifurcation of the two spectra: the perpendicular spectrum follows the Kolmogorov's slope, $E(k_{\perp})\sim k_{\perp}^{-5/3}$, while the parallel spectrum is steeper, $E(k_{\|})\sim k_{\|}^{-2}$. This result, initially seen in fast wind measured by \emph{Ulysses} \citep{horbury08} has been confirmed by several other studies \citep{podesta09a,luo10,wicks10,wicks11a,chen11a}. These magnetic field spectral scaling  observations 
provide an intriguing, if not unequivocal, connection to the Goldreich-Sridhar theory \citep{higdon84,goldreich95}.  It is important to notice that the spectral anisotropy, shown in Fig.~\ref{fig:inertial_range}, is only observed while  the local anisotropy analyses is used \citep{horbury08}. Such analysis  consists in following the magnetic field direction as it varies in space and scale, which may cause the measured spectra to contain higher order correlations \citep{matthaeus12}.

%%%%%%%% Discussion with referee and Rob W. %%%%%%%%%%%%%%

 The importance of the local field for the turbulence anisotropy analysis has been pointed out already in \citep{cho00,maron01,milano01}. 
The method proposed by \citep{horbury08}, and used by \citet{wicks10} in Fig.~\ref{fig:inertial_range},  is equivalent in some sense to the one used in \citet{milano01} for numerical simulations, but can appear contradictory with the requirement of the ergodic theorem (equivalence between space and time averaging)\footnote{In order to insure the equivalence between space and time averaging, the average should be taken over several correlation lengths, i.e. several energy injection lengths.}.  
However, there are practical implications that have to be considered: an individual packet of plasma passes a spacecraft once and never returns, meaning that the average magnetic field direction over many correlation lengths measured from a time series is not necessarily representative of the actual magnetic field direction at any point. Rather than taking simple time averages, here the local magnetic field direction (and local $\theta_{BV}$) to each fluctuation is measured,  and then fluctuations that have similar directions are averaged. 
Precisely, in Fig.~\ref{fig:inertial_range}, \citet{wicks10} used many hundreds of observations in each direction, so the ergodicity is met, but in a non-conventional way.

%%%%%%%%

Beyond the anisotropy of the fluctuations with respect to the magnetic field direction, \cite{boldyrev06} also suggested that the turbulence  can be anisotropic with respect to the local fluctuation direction -- and  that this anisotropy will be scale dependent. Remarkably,  in the solar wind observations  there is some recent evidence  for the scale-dependent alignment predicted by this theory at large scales \citep{podesta09e} and for the local 3D anisotropy small scales \citep{chen12b}.

The nature of imbalanced turbulence is also a topic of current interest. Alfv\'en waves can propagate parallel or anti-parallel to the magnetic field. Without the presence of both senses, the fluctuations are stable and will not decay. However, the level of imbalance is highly variable in the solar wind (fast wind is typically dominated by Alfv\'enic fluctuations propagating anti-sunward).

\subsubsection*{Velocity fluctuations}
\begin{figure}%[!htb] 
\centering
\includegraphics[width=8cm]{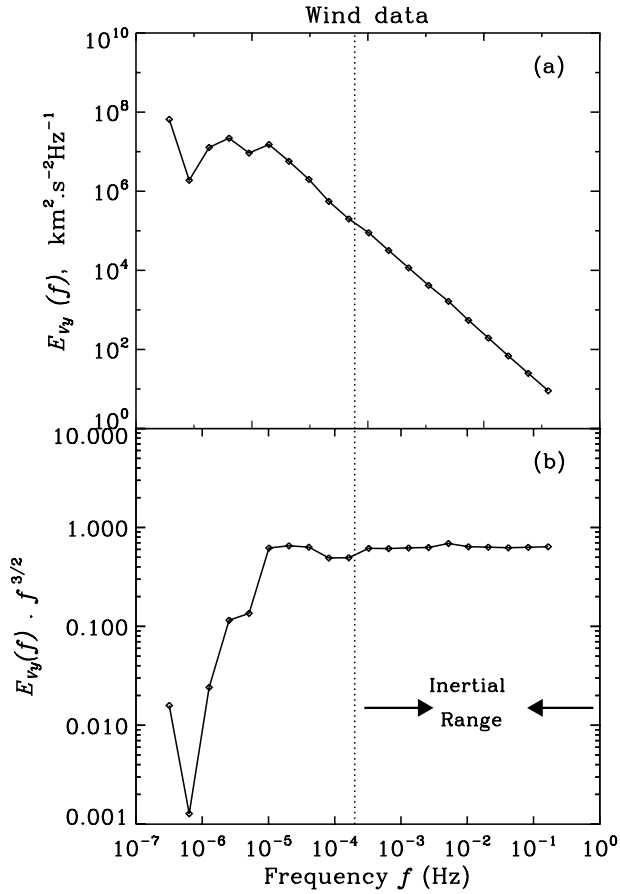}
\caption{(a)  Spectrum of velocity fluctuations of $V_y$ (GSE) component, measured by {\it Wind} as a function of the frequency in the spacecraft frame, the data have been published in \citep{salem09}. (b) Compensated  spectrum by an $f^{3/2}$ law: the resulting function is flat for $f>10^{-4}$~Hz.  Courtesy of C.~Salem. }
\label{fig:v_spec}
\end{figure}

 Velocity fluctuations in the solar wind appear to have a spectrum  significantly shallower than the magnetic field, with a spectral index near $-3/2$ \citep{grappin91,salem00,mangeney01,podesta07a,salem09,chen11b,chen13a,boldyrev11a,borovsky12_bv}. 
Fig.~\ref{fig:v_spec} shows (a) a velocity spectrum  and (b) a compensated spectrum with the $f^{3/2}$ function obtained from {\it Wind} measurements using the Haar wavelet technique  \citep{salem09}.  
Such a spectrum was predicted by the IK phenomenology for Alfv\'enic fluctuations propagating  in opposite directions along ${\bf B}$. However, in this model, both magnetic field and velocity spectra are expected to follow the $\sim k^{-3/2}$ power-law. 
  The difference of the solar wind inertial range with a pure Alfv\'enic turbulence described in the IK model (and with the anisotropic Goldreich-Sridhar model) is also an excess of magnetic energy with respect to the kinetic energy, see Fig.~8 in \citep{salem09}.  How can the  difference between the velocity and the magnetic spectra, and the excess of magnetic energy  in the solar wind, be explained? 
 Direct simulations of incompressible MHD usually show an excess of magnetic energy as well. It  has been attributed to a local dynamo effect which balances the linear Alfv\'en effect \citep{grappin83}. The difference between magnetic and kinetic energies is usually called in the literature ``residual energy''. The residual energy has been shown to follow a definite scaling which is related to the scaling of the total energy spectrum \citep{grappin83,muller05}, see also \citep{boldyrev09a,boldyrev12a,chen13a}. 

Another possible explanation of the difference between the observed magnetic and velocity spectra can be related to the presence of compressible fluctuations, not negligible for the energy exchange between scales.

\subsubsection*{Density fluctuations}
\begin{figure}%[!htb] 
\centering
\includegraphics[width=8cm]{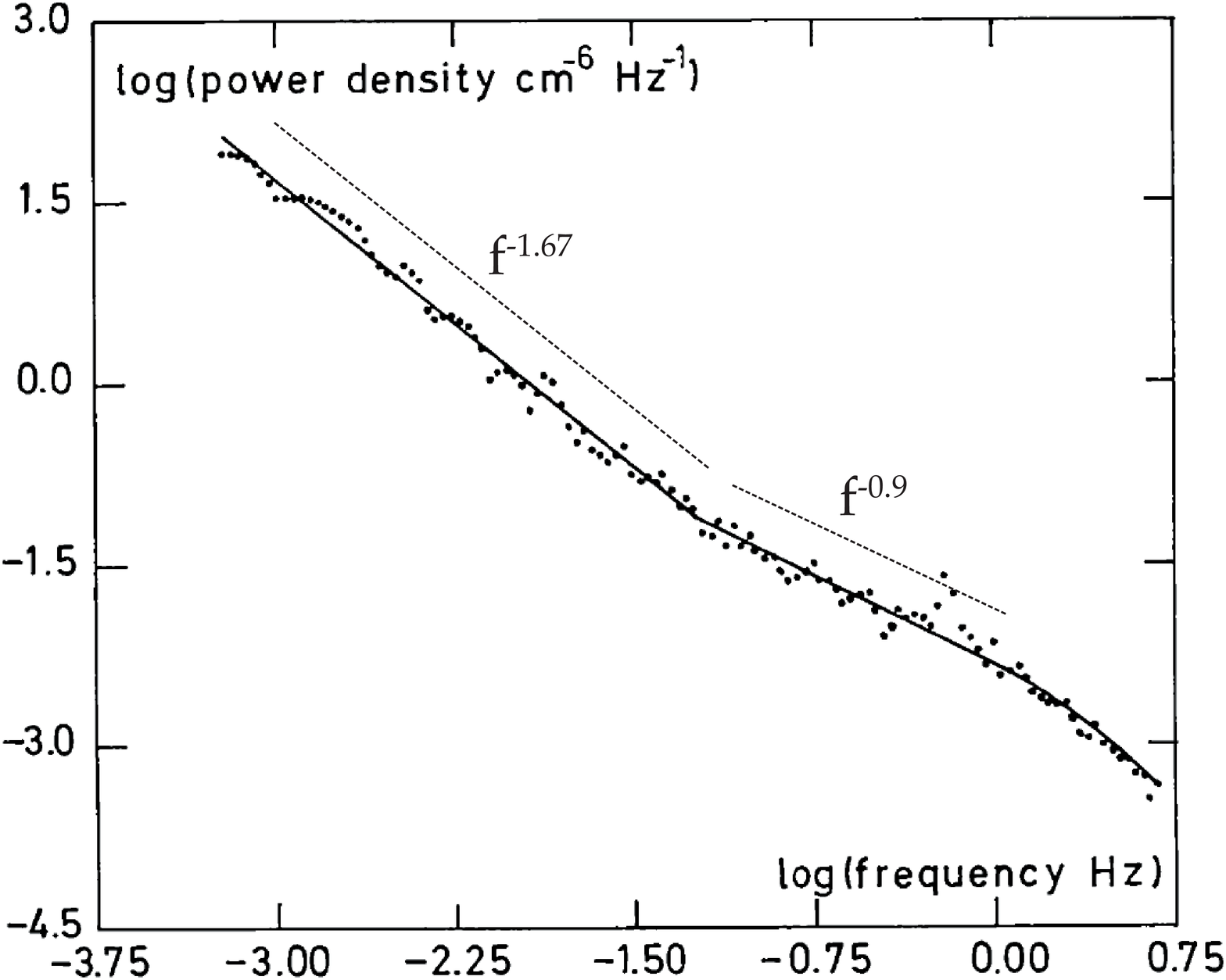}
\caption{Spectrum of electron density fluctuations $n_e$ measured by the {\it ISEE 1-2} spacecraft: two distinct power-laws are observed, the spectrum follows $\sim f^{-1.67\pm 0.05}$ within the frequency range $[10^{-3},6\cdot 10^{-2}]$~Hz, the spectrum is about $f^{-0.9\pm0.2}$ at $f>6\cdot 10^{-2}$~Hz. Around $1-2$~Hz the spectrum seems to change again, however,  this high frequency range is too narrow to make any firm conclusion (the maximal measured frequency is $5$~Hz). Figure from  \citep{celnikier83}. }
\label{fig:n_spec}
\end{figure}

Turbulent fluctuations within the inertial range are not only anisotropic in space (or in ${\bf k}$), but as well in their amplitudes with respect to ${\bf B}$. As we have discussed in the introduction, the non-compressive, Alfv\'enic turbulence dominates the solar wind at MHD scales, $\delta B_{\perp} \gg \delta B_{\|}$. Nevertheless, there is  a sub-dominant population of $\delta B_{\|}$ and density $\delta \rho$ fluctuations always present, with scaling properties suggestive of a turbulent cascade \citep{celnikier83,marsch90b,manoharan94,kellogg05,hnat05,issautier10,chen11b}. 
Fig.~\ref{fig:n_spec}   shows an example of an electron density spectrum measured by the {\it ISEE 1-2} satellites in the $[6\cdot 10^{-4},5]$~Hz frequency range  \citep{celnikier83}. At MHD scales, $f<10^{-1}$~Hz, the K41 scaling is observed.  At higher frequencies, i.e. around ion scales, one observes a spectrum flattening and then another steep spectrum. These high-frequency features will be discussed in more details in Section~3.

The origin of the compressible fluctuations  in the solar wind is not clear, as far as  fast and slow mode waves are strongly damped at most propagation angles. \citet{howes12} have recently argued, based on the dependence of the $\delta B_{\|}$-$\delta \rho$  correlation on the plasma beta $\beta$ (ratio of thermal  to magnetic pressure), that these fluctuations are slow mode and they appear to be anisotropic in wave-vectors \citep{he11a}. \citet{chen12b} measured the $\delta B_{\|}$ fluctuations to be more anisotropic than the Alfv\'enic component in the fast solar wind, suggesting this as a possible reason why they are not heavily damped \citep{schekochihin09}.  \citet{yao11} observe a clear anti-correlation between electron density and the magnetic field strength at different time scales (from 20~s to 1~h): the authors interpret their observations as multi-scale pressure-balanced structures  which may be stable in the solar wind.  This interpretation is consistent with the observation of intermittency in electron density fluctuations by the {\it Ulysses} spacecraft \citep{issautier10}.

\subsection{Intermittency}

\begin{figure}[!htb] 
\centering
\includegraphics[width=8cm]{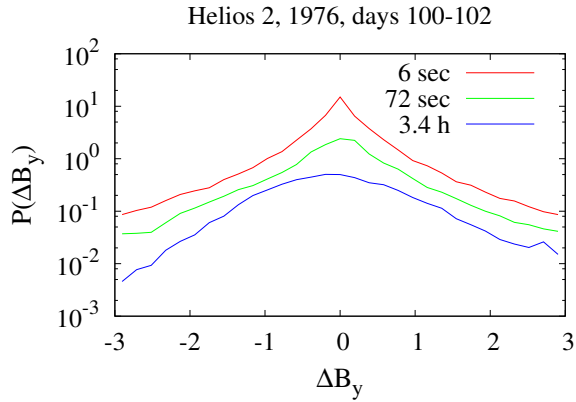}
\caption{Probability distribution functions (PDFs) of the tangential component of the standardized magnetic field fluctuations $\Delta B_y$ (in RTN coordinates) computed for three different time lags, as indicated in the legend. PDFs were estimated using 6 second \emph{Helios 2} data recorded in a stationary slow wind stream near $0.3$~AU on days 100 to 102 of 1976.  The data used here were published previously in \citet{bruno04}.}
\label{fig:pdf}
\end{figure}

In hydrodynamics, the amplitude of the fluctuations at a given scale -- and hence the local energy transfer rate -- is variable, a property known as intermittency,  i.e. turbulence  and its dissipation are  non-uniform in space  \citep{frisch95}. This results in the turbulence being bursty, which can be easily seen from the test of regularity of turbulent fluctuations \citep{mangeney12}. Usually, turbulent fluctuations at different time scales $\tau$ are approximated by increments calculated at these scales, $\delta y_{\tau}=y(t+\tau)-y(t)$. The time averages of these increments are called ``structure functions'' (for more details see the paper by \citet{dudokdewit13} in this book). In the presence of intermittency, the scaling of higher order moments of the structure functions diverges from the simple linear behaviour expected for non-intermittent, Gaussian fluctuations: in essence, at smaller scales, there are progressively more large jumps, as the turbulence generates small scale structures. This behaviour is also observed in the solar wind on MHD scales \citep{burlaga91,tu95,carbone95,sorriso-valvo99,veltri99a,veltri99b,salem00,mangeney01,bruno01,sorriso-valvo01,hnat03,veltri05,bruno05a,leubner05,jankovicova08,greco09,greco10,sorriso-valvo10}. Fig.~\ref{fig:pdf} shows probability distribution functions (PDF) of the tangential component of the  standardized magnetic field fluctuations $\Delta B_y=\delta B_y/\sigma(\delta B_y)$,  $\sigma$ being the standard deviation of $\delta B_y$ (in RTN coordinates\footnote{R is the radial direction, N is the normal to the ecliptic plane and T completes the direct frame.}) computed for three different time scales $\tau$. Intermittency results in the change of shape, from the large scale Gaussian to the small scale Kappa functions. 

 Intermittency is a crucial ingredient of turbulence. Being related to the full statistical properties of the fields, its characterization can give an important insight on the nature of turbulence and on possible dissipation mechanisms of turbulent energy. 

Note, as well, that as far as the third-order moment of fluctuations is related to the energy dissipation rate and is different from zero (see the K4/5 law, equation~(1)), turbulence must shows some non-gaussian features.  

Solar wind observations have shown that the intermittency of different fields can be remarkably different. In particular, it has been observed in several instances that the magnetic field is generally more intermittent than the velocity \citep{sorriso-valvo99,sorriso-valvo01}. The possibility that this implies that magnetic structures are passively convected by the velocity field has been discussed, but no clear evidence was established, so that this is still an open question \citep{bershadskii04,bruno07}. 

The use of data from {\it Helios 2} spacecraft, which  explored the inner heliosphere reaching about $0.3$ AU, has allowed to study the radial evolution of intermittency, and its dependency on the wind type (fast or slow) \citep{bruno03}. The fast wind has revealed an important increase of intermittency as the wind blows away from the Sun, while the slow wind is less affected by the radial distance $R$. This suggests that some evolution mechanism must be active in the fast solar wind. This could be either due to the slower development of turbulence in the fast wind, with respect to the slow wind, or to the presence of superposed uncorrelated Alfv\'enic fluctuations, which could hide the structures responsible for intermittency in the fast wind closer to the Sun. These uncorrelated Alfv\'enic fluctuations, ubiquitous in the fast wind, are indeed observed to decay with $R$, as suggested for example by a parametric instability model \citep{malara00,malara01,bruno03,bruno04}.

\begin{figure}
\centering
\includegraphics[width=8cm,angle=90]{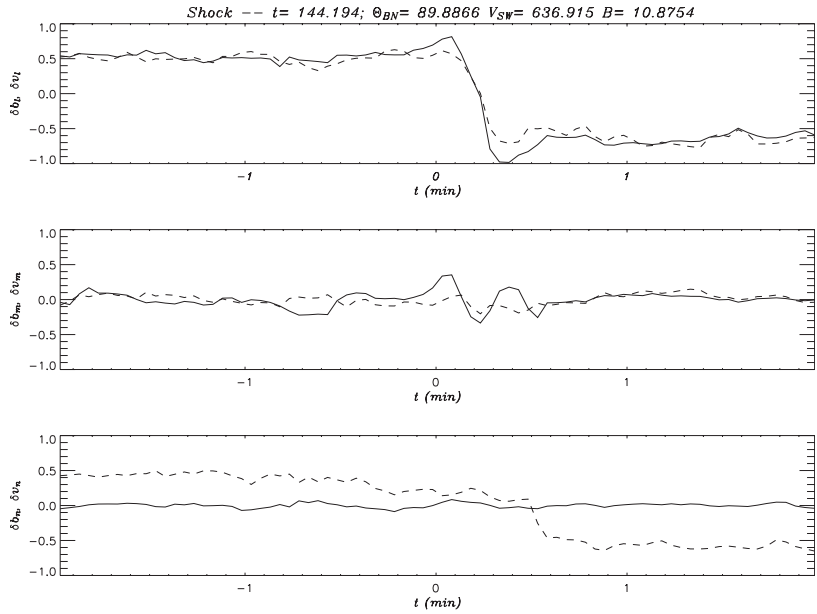}
\caption{Example of a coherent structure responsible for the non-Gaussian PDF tails in Fig.~\ref{fig:pdf} at small scales: a quasi-perpendicular shock wave at a time scale  of the order of $\tau=30$~sec. Measurements of $\delta {\bf B}$ in the local minimum variance frame (solid lines) and velocity fluctuations $\delta {\bf v}$ in the same frame  (dashed lines) as measured by {\it Wind} satellite in the fast solar wind  (courtesy of C.~Salem). }
\label{fig:struc}
\end{figure}
The ultimate responsible for emergence of intermittency are strong fluctuations of the fields with coupled phases over a finite range of scales. These are often referred to as coherent structures. 
Fig.~\ref{fig:struc} shows an example of a coherent structure responsible for the non-Gaussian PDF tails in Fig.~\ref{fig:pdf} at  small scales: a shock wave with its normal quasi-perpendicular to the local mean field \citep{veltri99a,salem00,veltri05}.   This kind of structures may be responsible for the dissipation of turbulent energy in the collisionless solar wind.

A complication in the solar wind is that sharp structures, discontinuities, are ubiquitous. Discontinuities typically involve a rotation in the magnetic field direction, and sometimes variations in  velocity, field magnitude and other plasma properties such as density and even temperature and composition \citep{owens11}. Parameters such as composition  do not change much  after the wind leaves the solar corona, so these might have been generated at its source. However, the vast majority of structures have no such signature: are these also part of the structure of the solar wind \citep{borovsky08}, or are they generated dynamically by the turbulence \citep{greco09,greco12}? These structures seem to be associated with enhanced temperature of the solar wind \citep{osman11a,wu13}, so they might represent a source of energy dissipation via reconnection or enhanced damping. Discontinuities, as sharp jumps, also contribute to the  intermittency of the solar wind turbulence. To what extent is the observed intermittency inherent to the plasma turbulence, therefore, as opposed to being an artifact of its generation in the corona? This is a currently unresolved issue and the topic of many recent works \citep{servidio11_npg,servidio12,zhdankin12_prl,borovsky12_mixing,osman12_kin,wu13,karimabadi13}.

\subsection{Energy Transfer Rate}

As we have mentioned in the introduction,  any turbulent flow is characterized by power-law energy spectra, presence of intermittency and linear dependence between the third order structure function and scale. This last property is the only exact result for hydrodynamic turbulence, known as the K4/5 law, see equation (\ref{eq:K4/5}). In plasmas, the incompressible MHD version of the K4/5 law has been obtained by \citet{politano98} by using the Elsasser fields ${\bf Z^\pm}(t)={\bf  v}(t)\pm{\bf  b}(t)/\sqrt{4\pi\rho}$  in place of velocity $\delta v$ in equation~(\ref{eq:K4/5}) (${\bf v}(t)$ and ${\bf b}(t)$ being the time dependent solar wind velocity and magnetic field).

The MHD equations can be conveniently written in terms of Elsasser variables ${\bf Z^{\pm}}$ as
\begin{equation}
\frac{\partial {\bf Z^{\pm}}}{\partial t} + \left({\bf Z^\mp} \cdot \nabla\right) {\bf Z^\pm} = -\nabla P + \eta^\prime\nabla^2 {\bf Z^{\pm}} \; ,
\label{mhd}
\end{equation}
where $P$ is the total pressure (magnetic plus kinetic), and $\eta^\prime=\eta=\nu$ is a dissipation coefficient\footnote{For simplicity, resistivity $\eta$ is assumed to be equal to viscosity $\nu$.}. Non-linear terms $\left({\bf Z^\mp} \cdot \nabla\right) {\bf Z^\pm}$ in equation~(\ref{mhd}) are responsible for the transfer of energy between fluctuations at different scales, originating the turbulent cascade and the typical Kolmogorov spectrum.
 The MHD version of the  K4/5 law for $\Delta Z^+$ is obtained by subtracting the equation (\ref{mhd}) for $Z^-$ from the one for $Z^+$, evaluated at two generic points separated by the scale $\ell=V_{sw}\tau$ along the flow direction, and then by multiplying the result by $\Delta Z^+$. 

This provides an evolution equation for the pseudo-energy flux\footnote{The pseudo-energy refers to the fact that the Elsasser fields, ${\bf Z^+}$ and  ${\bf Z^-}$, are pseudo-vectors. The pseudo-energy associated to each Elsasser variable, $\epsilon^{\pm}$, is not an invariant of the flow. An invariant of the flow is the total energy $(\epsilon^++\epsilon^-)/2$.}, which includes terms accounting for anisotropy, inhomogeneity and dissipation. Under the hypotheses of isotropy, local homogeneity and vanishing dissipation (i.e. within the inertial range, far from the dissipation scale), the simple linear relation can be retrieved in the stationary state \citep{politano98}:
\begin{equation}\label{eq:PP}
Y^\pm(\tau) = \left\langle |\Delta {\bf Z}^\pm(\tau,t)|^2\, \Delta Z^\mp_R(\tau,t)\right\rangle = {4 \over 3} \,\epsilon^\pm\, \ell,
\end{equation}
where  $Z^\mp_R$ is the radial component (i.e., along the mean solar wind flow ${\bf V}_{sw}$) of the Elsasser fields.  For a detailed description of the derivation, see e.g. \citep{danaila01,carbone09b}.

The turbulent cascade pseudo-energy fluxes~$\epsilon^{\pm}$ are defined as the trace of the dissipation rate tensors 
 $$\epsilon_{ij}^{\pm} = \eta \langle (\partial_i Z_i^{\pm}) (\partial_i Z_j^{\pm}) \rangle).$$ 
$\epsilon^{\pm}$ describe the energy transfer rate   and dissipation rate between the Elsasser field structures on scales within the inertial range of MHD turbulence.

\begin{figure}
\centering
\includegraphics[width=8cm]{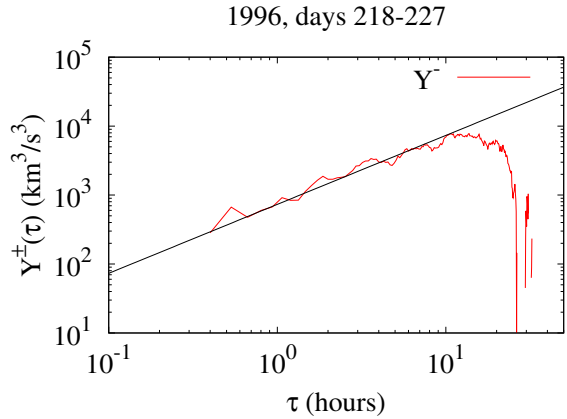}
\caption{The third order moment linear scaling law as evaluated in the 11 day  time interval starting on day 218 of 1996, during the high latitude scan of {\it Ulysses} spacecraft. The heliocentric distance was $4.2$~AU, the heliolatitude was $30^\circ$, and the mean wind speed of the sample was $735$~km/s. The linear fit predicted by the law (\ref{eq:PP}), is indicated. For this sample, the  pseudo-energy transfer rate is estimated to be $\epsilon^-=212$~J kg$^{-1}$s$^{-1}$. }
\label{fig:yaglom}
\end{figure}

The relation (\ref{eq:PP}) was first observed in numerical simulations of two dimensional MHD turbulence \citep{sorriso-valvo02,pietarila06}, and later in solar wind samples \citep{macbride05,sorriso-valvo07,macbride08}, despite the observational difficulties \citep{podesta09c} and the fact that solar wind turbulence is not isotropic (Section \ref{scalingandanisotropy}). An example of linear scaling from {\it Ulysses} high latitude data is shown in Fig.~\ref{fig:yaglom}.

The observation of the third order moment scaling is particularly important, since it suggests the presence of a (direct or inverse) turbulent cascade\footnote{The sign of the coefficient $\epsilon$ will give the direction of the cascade (i.e. the cascade is inverse for negative energy flux).} as the result of nonlinear interactions among fluctuations. It also suggests that solar wind turbulence is fully developed, as the dissipative effects have to be neglected in order to observe the linear scaling. It defines rigorously the extension of the inertial range, where a Kolmogorov like spectrum can be expected. In solar wind, the inertial range, as defined by the law of \citet{politano98}, equation~(\ref{eq:PP}), is found to be extremely variable, and can reach scales up to one day or even more \citep{sorriso-valvo07,marino12}, much larger than usually assumed following typical estimates from  the analysis of turbulent spectra. The variability  of the inertial range extension, i.e. the range of scales where the linear relation (\ref{eq:PP}) is observed,  is in agreement with earlier multifractal analysis of solar wind fluctuations \citep{burlaga93}. Moreover, recent results, obtained through conditioned analysis of solar wind fluctuations, have confirmed that, for high cross-helicity states, i.e. when $\langle{\bf v\cdot b}\rangle/(\langle v^2\rangle+\langle b^2\rangle)$ is high, the inertial range observed in the spectrum extends to such larger scales \citep{wicks13}.   It will be interesting as well to verify the influence of the solar wind expansion time $\tau_{exp}$ (in comparison with the non-linear time) on the extension of the inertial range (see our discussion in the introduction).

The third order moment law provides an experimental estimate of the mean energy transfer rates $\epsilon^{\pm}$, a measurement which is not possible otherwise, as the solar wind dissipation mechanisms (and so the viscosity $\eta$) are unknown. Solar wind energy transfer rates have been shown to lie between $\sim 0.1$~kJkg$^{-1}$s$^{-1}$ (in {\it Ulysses} high latitude fast wind data, far from the Earth) and up to $\sim 10$~kJkg$^{-1}$s$^{-1}$ in slow ecliptic wind at 1~AU \citep{sorriso-valvo07,marino08,macbride08,marino12,smith09}.
The rate of occurrence of the linear scaling in the solar wind time series, and the corresponding energy transfer rate, have been related to several solar wind parameters. For example, the energy transfer rate has been shown to anti-correlate with the cross-helicity level \citep{smith09,stawarz10,marino11,podesta11c,marino12}, confirming that alignment between velocity and magnetic field reduces the turbulent cascade, as expected for Alfv\'enic turbulence \citep{dobrowolny80,boldyrev06}.  Relationships with heliocentric distance and solar activity have also been pointed out, with controversial results \citep{marino11,marino12,coburn12}.

The estimation of the turbulent energy transfer rate has also shown that the electromagnetic turbulence may explain the observed solar wind non-adiabatic  profile of the  total proton temperature \citep{vasquez07,marino08,macbride08,stawarz09}. However,  this explanation does not take into account a possible ion temperature anisotropy,  known to be important in the solar wind (see Section 3.2). Indeed, the weakly collisional protons exhibit important temperature anisotropies (and complicated departures from a Maxwellian shape, \cite{marsch82}) and they have non double-adiabatic temperatures profiles. {\it Helios} observations indicate that protons need to be heated in the perpendicular direction from 0.3 to 1 AU, but   in the parallel direction they need to be cooled at 0.3~AU. This cooling rate gradually transforms to a heating rate at 1 AU \citep{hellinger11,hellinger13}. It is not clear if the turbulent cascade may cool the protons in the parallel direction (and transform this cooling to heating by 1~AU).

The phenomenological inclusion of possible contributions of density fluctuations to the turbulent energy transfer rate resulted in enhanced energy flux, providing a more efficient mechanism for the transport of energy to small scales \citep{carbone09}.

Anisotropic corrections to the third order law have also been explored using anisotropic models of solar wind turbulence \citep{macbride08,carbone09b,stawarz09,stawarz10,macbride10,osman11a}.

It is important to keep in mind that the solar wind expansion, the large scale velocity shears and the stream-stream interactions importantly affect the local turbulent cascade \citep{stawarz11,marino12}.  Their effect on the turbulent energy transfer rate needs to be further investigated \citep{wan09,hellinger13}.

\section{Turbulence at Kinetic Scales}

At 1~AU, the MHD scale cascade finishes in the vicinity of ion characteristic scales $\sim 0.1-0.3$~Hz in the spacecraft frame. Here the turbulent spectra of plasma parameters (magnetic and electric fields, density, velocity and temperature) change their shape, and  steeper  spectra are observed at larger wave-numbers or higher frequencies, e.g. \citep{leamon98a,bale05,alexandrova07,chen12a,safrankova13}.  There is a range of terminology used to describe this range, including ``dissipation range'', ``dispersion range'' and ``scattering range''. The possible physics taking place here includes dissipation of turbulent energy \citep{leamon98a,leamon99,leamon00,smith06a,schekochihin09,howes11a}, a further small scale turbulent cascade \citep{biskamp96,ghosh96,stawicki01,li01,galtier06b,alexandrova07,alexandrova08b,schekochihin09,howes11a,rudakov11,boldyrev12b} or a combination of both.

The transition from the MHD scale cascade to the small scale range is sometimes called the {\it ion spectral break} due to the shape of the magnetic field spectrum and to the scales at which it occurs. %at 1~AU. 
The physical processes responsible for the  break and the corresponding characteristic scale are under debate.  If the MHD scale cascade was filled with parallel propagating Alfv\'en waves, the break point would be at the ion cyclotron frequency $f_{ci}$, where the parallel Alfv\'en waves undergo the cyclotron damping.  The oblique kinetic Alfv\'en wave (KAW) turbulence is sensitive to the ion gyroradius $\rho_i$ \citep{schekochihin09,boldyrev12b} and the transition from MHD to Hall MHD occurs at the ion inertial length $\lambda_i$ \citep{galtier06b,servidio07,matthaeus08b,matthaeus10a}.

Recent {\it Cluster} measurements of magnetic fluctuations up to several hundred Hz in the solar wind \citep{alexandrova09a,sahraoui10,alexandrova12} show the presence of another spectral change at  electron scales. At scales smaller than electron scales, the plasma turbulence is expected to convert from electromagnetic to electrostatic (with the important scale being the Debye length, see, e.g., \citep{henri11}), but this is beyond the  scope of the present paper. 

The energy partitioning at kinetic scales, the spectral shape and the properties of the small scale cascade are important for understanding the dissipation of electromagnetic turbulence in collisionless  plasmas.

\subsection{Turbulence Around Ion Scales}

\begin{figure}
\centering
\includegraphics[width=7.cm]{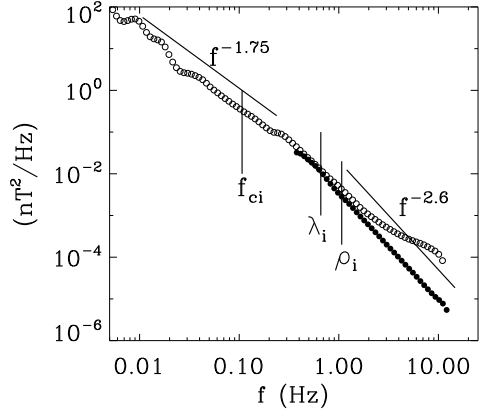}
\caption{Wavelet spectrum of magnetic fluctuations measured by {\it Cluster} in the solar wind  up to 12.5~Hz for the time interval analyzed in \citep{alexandrova08b}. The {\it Cluster}/FGM spectrum is represented by open circles, {\it Cluster}/Search Coil (STAFF-SC) spectrum, by filled circles. The characteristic ion scales are marked by vertical bars.}
\label{fig:spec-fgm-staff}
\end{figure}

Fig.~\ref{fig:spec-fgm-staff} shows an example of the solar wind magnetic field spectrum covering the end of the MHD inertial range and ion scales. The data are measured at 1~AU by {\it Cluster}/FGM (open circles) and {\it Cluster}/STAFF-SC (filled circles), which is more sensitive than FGM at high frequencies. One may conclude that the transition from the inertial range to another power-law spectrum is around ion scales, such as the ion cyclotron frequency $f_{ci}=0.1$~Hz, the ion inertial scale $\lambda_i$ corresponding to $f_{\lambda_i}=V_{sw}/(2\pi\lambda_i)\simeq 0.7$~Hz and the ion Larmor radius $\rho_i$ appearing at $f_{\rho_i}=V_{sw}/(2\pi\rho_i)\simeq 1$~Hz.  However, which of these ion scales is responsible for the spectral break is not evident from Fig.~\ref{fig:spec-fgm-staff}. 

\begin{figure}
\centering
\includegraphics[width=12.cm]{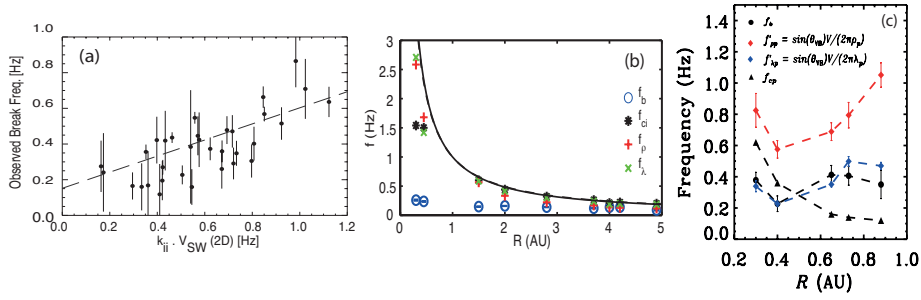}
\caption{(a) Observed  ion break frequency $f_b$ as a function of $f_{\lambda_i}=V_{sw}\sin\theta_{BV}/2\pi\lambda_i$, a correlation of 0.6 is observed \citep{leamon00}. (b) Radial evolution of $f_b$ compared with the radial evolution of $f_{ci}$, $f_{\rho}=V_{sw}/2\pi\rho_i$ and $f_{\lambda}=V_{sw}/2\pi\lambda_i$: none of the ion scales follow the break \citep{perri10b}. (c) Radial evolution of  $f_b$ (black dots) compared with $f_{ci}$ (black triangles), $f_{\rho_p}=\sin\theta_{BV}V_{sw}/2\pi\rho_p$ (red diamonds) and $f_{\lambda_p}=\sin\theta_{BV}V_{sw}/2\pi\lambda_p$ (blue diamonds) \citep{bourouaine12}.}
\label{fig:break}
\end{figure}

\citet{leamon00}  performed a statistical study of the spectral break values $f_b$  at 1~AU for different ion beta conditions, $\beta_i=nkT_i/(B^2/2\mu_0)\in [0.03,3]$\footnote{In this study, the authors used the statistical sample from \citep{leamon98a}, i.e.,  33 turbulent spectra up to $\sim 3$~Hz measured by {\it Wind} spacecraft within the slow and fast streams, $V_{sw} \in [300,700]$.}. The best correlation is found with the ion inertial length while taking into account the 2D nature of the turbulent fluctuations, i.e. $k_{\perp}\gg k_{\|}$, see Fig.~\ref{fig:break}(a).  A larger statistical sample of 960 spectra shows the dependence between $f_b$, and $f_{\lambda_i}\frac{B}{\delta B_b}$, where $\delta B_b/B$ is the relative amplitude of the fluctuations at the break scale \citep{markovskii08}. This result is still not explained. But, it is important to keep in mind that $\delta B_b/B$ is controlled by the ion instabilities in the solar wind when the  ion pressure is sufficiently anisotropic \citep{bale09}, see Section~\ref{sec:instab} for more details. 

A different approach has been used by \citet{perri10b}: the authors studied the radial evolution of the spectral break for distances $R\in [0.3, 5]$~AU. They showed that the ion break frequency is independent of the radial distance (see Fig.~\ref{fig:break}(b)). \citet{bourouaine12} explained this  result by the quasi-bidimensional topology of the turbulent fluctuations, i.e. $k_{\perp}\gg k_{\|}$. When this wave vector anisotropy is taken into account, the Doppler shifted frequency $2\pi f={\bf k\cdot V_{sw}}$ can be approximated by $kV_{sw}\sin\theta_{BV}$.  It appears that the ion inertial scale stays in the same range of frequencies as $f_b$, and a correlation of 0.7 is observed between $f_b$ and $f_{\lambda_i}=V\sin\theta_{BV}/2\pi\lambda_i$, see Fig.~\ref{fig:break}(c).

As we have discussed above, the transition to kinetic Alfv\'en turbulence happens at the ion gyroradius $\rho_i$ scale \citep{schekochihin09,boldyrev12a}, while the dispersive Hall effect becomes important  at the ion inertial length $\lambda_i$. Results of \citet{leamon00} and \citet{bourouaine12}  indicate, therefore, that the Hall effect may be responsible for the ion spectral break.  
Note that \citet{bourouaine12} analyzed {\it Helios} data only within fast solar wind streams with $\beta_i<1$, i.e. when $\lambda_i>\rho_i$\footnote{Ion plasma beta can be expressed in terms of ion scales: $\beta_i = 2\mu_0nkT_i/B^2=\rho_i^2/\lambda_i^2$}.  It is quite natural that the largest characteristic scale (or the smallest characteristic wave number) affects the spectrum first \citep{spangler90}. 
It will be interesting to verify these results for slow solar wind streams and high $\beta_i$ regimes.

Just above the break frequency, $f>f_b$, the spectra are quite variable. \citet{smith06a} show that  within a narrow frequency range $[0.4-0.8]$~Hz, the spectral index $\alpha$ varies  between $-4$ and $-2$.   This result was obtained using {\it ACE}/FGM measurements. However, one should be very careful while analyzing FGM data at frequencies higher than the ion break (i.e. at $f > 0.3$~Hz), 
where the digitalisation noise becomes important \citep{lepping95_wind,smith98_ace,balogh01_cluster}. For example, in Fig.~\ref{fig:spec-fgm-staff} the {\it Cluster}/FGM spectrum deviates from the STAFF spectrum at $f \geq 0.7$~Hz\footnote{The digitalisation noise at {\it Cluster}/FGM and at {\it ACE}/FGM is nearly the same, see \citep{smith98_ace,balogh01_cluster}.}.  

\begin{figure}
\centering
\includegraphics[width=8.cm]{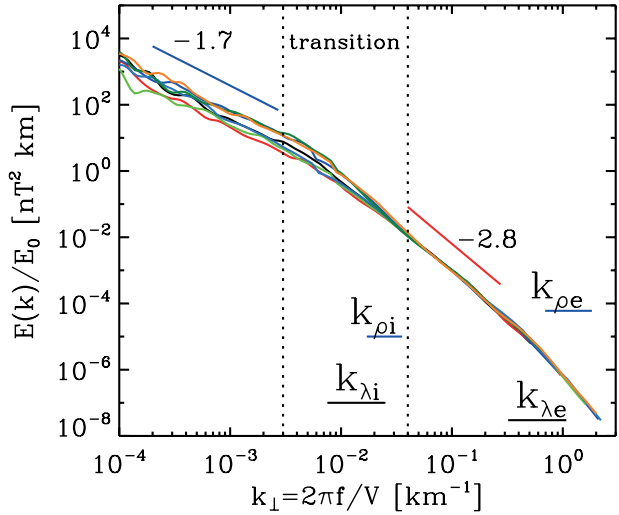}
\caption{7 solar wind spectra, analyzed in \citep{alexandrova09a,alexandrova10} under different plasma conditions as a function of the wave-vector $k_{\perp}$ perpendicular to the magnetic field. The spectra are superposed  with a normalization factor $E_0$ at scales smaller than all ion scales: one observes divergence of the spectra in the transition range around the ion scales $k_{\rho_i}$ and $k_{\lambda_i}$. }
\label{fig:spec-ii-scales}
\end{figure}
Fig.~\ref{fig:spec-ii-scales} shows several combined spectra, with {\it Cluster}/FGM data at low frequencies and {\it Cluster}/STAFF data at $f>f_b$.  The spectra are shown as a function of the wave-vector $k_{\perp}$\footnote{Cluster stays in the free solar wind not connected to the Earth's bow-shock, while the flow-to-field angle, $\theta_{BV}$, is quasi-perpendicular. Therefore, only $k_{\perp}$ wave vectors are well resolved.}. The spectra are superposed  at $k_{\perp} > k_{\rho_i}, \; k_{\lambda_i}$, i.e. at scales smaller than all ion scales:   while at these small scales all spectra follow the same law, around ion scales $k_{\rho_i}$ and $k_{\lambda_i}$ (named  here a {\it transition range}) one observes a divergence of the spectra.  The origin of this divergence is not completely clear. It is possible that ion damping, e.g. \citep{denskat83,sahraoui10},  a competition between the convective and Hall terms \citep{kiyani13} or ion anisotropy instabilities \citep{gary01,matteini07,matteini11,bale09} may be responsible for the spectral variability within the transition range.

One of the important properties of the transition range  is that the turbulent fluctuations become more compressible here \citep{leamon98a,alexandrova08b,hamilton08,turner11,salem12,kiyani13}. Let us define the level of compressibility of magnetic fluctuations as $\delta B_{\|}^2/\delta B_{tot}^2$, with $\delta B_{tot}^2$ being the total energy of the turbulent magnetic field fluctuations at the same scale as $\delta B_{\|}$ is estimated. If in the inertial range the level of compressibility is about 5\%, for $f>f_b$ it can reach 30\% and it depends on the  plasma beta $\beta_i$ \citep{alexandrova08b,hamilton08}. The increase of the compressibility at kinetic scales has been attributed to the compressive nature of kinetic Alfv\'en or whistler turbulence \citep{gary09,salem12,tenbarge12b}.  On the other hand, it can be described by the compressible Hall MHD  \citep{servidio07}. In particular, in the this framework, 
different levels of compressibility can  also explain the spectral index variations  in the transition range \citep{alexandrova07,alexandrova08b}.

\begin{figure}
\centering
\includegraphics[width=12.cm]{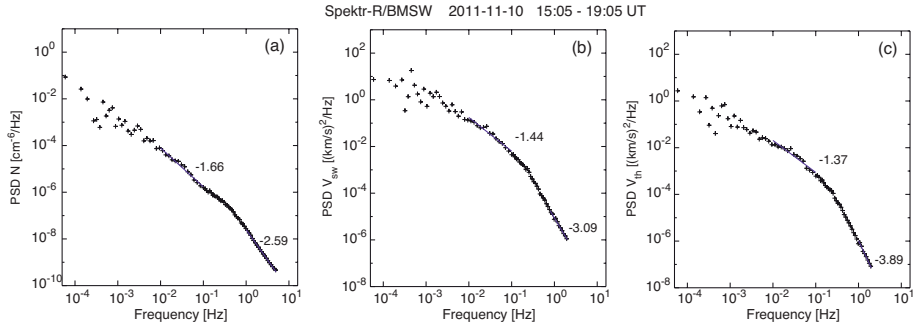}
\caption{  Spectra of ion moments, (a) density, (b) velocity, (c) ion thermal speed,  up to $\sim 3$~Hz as measured by {\it Spektr-R}/BMSW (Bright Monitor of Solar Wind) in the slow solar wind with $V_{sw}=365$~km/s and $\beta_p\simeq 0.2$.  Figure from \citet{safrankova13}. }
\label{fig:spec-v}
\end{figure}

The  flattening of the electron density spectrum from $\sim f^{-5/3}$ to $\sim f^{-1}$, seen in Fig.~\ref{fig:n_spec},  is observed  within the same range of scales as the increase of the magnetic compressibility. The shape of this flattening  is consistent with the transition between MHD scale Alfv\'enic turbulence and small scale KAW turbulence \citep{chandran09c,chen12c}. More recently, \citet{safrankova13} measured the ion density spectrum within the transition range, finding similar results, as expected from the quasi-neutrality condition. In addition, they showed the ion velocity and temperature spectra in this range to be steeper with slopes around $-3.4$.  An example of such spectra is shown in Fig.~\ref{fig:spec-v}. 

The transition range around ion scales is   also characterized by  magnetic fluctuations with quasi-perpendicular wave-vectors $k_{\perp} > k_{\|}$ and a plasma frame frequency close to zero \citep{sahraoui10,narita11,roberts13}. \citet{sahraoui10} interpret these observations as KAW turbulence, although \citet{narita11} found no clear dispersion relation. Magnetic fluctuations with nearly zero frequency and $k_{\perp}\gg k_{\|}$ can also be due to non-propagative coherent structures like current sheets \citep{veltri05,greco10,perri12a}, shocks \citep{salem00,veltri05,mangeney01}, current filaments \citep{rezeau93}, or Alfv\'en vortices propagating with a very slow phase speed $\sim 0.1 V_A$ in the plasma frame  \citep{pp92,alexandrova08a}.  Such vortices are known to be present within the ion transition range of the planetary magnetosheath turbulence, when ion beta is relatively low $\beta_i \leq 1$  \citep{alexandrova06,alexandrova08_grl}. 
 Recent {\it Cluster} observations in the fast solar wind suggest that the ion transition range can be populated with KAWs and Alfv\'en vortices \citep{roberts13}.

\begin{figure}
\centering
\includegraphics[scale=0.25]{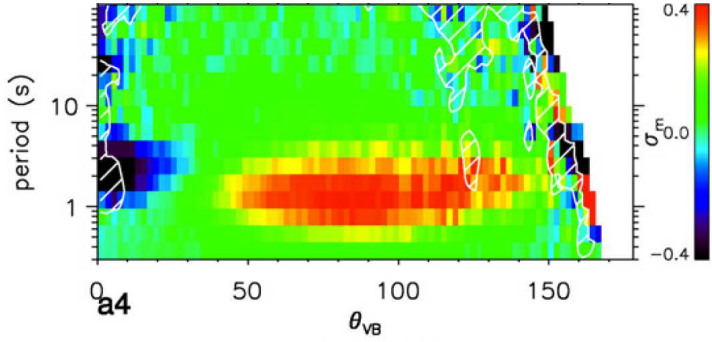}
\caption{Magnetic helicity $\sigma_m$ for an outward magnetic sector as measured by {\it STEREO} spacecraft as a function of time scale $\tau$~(s) and angle to the magnetic field $\theta_{VB}$ \citep{he11b}.}
\label{fig:he}
\end{figure}

As well as the spectrum of energy, the spectrum of magnetic helicity is also used to diagnose solar wind turbulence, and can tell us more details about the nature of the fluctuations \citep{matthaeus82c,howes10}. Magnetic helicity is defined as 
$\left<\mathbf{A}\cdot\mathbf{B}\right>$, where $\mathbf{B}=\nabla\times\mathbf{A}$, with $\mathbf{A}$ being the vector potential.  
It has been measured that at ion scales the magnetic helicity is anisotropic  \citep{he11b}.  Fig.~\ref{fig:he} shows the reduced magnetic helicity\footnote{i.e. the magnetic helicity measured along the satellite trajectory. } $\sigma_m$ as a function of the time scale and of the local flow-to-field angle $\theta_{BV}$.  The authors found that, at time scales corresponding to the ion scales (1 to 10~s), there was a significant positive (negative) magnetic helicity signature for inward (outward) directed magnetic field in the parallel direction (i.e. for $\theta_{BV}$ close to 0 or to 180). This is consistent with left-hand parallel propagating Alfven-ion-cyclotron waves. In the perpendicular direction, $\theta_{BV}\simeq 90^{\circ}$, they found a magnetic helicity signature of the opposite sense: positive (negative) for outward (inward) field, consistent with the right-hand polarization, inherent to both whistler and kinetic Alfv\'en waves. Outside the range of frequencies $(0.1-1)$~Hz, the magnetic helicity was generally zero. \citet{podesta11d} found the same result using {\it Ulysses} data and suggested the source of the parallel waves to be pressure anisotropy instabilities,  which we will now discuss in more details.

%%%%%%%%%%
%%
%%  Ion instability section - SDB - Feb 7 - START
%%
%%%%%%%%%%

\subsection{Ion scale instabilities driven by solar wind expansion and compression}\label{sec:instab}

The turbulent fluctuations, while cascading from the inertial range to the kinetic scales, will undergo strong kinetic effects in the vicinity of such ion scales as the ion skin depth or inertial scale $\lambda_i$, and near the thermal gyroradius $\rho_i$. At these small scales ion temperature anisotropy instabilities can occur \citep{gary01,marsch06,matteini07,matteini11,bale09}, and may remove energy from, or also inject it into, the turbulence.

\begin{figure}%[!htb] 
\centering
\includegraphics[width=12cm]{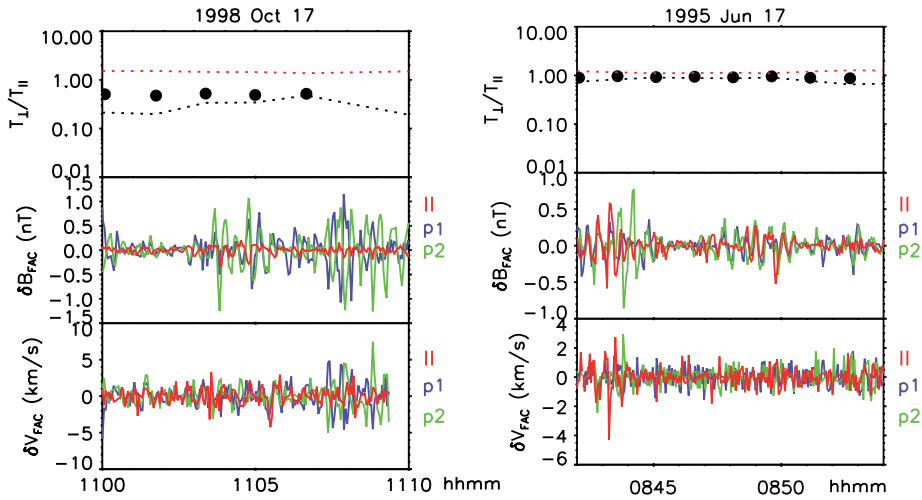}
\caption{ Left, time series data of measured proton temperature anisotropy (dots) and instability thresholds (top panel), of magnetic (2nd panel) and velocity (3rd panel) vector fluctuations in a field-aligned coordinate system (FAC), using 3 second measurements from the {\it Wind}/3DP instrument;  red lines indicate fluctuations parallel  to the mean field ${\bf B}$, $p1$ (violet) and $p2$ (green) represent the two perpendicular components.   As the measured proton anisotropy approaches the oblique firehose instability threshold (black dotted line in the top panel), Aflv\'enic-like fluctuations are excited and visible as perpendicular magnetic and velocity perturbations.  
 Right, the same format as left figure, but for the high ion beta regime, when the plasma conditions were close to both, mirror and firehose instability thresholds: both types of fluctuations, Aflv\'enic-like and compressive, are excited. }
\label{fig:bale_ts}
\end{figure}

\begin{figure}
\centering
\includegraphics[width=8cm]{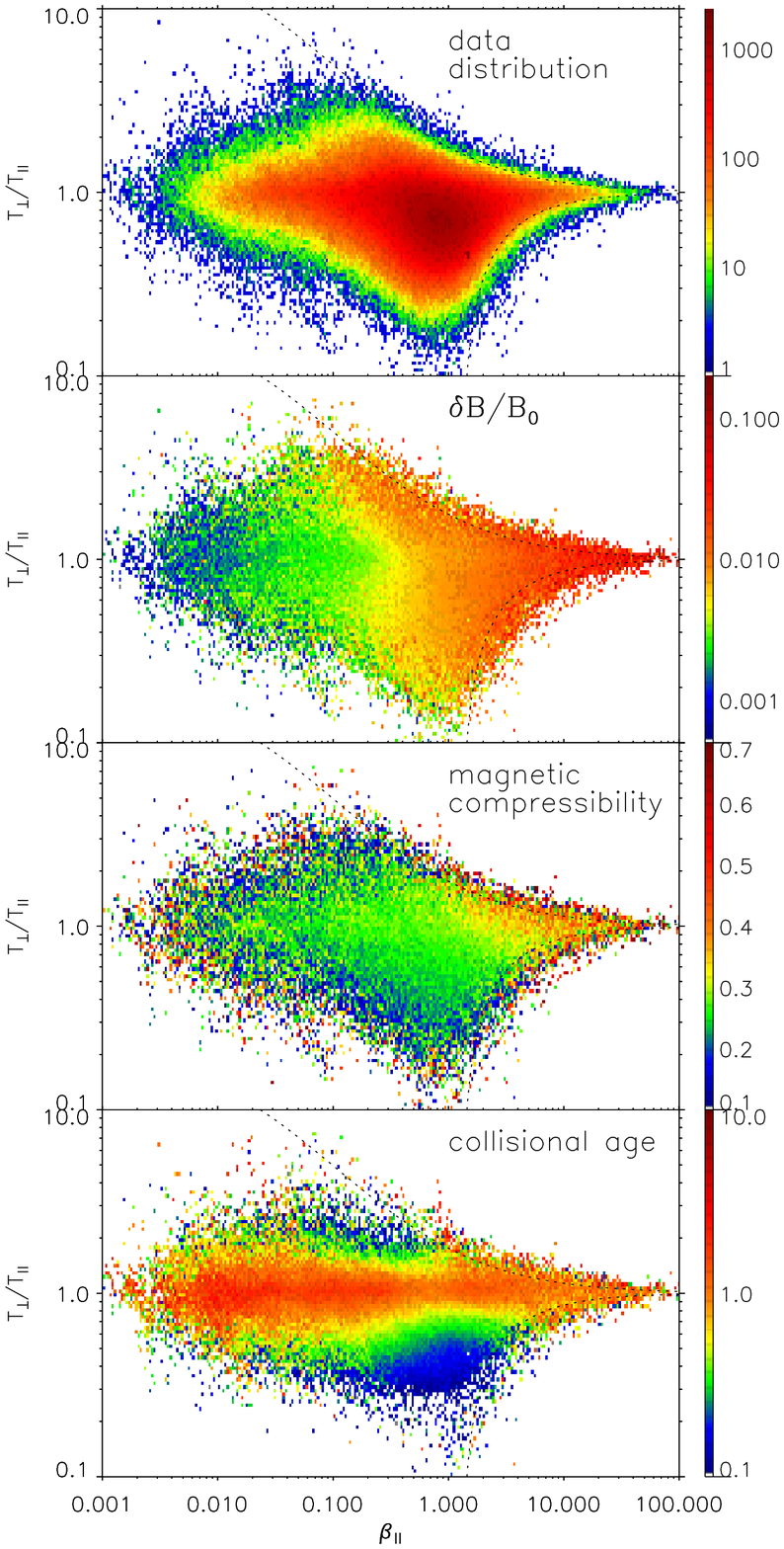}
\caption{Temperature anisotropy $T_\perp/T_{||}$ vs plasma parallel beta $\beta_{||}$ from \citep{bale09}.  The upper panel shows the constraint of plasma by the mirror (upper dashed line) and oblique firehose (lower dashed line) instabilities, as shown by \citet{hellinger06a}.  The second panel shows a statistical enhancement of magnetic fluctuations $\delta B/B$ (calculated at $f=0.3$~Hz, i.e. close to the ion spectral break) near the thresholds and at higher $\beta_{||}$.  The third panel shows the distribution of the
magnetic compressibility $\delta B_{||}/\delta B$ (at ion scales as well) and is consistent with mirror instability near that threshold.  The fourth panel shows the collisional age of the ions (i.e. the number of collisions suffered by a thermal ion between the Sun and the spacecraft at 1~AU) in the same parameter plane.}
\label{fig:bale_figure1}
\end{figure}

As the solar wind expands into space, mass flux conservation leads to a density profile that falls roughly as $1/R^2$ (beyond the solar wind acceleration region); the magnetic field decays similarly, although the solar rotation and frozen flux condition ensure an azimuthal component to the field.  If the solar wind plasma remains (MHD) fluid-like, then the double-adiabatic conditions (also called the Chew-Goldberger-Low or 'CGL') will apply \citep{CGL56} and will serve to modify adiabatically the plasma pressure components such that:

\noindent\begin{minipage}{0.5\linewidth}
\begin{equation}
\frac{p_{||} B^2}{\rho^3} = const
\end{equation}
\end{minipage}
\noindent\begin{minipage}{0.5\linewidth}
\begin{equation}
\frac{p_{\perp}}{\rho B} = const,
\end{equation}
\end{minipage}
 with $p_{\|,\perp}$ being the ion pressure along ($\|$) and perpendicular ($\perp$) to the mean field ${\bf B}$. 

Taken together, the CGL conditions suggest that an adiabatically transported fluid element should see its temperature ratio $T_\perp/T_{||}$ fall as approximately $1/R^2$ between 10 and 100$R_s$, as the solar wind expands outward ($R_s$ being the radius of the Sun). Therefore a parcel of plasma with an isotropic temperature ($T_\perp/T_{||} \sim 1$) at the edge of the solar wind acceleration region ($\sim 10 R_s$) will arrive at 1 AU in a highly anisotropic state $T_{||} \sim 100 T_\perp$, if it remains adiabatic.  Such a large temperature anisotropy  has never been observed in the solar wind because the CGL conditions do not take into account wave-particle interactions or kinetic effects, which can control plasma via different types of instabilities.

Several early  authors studied this possibility and looked for evidence of instability  \citep{gary76,gary96,kasper02phd,hellinger06a,matteini07,bourouaine10}.  
Relatively recent results using well-calibrated, statistical measurements from the {\it Wind} spacecraft have shown that the proton temperature  anisotropy $T_\perp/T_{||}$ is constrained by the $\beta_{||}$\footnote{Parallel ion beta is defined with the parallel ion temperature, $\beta_{\|}=nkT_{\|}/(B^2/2\mu_0 )$.}-dependent thresholds for the oblique firehose instability (for $T_\perp/T_{||} < 1$) and the mirror-mode instability (for $T_\perp/T_{||} > 1$) suggesting that the growth of ion-scale fluctuations acts to isotropize the plasma near the  thresholds \citep{gary93book}.  Indeed, a build-up of magnetic fluctuation power is observed near these thresholds \citep{bale09} and the fluctuations seen near the  mirror threshold and for $\beta_{||} > 1$ are compressive, as would be expected from the growth of mirror waves \citep{hasegawa69}.  Fig.~\ref{fig:bale_ts}~(left) shows time series  data of magnetic and velocity fluctuations as the solar wind approaches the oblique firehose instability threshold:  the top panel shows measurements  of the ion anisotropy (black dots) and the theoretical instability thresholds \citep{hellinger06a} as dotted lines.  When the solar wind approaches the  firehose threshold (black dotted line), enhanced fluctuation power is observed in the perpendicular components of the magnetic field and velocity, consistent with Alfv\'enic-like fluctuations excited by the firehose instability \citep{hellinger00,hellinger01}.  Fig.~\ref{fig:bale_ts}~(right) shows an example when the plasma conditions are close to both, mirror and firehose instability thresholds, and when both types of fluctuations, Alfv\'enic and compressive, are excited.

Fig.~\ref{fig:bale_figure1} is reproduced from \citep{bale09} and shows statistically the effect seen in Fig.~\ref{fig:bale_ts}.  One continuing puzzle here is the following:  the instability thresholds,  with the rate $\gamma\simeq 10^{-3}2\pi f_{ci}$, calculated by \citet{hellinger06a} suggest that the ion cyclotron instability should be unstable at values of $T_\perp/T_{||}$ lower than the mirror instability (at low $\beta_{\|}$), however there is no clear evidence in the data of an ion cyclotron limit. One reason for this may be that the mirror mode is non-propagating, and therefore more effective in pitch angle scattering.  In any case, this is unresolved.

The clear existence of instability-limited anisotropies, and the measurement of the associated ion-scale fluctuations, bring to light a very important question:  how much of the fluctuation power (magnetic, velocity, or other) measured near the ion scales in the solar wind is generated by instabilities, rather than driven by the turbulent cascade?  

\begin{figure}
\centering
\includegraphics[width=8cm]{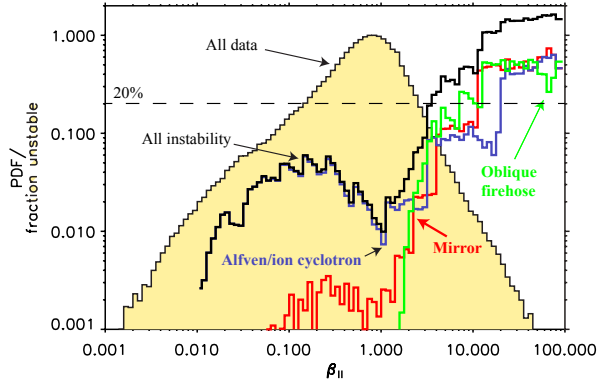}
\caption{Probability distribution of parallel ion beta $\beta_{\|}$ in the data set analyzed by \citet{bale09}. The total distribution is shown in yellow; the most probable value of $\beta_{||}$ in the solar wind is around 0.8.  The various  coloured lines show the normalized histograms of occurrence of data at   and beyond  a certain threshold for different types of instabilities, as calculated by \citet{hellinger06a},  the black line gives the sum of all coloured histograms: at high $\beta_{\|}$, more than 20\% of the solar wind is unstable.}
\label{fig:bale_PDF}
\end{figure}

Fig.~\ref{fig:bale_PDF} shows the probability distribution (see the yellow histogram) of parallel ion beta $\beta_{\|}$, using the {\it Wind} dataset described in \citep{bale09}. The coloured lines show the cumulative distribution of  ``unstable" measurements,  i.e. data points around and beyond the theoretical instability thresholds indicated in Fig.~\ref{fig:bale_figure1} by dotted lines.  The black line gives the sum of all colored histograms.  For solar wind intervals with $\beta_{\|} \ge \sim 3$, more than 20\% of the intervals would be unstable.  However, the magnetic field fluctuation measurements, shown in Fig.~\ref{fig:bale_figure1}, suggest that the power is enhanced well before the thresholds -- hence the effect may be much larger.

It seems that  the magnetic and velocity fluctuation power is injected near the ion scales by instabilities, whose energy source is solar wind expansion or compression, and that this effect is dependent on the plasma $\beta$.  These quasi-linear ion instabilities co-exist with the non-linear turbulent cascade in the solar wind. Therefore, if the goal is to study cascade physics, care must be taken when studying ion scale fluctuations, to be certain that the plasma is very near to isotropic $T_\perp/T_{||} \sim 1$  to avoid the quasi-linear ion instabilities.  Interestingly, the bottom panel of Fig.~\ref{fig:bale_figure1},  which shows the collisional age of protons\footnote{The collisional age is defined as $\tau_{coll}=\nu_{pp}R/V_{sw}$, the Coulomb proton-proton collision frequency $\nu_{pp}$ multiplied by the transit time (or expansion time) from the Sun to 1~AU and is an estimate of the number of binary collisions in each plasma parcel during transit from the Sun to the spacecraft.}, demonstrates that the condition $T_\perp/T_{||} \sim 1$ corresponds to  a solar wind plasma that is collisionally well-processed  (`old') and so remains `fluid-like', rather than kinetic.  The measurements of `kinetic' turbulence must be qualified by considering the particle pressure anisotropies, and relative drifts  between protons and $\alpha$-particles and protons and electrons \citep{chen13a,perrone13}. 

%%%%%%%%%%
%%
%%  Ion instability section - SDB - Feb 7 - END
%%
%%%%%%%%%%

\subsection{Small Scale Inertial Range Between Ion and Electron Scales, and Dissipation at Electron Scales}\label{sec:electron-sacles}

As far as the turbulent cascade crosses the ion scales and before reaching the electron scales  (the satellite frequencies being $3\leq f \leq 30$~Hz), magnetic spectra follow $\sim k_{\perp}^{-2.8}$ \citep{alexandrova09a,chen10b,sahraoui10}, see Fig.~\ref{fig:spec-ii-scales}. This spectral shape seems to be independent of the local plasma parameters, as far as the angle between the flow and the field $\theta_{BV}$ is quasi-perpendicular \citep{alexandrova09a,alexandrova12}. 

\begin{figure}
\centering
\includegraphics[width=8cm]{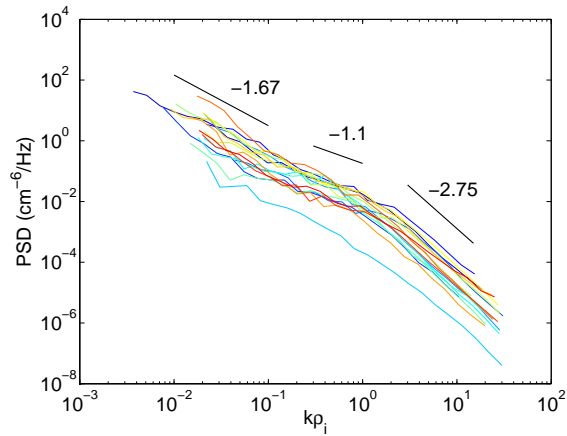}
\caption{17 electron density spectra normalized in scale to the ion gyroradius, showing a flattening at ion scales $\sim (k\rho_i)^{-1}$, as in Fig.~\ref{fig:n_spec}, and a slope close to --2.75 between ion and electron scales \citep{chen12c} in agreement with the magnetic spectrum at these scales, see Fig.~\ref{fig:spec-ii-scales}.}
\label{fig:density_spec_chen12}
\end{figure}

The electron density spectrum between ion and electron scales was measured by \citet{chen12a,chen12c} using the high frequency spacecraft potential on \emph{ARTEMIS}. Fig.~\ref{fig:density_spec_chen12} shows 17 electron density spectra normalized to the ion gyroradius,   measured for $\theta_{BV}>45^{\circ}$ in the solar wind. At large scales, the spectra are in agreement with previous observations (see Fig.~\ref{fig:n_spec}). At small scales, for $k\rho_i \geq 3$ the electron density spectra follow the $\sim k^{-2.75}$ power-law, which is close to the typical value of --2.8 found in the magnetic field spectrum. 

The observations of well defined power-laws in magnetic and density spectra   between ion and electron scales suggest that at these scales there is a small scale inertial range \citep{alexandrova07,alexandrova08b,alexandrova09a,kiyani09a,chen10b,sahraoui10,chen12a} or an electron inertial range \citep{smith12}.  

Kolmogorov arguments for Electron MHD lead to a $\sim k^{-7/3}$ magnetic energy spectrum  \citep{biskamp96,biskamp99a,cho04}. More recent theories of strong KAW turbulence also predict a --7/3 spectrum for both density and magnetic field \citep{schekochihin09}. The fact that the observed spectra are typically steeper than this has been explained in several ways, including electron Landau damping \citep{howes11a}, compressibility effect \citep{alexandrova07}  and an intermittency correction resulting in a spectral index of --8/3 \citep{boldyrev12b}.  The same spectral index of $-8/3$ can be also obtained in quasi-bidimentional strong Electron MHD turbulence ($k_{\perp} \gg k_{\|}$) when parallel cascade is weak  \citep{galtier05a}. A model of \citet{rudakov11}  of KAW turbulence with nonlinear scattering of waves by plasma particles gives spectral index between 2 and 3. 

\begin{figure}
\centering
\includegraphics[scale=0.2]{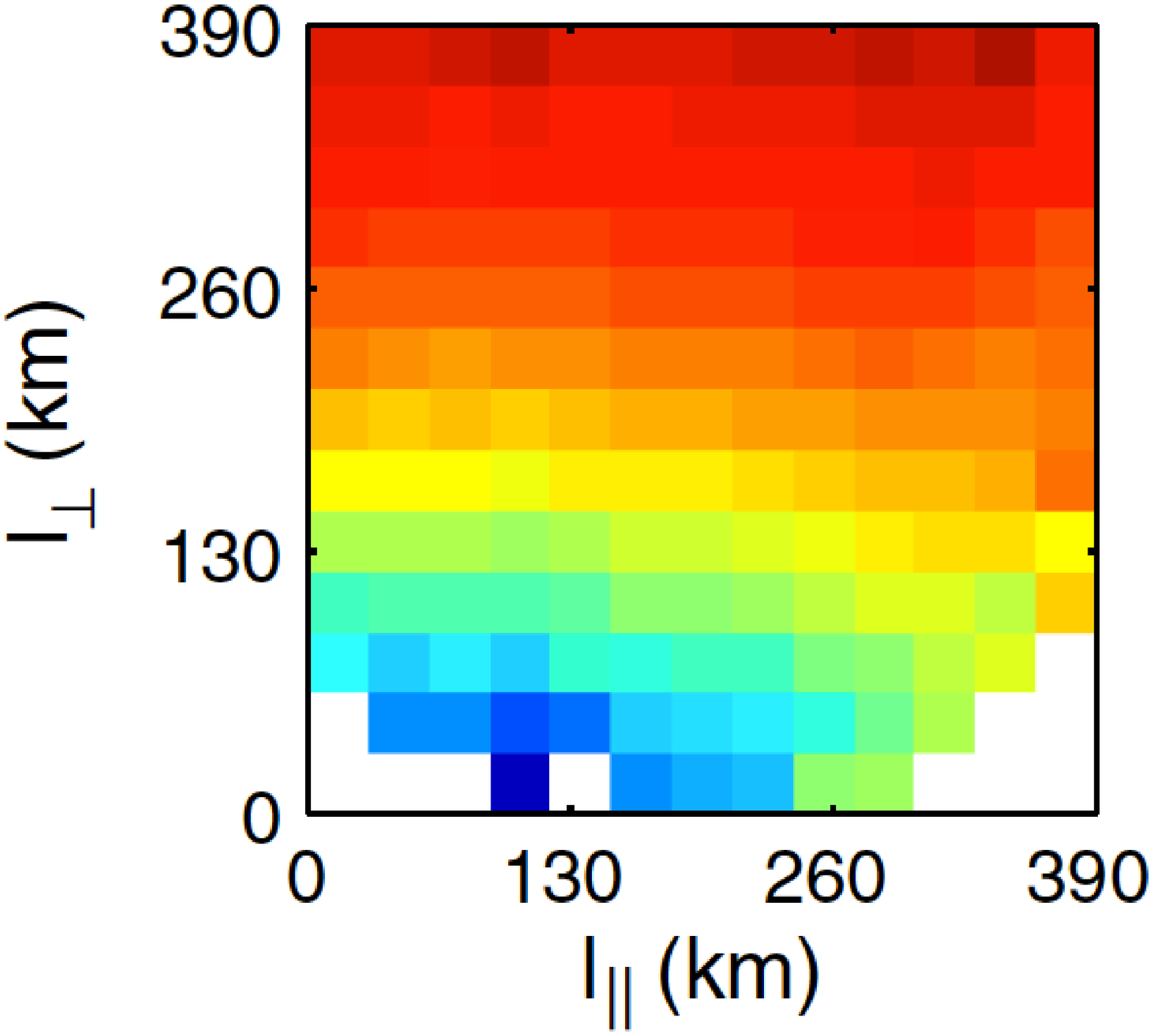}
\hspace{5mm}
\includegraphics[scale=0.15]{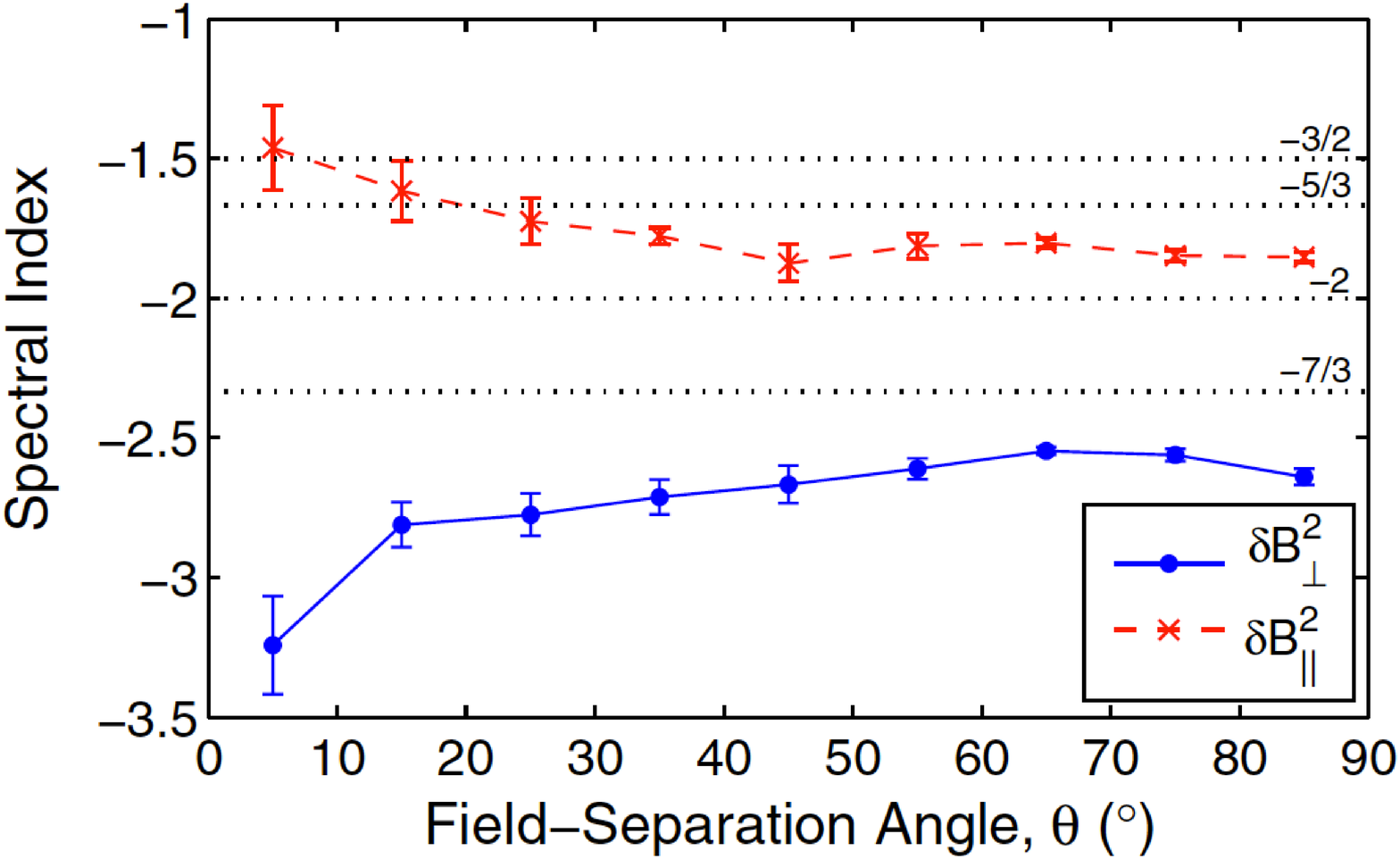}
\caption{Left: power in $\delta B_\perp$ (in color) as a function of parallel and perpendicular scale between ion and electron scales \citep{chen10b}. Right: spectral index as a function of angle $\theta_{B}$  (the angle between ${\bf B}$ and the separation vector between {\it Cluster} satellites) for the perpendicular $\delta B_{\perp}$ and parallel $\delta B_{\|}$ field components \citep{chen10b}.}
\label{fig:anisotropy}
\end{figure}

As we have mentioned, the magnetic and density spectra of  Fig.~\ref{fig:spec-ii-scales} and Fig.~\ref{fig:density_spec_chen12} are measured for quasi-perpendicular $\theta_{BV}$. Varying this angle, one may resolve turbulent fluctuations with different ${\bf k}$, as discussed in Section 2.1. \citet{chen10b} used a multi-spacecraft technique to measure the wavevector anisotropy of the turbulence between ion and electron scales  (up to $\sim 10$~Hz) using two-point structure functions. They found the turbulence to be anisotropic in the same sense as in the MHD scale cascade, with $k_{\perp}>k_{\|}$, corresponding to ``eddies'' elongated along the local mean field direction (Fig.~\ref{fig:anisotropy}, left). They also found the spectral index of the perpendicular magnetic fluctuations $\delta B_{\perp}$ to become steeper  for small $\theta_{B}$ (the angle between ${\bf B}$ and the separation vector between {\it Cluster} satellites), i.e. for ${\bf k}$ parallel to ${\bf B}$ (Fig.~\ref{fig:anisotropy}, right), suggestive of strong whistler or kinetic Alfv\'en turbulence \citep{cho04,schekochihin09,chen10a,boldyrev12b}. 
Note that two-point structure functions cannot resolve spectral indices steeper than $-3$, e.g. \citep{abry95,abry09,chen10b}.  So, it is possible that the parallel spectral index of $\delta B_{\perp}$ is steeper than what is shown in Fig.~\ref{fig:anisotropy}~(right).

Recently, \citet{turner11} studied anisotropy of the magnetic fluctuations up to $\sim 20$~Hz. The authors used the reference frame based on the mean magnetic field and velocity, which allow to check the axisymmetry and importance of the Doppler shift for $k_{\perp}$ fluctuations \citep{bieber96}. The authors found that the spectrum of magnetic fluctuations in the direction perpendicular to the velocity vector in the plane perpendicular to ${\bf B}$, ${\bf V}_{sw\perp}$, is higher than the spectrum of $\delta B$ along ${\bf V}_{sw\perp}$.  This is consistent with a turbulence with $k_{\perp}\gg k_{\|}$, where the fluctuations with ${\bf k}$ along ${\bf V}_{sw\perp}$ are more affected by the Doppler shift than the fluctuations with ${\bf k}$ perpendicular to ${\bf V}_{sw\perp}$. These results are also in agreement with the magnetosheath observations between ion and electron scales \citep{alexandrova08c}.

%%%%%%%%% e-dissipation range  %%%%%%%%%%%%

What happens at smaller scales? Several authors have suggested that the electromagnetic turbulent  cascade in the solar wind dissipates at  electron scales. These scales are usually called {\it electron dissipation range}, e.g. \citep{smith12}.

\begin{figure}
\centering
\includegraphics[width=8cm]{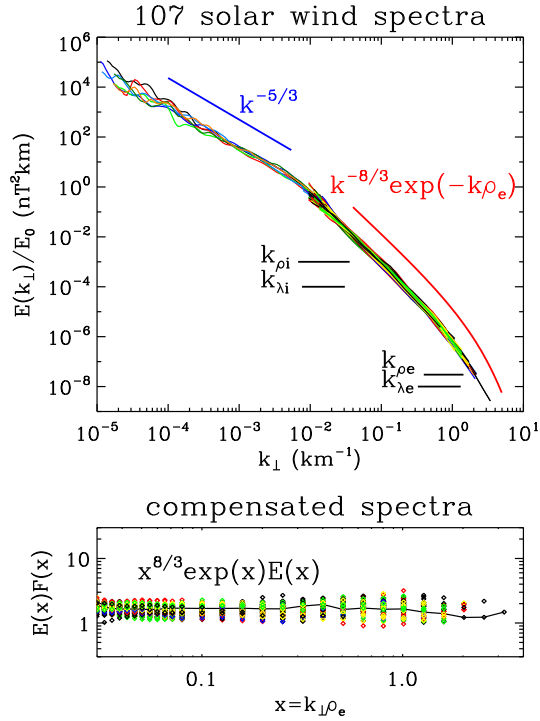}
\caption{100 magnetic field spectra in the kinetic range to a fraction of electron scales and 7 magnetic field spectra covering fluid and kinetic scales, with spectra compensated to $(k_{\perp} \rho_e)^{8/3}\exp(k_{\perp}\rho_e)$ in the lower panel \citep{alexandrova12}.}
\label{fig:olga12}
\end{figure}
Fig.~\ref{fig:olga12}  is reproduced from \citep{alexandrova12}. The upper panel shows a number of magnetic field spectra measured under different plasma conditions: 100 spectra  from ion scales to a fraction of electron scales, and 7 spectra measured from the MHD range to a fraction of electron scales. At scales smaller than the ion scales ($k_{\perp}>k_{\rho i},k_{\lambda i}$), all the spectra can be described by one algebraic function covering electron inertial and dissipation ranges, 
\begin{equation}\label{eq:exp-model-general}
E(k_{\perp}) = E_0 k_{\perp}^{-\alpha}\exp{(-k_{\perp}\ell_d)}
\end{equation} 
where $\alpha\simeq 8/3$ and where $\ell_d$ is found to be related to the electron Larmor radius $\rho_e$, with a correlation coefficient of 0.7.  This law is independent of the solar wind properties, slow or fast, and of ion and electron plasma beta, indicating the universality of the turbulent cascade at electron scales. The compensated 100 spectra with the  $k_{\perp}^{8/3}\exp{(k_{\perp}\rho_e)}$--function are shown in the bottom panel of Fig.~\ref{fig:olga12}: they are flat over about 2 decades confirming the choice of the model function 
\begin{equation}\label{eq:exp-model}
E(k_{\perp})=E_0 k_{\perp}^{-8/3}\exp(-k_{\perp}\rho_e)
\end{equation} 
to describe solar wind spectrum at such small scales.

 It is interesting that a similar curved spectrum is expected in the Interstellar Medium turbulence, but at ion scales \citep{spangler90,haverkorn13}.

\begin{figure}
\centering
\includegraphics[width=12cm]{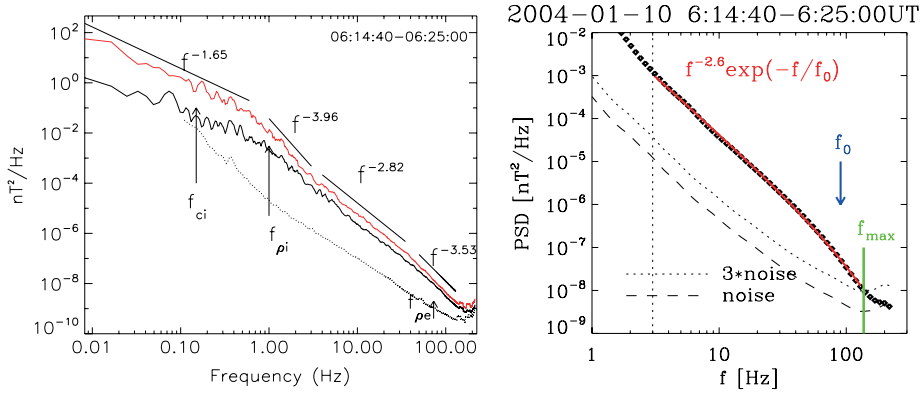} 
\caption{Left: Magnetic spectrum from \citep{sahraoui10}, compared with $\sim f^{-2.8}$ for $4\leq f \leq 35$~Hz and with $\sim f^{-3.5}$ for $50 \leq f \leq 120$~Hz, the break frequency is around 40~Hz. Right: a zoom on the high frequency part of the spectrum on the left, fitted with  $\sim f^{-2.6}\exp{(-f/f_0)}$,  the exponential cut-off frequency $f_0=90$~Hz close to the  Doppler shifted $\rho_e$, $f_0\simeq f_{\rho_e}=V_{sw}/2\pi\rho_e$. This last fitting function is equivalent to the model~(\ref{eq:exp-model}) for wave vectors. }
\label{fig:s10_exp}
\end{figure}

Another  description of the spectrum within the electron inertial and dissipation ranges was proposed by \citet{sahraoui10}. It consists of two power-laws separated by a break,   see Fig.~\ref{fig:s10_exp}~(left).  This double-power-law model can be formulated as 
\begin{equation}\label{eq:break}
\tilde{E}(k_{\perp})=A_1k_{\perp}^{-\alpha_1}(1-H(k_{\perp}-k_b)) + A_2k_{\perp}^{-\alpha_2}H(k_{\perp}-k_b),
\end{equation}
$H(k_{\perp}-k_b)$ being the Heaviside function, $k_b$ the wave number of the break, $A_{1,2}$ the amplitudes of the two power-law functions with spectral indices $\alpha_{1,2}$ on both sides of $k_b$.  This model has five free parameters. A statistical study of the solar wind  magnetic spectra at high frequencies ($f>3$~Hz) shows that $\alpha_1$ does not vary a lot,  $\alpha_1=2.86\pm0.08$ \citep{alexandrova12}. Then the amplitudes $A_1$ and $A_2$ are equal at the break point. Therefore we can fix two of the five parameters of  model (\ref{eq:break}). This model  has thus three free parameters, $A_1$, $\alpha_2$ and $k_b$ (in comparison with one free parameter, $E_0$, in equation~(\ref{eq:exp-model})).

Fig.~\ref{fig:s10_exp}(left) shows the frequency spectrum from  \citep{sahraoui10}, compared at high frequencies\footnote{{\it Cluster}/Staff-SC measurements in the burst mode.}, $f>3$~Hz, with the double power-law model~(\ref{eq:break}) with $\alpha_1\simeq 2.8$, $\alpha_2 \simeq 3.5$ and the spectral break at $f_b\simeq 40$~Hz.  Fig.~\ref{fig:s10_exp}(right) shows the total power spectral density for the same dataset fitted with the exponential model~(\ref{eq:exp-model-general}),  which can be written for frequency spectrum as $\sim f^{-\alpha}\exp(-f/f_0)$. The parameters of the fit are $\alpha \simeq 8/3$ and  the exponential cut-off frequency $f_0= 90$~Hz, which is close to the Doppler shifted electron gyro-radius $\rho_e$ for this time interval.  Therefore, the model~(\ref{eq:exp-model}) can be applied in this particular case as well. 

In the statistical study by \citet{alexandrova12}, the authors  concluded that  model function~(\ref{eq:exp-model}) describes all observed spectra, while the double-power-law model~(\ref{eq:break})   cannot describe a large part of the observed spectra.  Indeed  the unique determination of the spectral break  $k_b$  with $A_1=A_2$ at the break is not always possible because of the spectral curvature, and for low intensity spectra there are not enough data points to allow a good determination of $\alpha_2$.

\begin{figure}
\centering
\includegraphics[width=12cm]{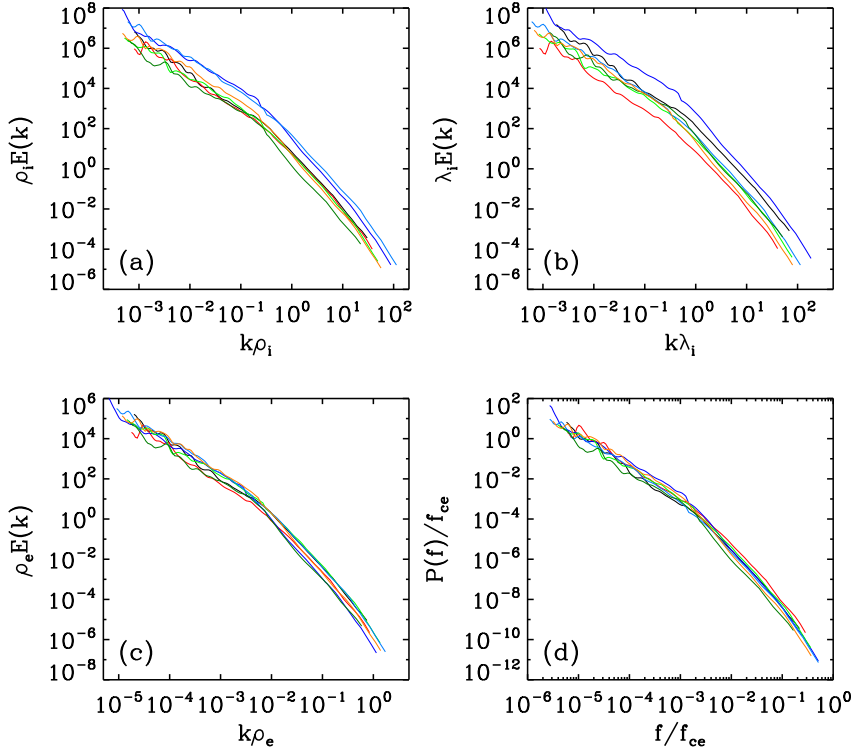}
\caption{Universal Kolmogorov function $\propto \ell_d E(k)$ for hypothesized
dissipation scales $\ell_d$ as a function of (a) $k\rho_i$, (b) $k\lambda_i$, (c) $k\rho_e$ and (d) $f/f_{ce}$. Figure from \citep{alexandrova09a}, corrected for 3 STAFF-SA frequencies, as explained in \citep{alexandrova12}.}
\label{fig:univ-K-function}
\end{figure}

 The equivalence between the electron gyro-radius $\rho_e$, in the solar wind turbulence, and the dissipation scale $\ell_d$, in the usual fluid turbulence, can be seen from Fig.~\ref{fig:univ-K-function} where the Universal Kolmogorov Function $E(k)\ell_d/\eta^2$ is shown as a function of $k\ell_d$ \citep{frisch95,Davidson04book}, for three different candidates for the dissipation scale $\ell_d$, namely for $\rho_i$, $\lambda_i$ and $\rho_e$; and for one time characteristic scale, namely the electron gyro-period $f_{ce}^{-1}$. For simplicity, the kinematic viscosity $\eta$  is assumed to be constant, despite the varying plasma conditions. One can see that the $\rho_i$ and $\lambda_i$ normalizations are not efficient to collapse the spectra together. Normalization on $\lambda_e$ gives the same result as for $\lambda_i$.  At the same time, the normalizations on $\rho_e$ and $f_{ce}$ bring the spectra close to each other, as expected while normalizing by $\ell_d$.  In addition to the spectral analysis presented in Fig.~\ref{fig:olga12}, the Universal Kolmogorov Function normalization gives an independent confirmation that the spatial scale which may play the role of the dissipation scale, in the weakly collisional solar wind, is the electron gyro-radius $\rho_e$.

\begin{figure}
\centering
\includegraphics[width=12cm]{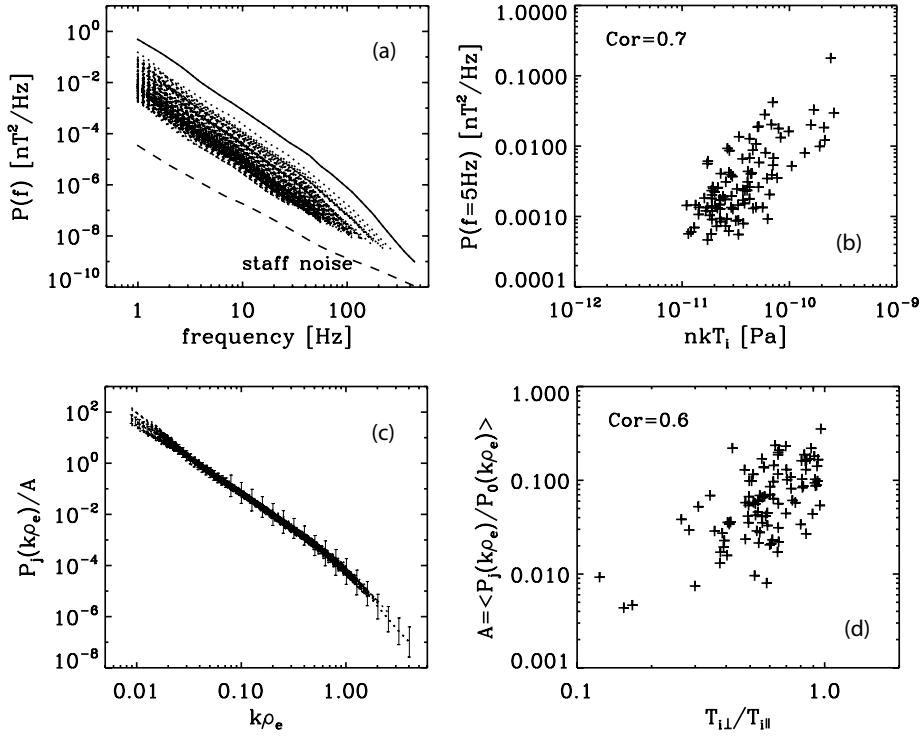}
\caption{(a) The 100 magnetic frequency spectra measured by {\it Cluster}/STAFF in the solar wind for $f>1$~Hz, analyzed in \citep{alexandrova12}; (b) Intensity of the frequency spectra at a fixed frequency $f=5$~Hz as a function of the ion thermal pressure $nkT_i$; (c) The same spectra as in panel (a) but shown as a function of $k\rho_e$ and superposed using an amplitude factor $A$ (equivalent to $E_0$ in Fig.~\ref{fig:olga12}); (d) The amplitude $A$ as a function of the ion temperature anisotropy $T_{i\perp}/T_{i\|}$.   Figure from \citep{alexandrova11}.}
\label{fig:amp}
\end{figure}

 It is important to mention, that the amplitude parameter $E_0$ of the exponential model~(\ref{eq:exp-model})  is found to be related to the solar wind plasma parameters \citep{alexandrova11}, see Fig.~\ref{fig:amp}. The amplitude of the raw frequency spectra is well correlated with  the ion thermal pressure $nkT_i$  (Fig.~\ref{fig:amp}, upper line):  this is similar to the amplitude of the inertial range spectrum, which is found to be correlated to the ion thermal speed \citep{grappin90}. The amplitude of the $k$-spectra, as well as the amplitude of the normalised $k\rho_e$-spectra,  appears to depend on the ion temperature anisotropy  (Fig.~\ref{fig:amp}, lower line). This last result suggests that the ion instabilities present around the ion break scale may indeed inject or remove energy from the cascade  (see  our discussion in Section~3.2). Therefore, the scales around the ion break  (or `transition range', see Fig.~\ref{fig:spec-ii-scales}) may be seen, in part, as the energy injection scales for the small scale inertial range. 

In usual fluid turbulence, the far dissipation range is described by $E(k) \sim k^{3}\exp(-c k\ell_d)$ (with $c\simeq 7$) \citep{chen93}.  The exponential tail is due to the resistive damping rate $\gamma \propto k^2$ valid in a collisional fluid. In  the collisionless plasma of the solar wind there is no resistive damping, and thus the observation of the exponential spectrum within the electron dissipation range  deserves an explanation. 

\citet{howes11c} consider a model (``weakened cascade model'') which includes the nonlinear transfer of energy from large to small scales in Fourier space and the damping of kinetic Alfv\'en waves. The spectral laws are respectively $E_k \propto k_\bot^{-5/3}$ at large scales and $E_k \propto k_\bot^{-7/3}$  between ion and electron scales. The damping becomes important at electron Larmor radius $\rho_e$ scale. It is obtained by linearising the Vlasov-Maxwell equations in the gyrokinetic limit ($k_\| \ll k_\bot$, with frequencies $f \ll f_{ci}$). 
For $k_\bot \rho_i \gg 1$ it has the form  $\gamma \propto k_\| k_\bot^2$. Taking into account the  assumption of critical balance $\tau_{nl}=\tau_A$ (i.e. $k_{\perp}v=k_\| V_A$) \citep{goldreich95}, and the spectral index $-7/3$ (i.e. $v\sim k_\bot^{-2/3}$),  one gets $k_\| \propto k_\bot^{1/3}$. Therefore, the damping term takes the form  $\gamma \propto  k_\bot^{2+1/3}$.  The exponent of the damping rate is thus very close to the $k^2$ scaling of the Laplacian viscous term, which is known to lead in hydrodynamical turbulence to an exponential tail in the dissipation range. Indeed, when taking into account the damping term, \citet{howes11c} obtain numerically a final curved tail at scales smaller than electron scales. Superficially, this spectrum thus resembles the analytic form which we have found to be valid to describe  the solar wind turbulence, equation~(\ref{eq:exp-model}).

As we have just seen,  the model of  \citet{howes11c} assumes the  $k_{\perp}\gg k_{\|}$--anisotropy and very low frequencies $f\ll f_{ci}$. Present multi-satellite observations can not cover the electron inertial and dissipation ranges at scales smaller than the smallest satellite separation $\sim 100$~km. Only the one-satellite technique of \citet{bieber96} can be used. A first attempt to determine the distribution of wave-vectors ${\bf k}$ of the electromagnetic fluctuations within the electron inertial and dissipation ranges (for the observed frequencies $[8,500]$~Hz) was carried out in the magnetosheath by \citet{mangeney06}.
They show that the wavevectors ${\bf k}$ of the electromagnetic fluctuations are distributed within the plane nearly perpendicular to the mean field ${\bf B}$, with an angle of $\sim \pm5^{\circ}$ around this plane.   However, the authors have not found any agreement between the observed properties of magnetic fluctuations and KAW turbulence.

The nature of  turbulence between ion and electron scales is still under debate.  As with the MHD scale cascade, there are a number of observational and theoretical works, which identify turbulent fluctuations at small scales as having properties of linear wave modes \citep[e.g.,][]{denskat83,goldstein94,ghosh96,biskamp96,leamon98a,biskamp99a,cho04,bale05,galtier06b,sahraoui10,howes06,howes08a,schekochihin09,gary09,chandran09c,sahraoui09,chen10a,sahraoui10,howes11b,salem12,klein12,boldyrev12b,chen12c}.  A recent analysis by   \citet{chen13b}  showed that the ratio of density to magnetic fluctuations in the range between ion and electron scales is very close to that expected for kinetic Alfv\'en waves, and not whistler waves, and concluded that the fluctuations in this range are predominantly strong kinetic Alfv\'en turbulence. The precise interplay between linear and non-linear physics is an important unsolved problem in plasma turbulence.

Solar wind observations  and numerical simulations show that the fluctuations at kinetic scales  have non-Gaussian distributions, indicating the presence of intermittency \citep{alexandrova07,alexandrova08b,kiyani09a,wan12prl,wu13}. Recently, small scale coherent current sheets have been identified at scales close to electron scales \citep{perri12a}. These features are consistent with strong, rather than weak (or wave) turbulence. The properties of the intermittency at small scales are not clear at the moment. There are two contradictory observations: (i) \citet{alexandrova08b} show a scale dependent deviation from Gaussianity of the PDFs of the magnetic fluctuations $\delta B_R$ (along the solar wind flow); (ii) \citet{kiyani09a} show observations suggesting a scale-invariance within the small scales. Further work is needed to understand this discrepancy.

\section{Discussion}
In this paper we have discussed solar wind turbulence observations in a large range of scales:  from the   MHD  scales to the electron characteristic scales. 

At MHD scales, within the inertial range, the solar wind turbulence presents several general characteristics inherent to fully developed fluid turbulence: (i) energy spectra of different plasma parameters have well-defined power-laws; (ii) the probability distribution functions deviate from a Gaussian distribution, indicating stronger gradients  at smaller scales (intermittency); (iii) the third order moments of turbulent fluctuations have the linear dependence on scale (the proportionality coefficient giving   the energy transfer rate). 
The anisotropy of turbulence with respect to a mean magnetic field is shown to be important: the turbulence develops mostly  in the plane perpendicular to  ${\bf B}$, i.e. with $k_{\perp} \gg k_{\|}$. The perpendicular magnetic spectrum follows $\sim k_{\perp}^{-5/3}$ scaling, while the parallel spectrum is steeper $\sim k_{\|}^{-2}$. 
The dominant fluctuations are Afv\'enic in nature, i.e. $\delta B_{\perp} > \delta B_{\|}$, however, the velocity spectrum has a spectral slope of $-3/2$ and it does not follow the magnetic spectrum. 
There is a small fraction of the turbulent energy in compressible fluctuations. It is not clear whether they behave as a passive contaminant as in compressible neutral fluid turbulence or they are an active component of the turbulence in the solar wind. In other words, is it possible to describe these compressible fluctuations independently of the dominant Alfv\'enic cascade, or are they  inherently coupled? This question is a matter of debate. 

The MHD inertial range ends at ion characteristic scales. Here, different kinetic effects may take place and inject or remove energy from the turbulent cascade.  In particular, the large scale energy reservoir related to the solar wind spherical expansion may be released into fluctuations, throughout instabilities, like mirror and oblique firehose instabilities, which becomes important for high ion betas ($\beta_i>3$).  Then these fluctuations may interact with particles and dissipate, or participate to the turbulent cascade at smaller scales. 
At lower  beta, the plasma is stable and more or less isotropic: no additional energy is expected to arrive to the turbulent cascade. However, the exact energy partition between fluid and kinetic degrees of freedom at ion scales is still under debate.  Around ion scales magnetic spectra are variable, and the compressibility increases. A spectral break seems to appear at the ion inertial scale, suggesting that dispersive effects (Hall effect) become significant. 

Between ion and electron scales, a small scale turbulent cascade seems to be established. This cascade is characterised by a $k_{\perp} \gg k_{\|}$ anisotropy, as the MHD cascade. The $k_{\perp}$ magnetic and density spectra have a power-law shape with $\simeq -2.8$ spectral index. Fluctuations are more compressible than within the MHD inertial range and this compressibility seems to depend on the local plasma parameters, like the plasma $\beta$.  Magnetic fluctuations are non-Gaussian, indicating the presence of the intermittency.      

Approaching electron scales, the fluctuations are no more self-similar: the spectrum is no more a power-law, but an exponential cut-off is observed indicating an onset of dissipation. The dissipation range spectrum is observed to have a general shape. One algebraic function $\sim k_{\perp}^{-8/3}\exp(-k_{\perp}\rho_e)$ describes well the whole spectrum covering the small scale inertial range and the dissipation range.   

The nature of the small scale cascade between ion and electron scales and the dissipation mechanism at electron scales are still under debate.  The model of \citet{howes11c}  can describe the observed exponential cut-off. The dissipation mechanism in this model is based on  a quasi-linear description of the Landau damping of kinetic Alfv\'en waves onto electrons. Whether such description can apply on the solar wind observations is however under debate because of the presence of a significant degree of intermittency at kinetic scales.

To build a realistic model of the dissipation in the solar wind we need still  to resolve an open question on the nature of the turbulent fluctuations: is it a mixture of linear waves or is it a strong turbulence with dissipation restricted to intermittent coherent structures?  What is the topology of these structures -- current sheets, shocks, solitons or coherent vortices? 

\begin{acknowledgements}
We all thank the International Space Science Institute (ISSI, Bern) for hospitality. SDB, CHKC, TH and LSV acknowledge the Marie Curie Project FP7 PIRSES-2010-269297 - ``Turboplasmas''. OA thanks Andr\'e Mangeney, Roland Grappin, Nicole Meyer, Robert Wicks and Petr Hellinger for discussions, Chadi Salem, Silvia Perri, William Matthaeus and Sofiane Bourouaine for providing figures, and Catherine Lacombe for reading this manuscript. 
\end{acknowledgements}

\bibliographystyle{aps-nameyear}      
\bibliography{bibliography}

\begin{thebibliography}{223}
% BibTex style file: aps.bst (nameyear), 2011-02-21
\ifx \bisbn   \undefined \def \bisbn  #1{ISBN #1}\fi
\ifx \binits  \undefined \def \binits#1{#1} \fi
\ifx \bauthor  \undefined \def \bauthor#1{#1} \fi
\ifx \bjtitle  \undefined \def \bjtitle#1{\textrm{#1}}\fi
\ifx \batitle  \undefined \def \batitle#1{#1} \fi
\ifx \bctitle  \undefined \def \bctitle#1{#1} \fi
\ifx \bvolume  \undefined \def \bvolume#1{\textbf{#1}}\fi
\ifx \byear  \undefined \def \byear#1{#1} \fi
\ifx \bissue  \undefined \def \bissue#1{#1} \fi
\ifx \bfpage  \undefined \def \bfpage#1{#1} \fi
\ifx \blpage  \undefined \def \blpage #1{#1} \fi
\ifx \burl  \undefined \def \burl#1{#1} \fi
\ifx \doiurl  \undefined \def \doiurl#1{#1} \fi
\ifx \betal  \undefined \def \betal{et al.} \fi
\ifx \binstitute  \undefined \def \binstitute#1{#1} \fi
\ifx \beditor  \undefined \def \beditor#1{#1} \fi
\ifx \bpublisher  \undefined \def \bpublisher#1{#1} \fi
\ifx \bbtitle  \undefined \def \bbtitle#1{\textit{#1}} \fi
\ifx \bedition  \undefined \def \bedition#1{#1} \fi
\ifx \bseriesno  \undefined \def \bseriesno#1{#1} \fi
\ifx \blocation  \undefined \def \blocation#1{#1} \fi
\ifx \bsertitle  \undefined \def \bsertitle#1{#1} \fi
\ifx \bsnm \undefined \def \bsnm#1{#1} \fi
\ifx \bsuffix \undefined \def \bsuffix#1{#1} \fi
\ifx \bparticle \undefined \def \bparticle#1{#1} \fi
\ifx \barticle \undefined \def \barticle#1{#1} \fi
\ifx \botherref \undefined \def \botherref #1{#1} \fi
\ifx \url \undefined \def \url#1{#1} \fi
\ifx \bchapter \undefined \def \bchapter#1{#1} \fi
\ifx \bbook \undefined \def \bbook#1{#1} \fi
\ifx \bcomment \undefined \def \bcomment#1{#1} \fi
\ifx \oauthor \undefined \def \oauthor#1{#1} \fi
\ifx \citeauthoryear \undefined \def \citeauthoryear#1{#1} \fi
\ifx \texttildelow  \undefined \def \texttildelow{\symbol{126}} \fi
\def \endbibitem {}
\ifx \bconflocation  \undefined \def \bconflocation#1{#1} \fi

\bibitem[\protect\citeauthoryear{Abry et~al.}{1995}]{abry95}
\begin{botherref}
\oauthor{\binits{P.} \bsnm{Abry}},
\oauthor{\binits{P.} \bsnm{Gon{\c c}alves}},
\oauthor{\binits{P.} \bsnm{Flandrin}},
Wavelets, spectrum analysis and 1/f processes.
Wavelets and statistics
(1995).
Lecture Notes in Statistics.
\url{http://perso.ens-lyon.fr/paulo.goncalves/pub/lns95.pdf}
\end{botherref}
\endbibitem

\bibitem[\protect\citeauthoryear{Abry et~al.}{2009}]{abry09}
\begin{bbook}
\bauthor{\binits{P.} \bsnm{Abry}},
\bauthor{\binits{P.} \bsnm{Gon{\c c}alves}},
\bauthor{\binits{J.} \bsnm{L{\'e}vy~V{\'e}hel}},
\bbtitle{Scaling, {F}ractals and {W}avelets}.
\bsertitle{Digital signal and image processing series}
(\bpublisher{ISTE -- John Wiley \& Sons, Inc.},
\blocation{London (UK)}, \byear{2009})
\end{bbook}
\endbibitem

\bibitem[\protect\citeauthoryear{{Alexandrova}}{2008}]{alexandrova08a}
\begin{barticle}
\bauthor{\binits{O.} \bsnm{{Alexandrova}}},
\batitle{{Solar wind vs magnetosheath turbulence and Alfv{\'e}n vortices}}.
\bjtitle{\npg}
\bvolume{15},
\bfpage{95}--\blpage{108}
(\byear{2008})
\end{barticle}
\endbibitem

\bibitem[\protect\citeauthoryear{{Alexandrova} and
  {Saur}}{2008}]{alexandrova08_grl}
\begin{barticle}
\bauthor{\binits{O.} \bsnm{{Alexandrova}}},
\bauthor{\binits{J.} \bsnm{{Saur}}},
\batitle{{Alfv{\'e}n vortices in Saturn's magnetosheath: Cassini
  observations}}.
\bjtitle{\grl}
\bvolume{35},
\bfpage{15102}
(\byear{2008}).
doi:\doiurl{10.1029/2008GL034411}
\end{barticle}
\endbibitem

\bibitem[\protect\citeauthoryear{{Alexandrova} et~al.}{2008}]{alexandrova08c}
\begin{barticle}
\bauthor{\binits{O.} \bsnm{{Alexandrova}}},
\bauthor{\binits{C.} \bsnm{{Lacombe}}},
\bauthor{\binits{A.} \bsnm{{Mangeney}}},
\batitle{{Spectra and anisotropy of magnetic fluctuations in the Earth's
  magnetosheath: Cluster observations}}.
\bjtitle{\ang}
\bvolume{26},
\bfpage{3585}--\blpage{3596}
(\byear{2008})
\end{barticle}
\endbibitem

\bibitem[\protect\citeauthoryear{{Alexandrova} et~al.}{2006}]{alexandrova06}
\begin{barticle}
\bauthor{\binits{O.} \bsnm{{Alexandrova}}},
\bauthor{\binits{A.} \bsnm{{Mangeney}}},
\bauthor{\binits{M.} \bsnm{{Maksimovic}}},
\bauthor{\binits{N.} \bsnm{{Cornilleau-Wehrlin}}},
\bauthor{\binits{J.-M.} \bsnm{{Bosqued}}},
\bauthor{\binits{M.} \bsnm{{Andr{\'e}}}},
\batitle{{Alfv{\'e}n vortex filaments observed in magnetosheath downstream of a
  quasi-perpendicular bow shock}}.
\bjtitle{\jgr}
\bvolume{111}(\bissue{A10}),
\bfpage{12208}
(\byear{2006}).
doi:\doiurl{10.1029/2006JA011934}
\end{barticle}
\endbibitem

\bibitem[\protect\citeauthoryear{{Alexandrova} et~al.}{2007}]{alexandrova07}
\begin{barticle}
\bauthor{\binits{O.} \bsnm{{Alexandrova}}},
\bauthor{\binits{V.} \bsnm{{Carbone}}},
\bauthor{\binits{P.} \bsnm{{Veltri}}},
\bauthor{\binits{L.} \bsnm{{Sorriso-Valvo}}},
\batitle{{Solar wind Cluster observations: Turbulent spectrum and role of Hall
  effect}}.
\bjtitle{\pss}
\bvolume{55},
\bfpage{2224}--\blpage{2227}
(\byear{2007}).
doi:\doiurl{10.1016/j.pss.2007.05.022}
\end{barticle}
\endbibitem

\bibitem[\protect\citeauthoryear{{Alexandrova} et~al.}{2008}]{alexandrova08b}
\begin{barticle}
\bauthor{\binits{O.} \bsnm{{Alexandrova}}},
\bauthor{\binits{V.} \bsnm{{Carbone}}},
\bauthor{\binits{P.} \bsnm{{Veltri}}},
\bauthor{\binits{L.} \bsnm{{Sorriso-Valvo}}},
\batitle{{Small-Scale Energy Cascade of the Solar Wind Turbulence}}.
\bjtitle{\apj}
\bvolume{674},
\bfpage{1153}--\blpage{1157}
(\byear{2008}).
doi:\doiurl{10.1086/524056}
\end{barticle}
\endbibitem

\bibitem[\protect\citeauthoryear{{Alexandrova} et~al.}{2009}]{alexandrova09a}
\begin{barticle}
\bauthor{\binits{O.} \bsnm{{Alexandrova}}},
\bauthor{\binits{J.} \bsnm{{Saur}}},
\bauthor{\binits{C.} \bsnm{{Lacombe}}},
\bauthor{\binits{A.} \bsnm{{Mangeney}}},
\bauthor{\binits{J.} \bsnm{{Mitchell}}},
\bauthor{\binits{S.J.} \bsnm{{Schwartz}}},
\bauthor{\binits{P.} \bsnm{{Robert}}},
\batitle{{Universality of Solar-Wind Turbulent Spectrum from MHD to Electron
  Scales}}.
\bjtitle{\prl}
\bvolume{103}(\bissue{16}),
\bfpage{165003}
(\byear{2009}).
doi:\doiurl{10.1103/PhysRevLett.103.165003}
\end{barticle}
\endbibitem

\bibitem[\protect\citeauthoryear{{Alexandrova} et~al.}{2010}]{alexandrova10}
\begin{barticle}
\bauthor{\binits{O.} \bsnm{{Alexandrova}}},
\bauthor{\binits{J.} \bsnm{{Saur}}},
\bauthor{\binits{C.} \bsnm{{Lacombe}}},
\bauthor{\binits{A.} \bsnm{{Mangeney}}},
\bauthor{\binits{S.J.} \bsnm{{Schwartz}}},
\bauthor{\binits{J.} \bsnm{{Mitchell}}},
\bauthor{\binits{R.} \bsnm{{Grappin}}},
\bauthor{\binits{P.} \bsnm{{Robert}}},
\batitle{{Solar wind turbulent spectrum from MHD to electron scales}}.
\bjtitle{Twelfth International Solar Wind Conference}
\bvolume{1216},
\bfpage{144}--\blpage{147}
(\byear{2010}).
doi:\doiurl{10.1063/1.3395821}
\end{barticle}
\endbibitem

\bibitem[\protect\citeauthoryear{{Alexandrova} et~al.}{2011}]{alexandrova11}
\begin{botherref}
\oauthor{\binits{O.} \bsnm{{Alexandrova}}},
\oauthor{\binits{C.} \bsnm{{Lacombe}}},
\oauthor{\binits{A.} \bsnm{{Mangeney}}},
\oauthor{\binits{R.} \bsnm{{Grappin}}},
{Fluid-like dissipation of magnetic turbulence at electron scales in the solar
  wind}.
arXiv:1111.5649v1, in preparation
(2011)
\end{botherref}
\endbibitem

\bibitem[\protect\citeauthoryear{{Alexandrova} et~al.}{2012}]{alexandrova12}
\begin{barticle}
\bauthor{\binits{O.} \bsnm{{Alexandrova}}},
\bauthor{\binits{C.} \bsnm{{Lacombe}}},
\bauthor{\binits{A.} \bsnm{{Mangeney}}},
\bauthor{\binits{R.} \bsnm{{Grappin}}},
\bauthor{\binits{M.} \bsnm{{Maksimovic}}},
\batitle{{Solar Wind Turbulent Spectrum at Plasma Kinetic Scales}}.
\bjtitle{\apj}
\bvolume{760},
\bfpage{121}
(\byear{2012}).
doi:\doiurl{10.1088/0004-637X/760/2/121}
\end{barticle}
\endbibitem

\bibitem[\protect\citeauthoryear{{Bale} et~al.}{2005}]{bale05}
\begin{barticle}
\bauthor{\binits{S.D.} \bsnm{{Bale}}},
\bauthor{\binits{P.J.} \bsnm{{Kellogg}}},
\bauthor{\binits{F.S.} \bsnm{{Mozer}}},
\bauthor{\binits{T.S.} \bsnm{{Horbury}}},
\bauthor{\binits{H.} \bsnm{{Reme}}},
\batitle{{Measurement of the Electric Fluctuation Spectrum of
  Magnetohydrodynamic Turbulence}}.
\bjtitle{\prl}
\bvolume{94}(\bissue{21}),
\bfpage{215002}
(\byear{2005}).
doi:\doiurl{10.1103/PhysRevLett.94.215002}
\end{barticle}
\endbibitem

\bibitem[\protect\citeauthoryear{{Bale} et~al.}{2009}]{bale09}
\begin{barticle}
\bauthor{\binits{S.D.} \bsnm{{Bale}}},
\bauthor{\binits{J.C.} \bsnm{{Kasper}}},
\bauthor{\binits{G.G.} \bsnm{{Howes}}},
\bauthor{\binits{E.} \bsnm{{Quataert}}},
\bauthor{\binits{C.} \bsnm{{Salem}}},
\bauthor{\binits{D.} \bsnm{{Sundkvist}}},
\batitle{{Magnetic Fluctuation Power Near Proton Temperature Anisotropy
  Instability Thresholds in the Solar Wind}}.
\bjtitle{\prl}
\bvolume{103},
\bfpage{211101}
(\byear{2009}).
doi:\doiurl{10.1103/PhysRevLett.103.211101}
\end{barticle}
\endbibitem

\bibitem[\protect\citeauthoryear{{Balogh} et~al.}{2001}]{balogh01_cluster}
\begin{barticle}
\bauthor{\binits{A.} \bsnm{{Balogh}}},
\bauthor{\binits{C.M.} \bsnm{{Carr}}},
\bauthor{\binits{M.H.} \bsnm{{Acu{\~n}a}}},
\bauthor{\binits{M.W.} \bsnm{{Dunlop}}},
\bauthor{\binits{T.J.} \bsnm{{Beek}}},
\bauthor{\binits{P.} \bsnm{{Brown}}},
\bauthor{\binits{K.-H.} \bsnm{{Forna{\c c}on}}},
\bauthor{\binits{E.} \bsnm{{Georgescu}}},
\bauthor{\binits{K.-H.} \bsnm{{Glassmeier}}},
\bauthor{\binits{J.} \bsnm{{Harris}}},
\bauthor{\binits{G.} \bsnm{{Musmann}}},
\bauthor{\binits{T.} \bsnm{{Oddy}}},
\bauthor{\binits{K.} \bsnm{{Schwingenschuh}}},
\batitle{{The Cluster Magnetic Field Investigation: overview of in-flight
  performance and initial results}}.
\bjtitle{Annales Geophysicae}
\bvolume{19},
\bfpage{1207}--\blpage{1217}
(\byear{2001}).
doi:\doiurl{10.5194/angeo-19-1207-2001}
\end{barticle}
\endbibitem

\bibitem[\protect\citeauthoryear{{Bershadskii} and
  {Sreenivasan}}{2004}]{bershadskii04}
\begin{barticle}
\bauthor{\binits{A.} \bsnm{{Bershadskii}}},
\bauthor{\binits{K.R.} \bsnm{{Sreenivasan}}},
\batitle{{Intermittency and the Passive Nature of the Magnitude of the Magnetic
  Field}}.
\bjtitle{\prl}
\bvolume{93}(\bissue{6}),
\bfpage{064501}
(\byear{2004}).
doi:\doiurl{10.1103/PhysRevLett.93.064501}
\end{barticle}
\endbibitem

\bibitem[\protect\citeauthoryear{{Bieber} et~al.}{1996}]{bieber96}
\begin{barticle}
\bauthor{\binits{J.W.} \bsnm{{Bieber}}},
\bauthor{\binits{W.} \bsnm{{Wanner}}},
\bauthor{\binits{W.H.} \bsnm{{Matthaeus}}},
\batitle{{Dominant two-dimensional solar wind turbulence with implications for
  cosmic ray transport}}.
\bjtitle{\jgr}
\bvolume{101},
\bfpage{2511}--\blpage{2522}
(\byear{1996}).
doi:\doiurl{10.1029/95JA02588}
\end{barticle}
\endbibitem

\bibitem[\protect\citeauthoryear{{Biskamp}}{1993}]{biskamp93}
\begin{bbook}
\bauthor{\binits{D.} \bsnm{{Biskamp}}},
\bbtitle{Nonlinear Magnetohydrodynamics}
\byear{1993}
\end{bbook}
\endbibitem

\bibitem[\protect\citeauthoryear{{Biskamp} et~al.}{1996}]{biskamp96}
\begin{barticle}
\bauthor{\binits{D.} \bsnm{{Biskamp}}},
\bauthor{\binits{E.} \bsnm{{Schwarz}}},
\bauthor{\binits{J.F.} \bsnm{{Drake}}},
\batitle{{Two-Dimensional Electron Magnetohydrodynamic Turbulence}}.
\bjtitle{\prl}
\bvolume{76},
\bfpage{1264}--\blpage{1267}
(\byear{1996}).
doi:\doiurl{10.1103/PhysRevLett.76.1264}
\end{barticle}
\endbibitem

\bibitem[\protect\citeauthoryear{{Biskamp} et~al.}{1999}]{biskamp99a}
\begin{barticle}
\bauthor{\binits{D.} \bsnm{{Biskamp}}},
\bauthor{\binits{E.} \bsnm{{Schwarz}}},
\bauthor{\binits{A.} \bsnm{{Zeiler}}},
\bauthor{\binits{A.} \bsnm{{Celani}}},
\bauthor{\binits{J.F.} \bsnm{{Drake}}},
\batitle{{Electron magnetohydrodynamic turbulence}}.
\bjtitle{\pop}
\bvolume{6},
\bfpage{751}--\blpage{758}
(\byear{1999}).
doi:\doiurl{10.1063/1.873312}
\end{barticle}
\endbibitem

\bibitem[\protect\citeauthoryear{{Boldyrev}}{2006}]{boldyrev06}
\begin{barticle}
\bauthor{\binits{S.} \bsnm{{Boldyrev}}},
\batitle{{Spectrum of Magnetohydrodynamic Turbulence}}.
\bjtitle{\prl}
\bvolume{96}(\bissue{11}),
\bfpage{115002}
(\byear{2006}).
doi:\doiurl{10.1103/PhysRevLett.96.115002}
\end{barticle}
\endbibitem

\bibitem[\protect\citeauthoryear{{Boldyrev} and {Perez}}{2009}]{boldyrev09a}
\begin{barticle}
\bauthor{\binits{S.} \bsnm{{Boldyrev}}},
\bauthor{\binits{J.C.} \bsnm{{Perez}}},
\batitle{{Spectrum of Weak Magnetohydrodynamic Turbulence}}.
\bjtitle{\prl}
\bvolume{103}(\bissue{22}),
\bfpage{225001}
(\byear{2009}).
doi:\doiurl{10.1103/PhysRevLett.103.225001}
\end{barticle}
\endbibitem

\bibitem[\protect\citeauthoryear{{Boldyrev} and {Perez}}{2012}]{boldyrev12b}
\begin{barticle}
\bauthor{\binits{S.} \bsnm{{Boldyrev}}},
\bauthor{\binits{J.C.} \bsnm{{Perez}}},
\batitle{{Spectrum of Kinetic-Alfv{\'e}n Turbulence}}.
\bjtitle{\apjl}
\bvolume{758},
\bfpage{44}
(\byear{2012}).
doi:\doiurl{10.1088/2041-8205/758/2/L44}
\end{barticle}
\endbibitem

\bibitem[\protect\citeauthoryear{{Boldyrev} et~al.}{2012}]{boldyrev12a}
\begin{botherref}
\oauthor{\binits{S.} \bsnm{{Boldyrev}}},
\oauthor{\binits{J.C.} \bsnm{{Perez}}},
\oauthor{\binits{Y.} \bsnm{{Wang}}},
{Residual Energy in Weak and Strong MHD Turbulence}.
ASP Conf.~Ser.~(submitted)
(2012)
\end{botherref}
\endbibitem

\bibitem[\protect\citeauthoryear{{Boldyrev} et~al.}{2011}]{boldyrev11a}
\begin{barticle}
\bauthor{\binits{S.} \bsnm{{Boldyrev}}},
\bauthor{\binits{J.C.} \bsnm{{Perez}}},
\bauthor{\binits{J.E.} \bsnm{{Borovsky}}},
\bauthor{\binits{J.J.} \bsnm{{Podesta}}},
\batitle{{Spectral Scaling Laws in Magnetohydrodynamic Turbulence Simulations
  and in the Solar Wind}}.
\bjtitle{\apjl}
\bvolume{741},
\bfpage{19}
(\byear{2011}).
doi:\doiurl{10.1088/2041-8205/741/1/L19}
\end{barticle}
\endbibitem

\bibitem[\protect\citeauthoryear{{Borovsky}}{2008}]{borovsky08}
\begin{barticle}
\bauthor{\binits{J.E.} \bsnm{{Borovsky}}},
\batitle{{Flux tube texture of the solar wind: Strands of the magnetic carpet
  at 1 AU?}}
\bjtitle{\jgr}
\bvolume{113}(\bissue{A12}),
\bfpage{8110}
(\byear{2008}).
doi:\doiurl{10.1029/2007JA012684}
\end{barticle}
\endbibitem

\bibitem[\protect\citeauthoryear{{Borovsky}}{2012a}]{borovsky12_mixing}
\begin{barticle}
\bauthor{\binits{J.E.} \bsnm{{Borovsky}}},
\batitle{{Looking for evidence of mixing in the solar wind from 0.31 to 0.98
  AU}}.
\bjtitle{Journal of Geophysical Research (Space Physics)}
\bvolume{117}(\bissue{A16}),
\bfpage{6107}
(\byear{2012}a).
doi:\doiurl{10.1029/2012JA017525}
\end{barticle}
\endbibitem

\bibitem[\protect\citeauthoryear{{Borovsky}}{2012b}]{borovsky12_bv}
\begin{barticle}
\bauthor{\binits{J.E.} \bsnm{{Borovsky}}},
\batitle{{The velocity and magnetic field fluctuations of the solar wind at 1
  AU: Statistical analysis of Fourier spectra and correlations with plasma
  properties}}.
\bjtitle{Journal of Geophysical Research (Space Physics)}
\bvolume{117}(\bissue{A16}),
\bfpage{5104}
(\byear{2012}b).
doi:\doiurl{10.1029/2011JA017499}
\end{barticle}
\endbibitem

\bibitem[\protect\citeauthoryear{{Bourouaine} et~al.}{2010}]{bourouaine10}
\begin{barticle}
\bauthor{\binits{S.} \bsnm{{Bourouaine}}},
\bauthor{\binits{E.} \bsnm{{Marsch}}},
\bauthor{\binits{F.M.} \bsnm{{Neubauer}}},
\batitle{{Correlations between the proton temperature anisotropy and transverse
  high-frequency waves in the solar wind}}.
\bjtitle{\grl}
\bvolume{37},
\bfpage{14104}
(\byear{2010}).
doi:\doiurl{10.1029/2010GL043697}
\end{barticle}
\endbibitem

\bibitem[\protect\citeauthoryear{{Bourouaine} et~al.}{2011}]{bourouaine11}
\begin{barticle}
\bauthor{\binits{S.} \bsnm{{Bourouaine}}},
\bauthor{\binits{E.} \bsnm{{Marsch}}},
\bauthor{\binits{F.M.} \bsnm{{Neubauer}}},
\batitle{{Temperature anisotropy and differential streaming of solar wind ions.
  Correlations with transverse fluctuations}}.
\bjtitle{\aap}
\bvolume{536},
\bfpage{39}
(\byear{2011}).
doi:\doiurl{10.1051/0004-6361/201117866}
\end{barticle}
\endbibitem

\bibitem[\protect\citeauthoryear{{Bourouaine} et~al.}{2012}]{bourouaine12}
\begin{barticle}
\bauthor{\binits{S.} \bsnm{{Bourouaine}}},
\bauthor{\binits{O.} \bsnm{{Alexandrova}}},
\bauthor{\binits{E.} \bsnm{{Marsch}}},
\bauthor{\binits{M.} \bsnm{{Maksimovic}}},
\batitle{{On Spectral Breaks in the Power Spectra of Magnetic Fluctuations in
  Fast Solar Wind between 0.3 and 0.9 AU}}.
\bjtitle{\apj}
\bvolume{749},
\bfpage{102}
(\byear{2012}).
doi:\doiurl{10.1088/0004-637X/749/2/102}
\end{barticle}
\endbibitem

\bibitem[\protect\citeauthoryear{{Bruno} and {Carbone}}{2005}]{bruno05a}
\begin{barticle}
\bauthor{\binits{R.} \bsnm{{Bruno}}},
\bauthor{\binits{V.} \bsnm{{Carbone}}},
\batitle{{The Solar Wind as a Turbulence Laboratory}}.
\bjtitle{\lrsp}
\bvolume{2},
\bfpage{4}
(\byear{2005})
\end{barticle}
\endbibitem

\bibitem[\protect\citeauthoryear{Bruno et~al.}{2001}]{bruno01}
\begin{barticle}
\bauthor{\binits{R.} \bsnm{Bruno}},
\bauthor{\binits{V.} \bsnm{Carbone}},
\bauthor{\binits{P.} \bsnm{Veltri}},
\bauthor{\binits{E.} \bsnm{Pietropaolo}},
\bauthor{\binits{B.} \bsnm{Bavassano}},
\batitle{Identifying intermittency events in the solar wind}.
\bjtitle{Planetary and Space Science}
\bvolume{49}(\bissue{12}),
\bfpage{1201}--\blpage{1210}
(\byear{2001}).
\bcomment{<ce:title>Nonlinear Dynamics and Fraactals in Space</ce:title>}.
doi:\doiurl{10.1016/S0032-0633(01)00061-7}.
\burl{http://www.sciencedirect.com/science/article/pii/S0032063301000617}
\end{barticle}
\endbibitem

\bibitem[\protect\citeauthoryear{{Bruno} et~al.}{2003}]{bruno03}
\begin{barticle}
\bauthor{\binits{R.} \bsnm{{Bruno}}},
\bauthor{\binits{V.} \bsnm{{Carbone}}},
\bauthor{\binits{L.} \bsnm{{Sorriso-Valvo}}},
\bauthor{\binits{B.} \bsnm{{Bavassano}}},
\batitle{{Radial evolution of solar wind intermittency in the inner
  heliosphere}}.
\bjtitle{Journal of Geophysical Research (Space Physics)}
\bvolume{108},
\bfpage{1130}
(\byear{2003}).
doi:\doiurl{10.1029/2002JA009615}
\end{barticle}
\endbibitem

\bibitem[\protect\citeauthoryear{{Bruno} et~al.}{2004}]{bruno04}
\begin{barticle}
\bauthor{\binits{R.} \bsnm{{Bruno}}},
\bauthor{\binits{V.} \bsnm{{Carbone}}},
\bauthor{\binits{L.} \bsnm{{Primavera}}},
\bauthor{\binits{F.} \bsnm{{Malara}}},
\bauthor{\binits{L.} \bsnm{{Sorriso-Valvo}}},
\bauthor{\binits{B.} \bsnm{{Bavassano}}},
\bauthor{\binits{P.} \bsnm{{Veltri}}},
\batitle{{On the probability distribution function of small-scale
  interplanetary magnetic field fluctuations}}.
\bjtitle{Annales Geophysicae}
\bvolume{22},
\bfpage{3751}--\blpage{3769}
(\byear{2004}).
doi:\doiurl{10.5194/angeo-22-3751-2004}
\end{barticle}
\endbibitem

\bibitem[\protect\citeauthoryear{{Bruno} et~al.}{2007}]{bruno07}
\begin{barticle}
\bauthor{\binits{R.} \bsnm{{Bruno}}},
\bauthor{\binits{R.} \bsnm{{D'Amicis}}},
\bauthor{\binits{B.} \bsnm{{Bavassano}}},
\bauthor{\binits{V.} \bsnm{{Carbone}}},
\bauthor{\binits{L.} \bsnm{{Sorriso-Valvo}}},
\batitle{{Magnetically dominated structures as an important component of the
  solar wind turbulence}}.
\bjtitle{\ang}
\bvolume{25},
\bfpage{1913}--\blpage{1927}
(\byear{2007})
\end{barticle}
\endbibitem

\bibitem[\protect\citeauthoryear{{Burlaga}}{1991}]{burlaga91}
\begin{barticle}
\bauthor{\binits{L.F.} \bsnm{{Burlaga}}},
\batitle{{Intermittent turbulence in the solar wind}}.
\bjtitle{\jgr}
\bvolume{96},
\bfpage{5847}--\blpage{5851}
(\byear{1991}).
doi:\doiurl{10.1029/91JA00087}
\end{barticle}
\endbibitem

\bibitem[\protect\citeauthoryear{Burlaga}{1993}]{burlaga93}
\begin{barticle}
\bauthor{\binits{L.F.} \bsnm{Burlaga}},
\batitle{Intermittent turbulence in large-scale velocity fluctuations at 1 au
  near solar maximum}.
\bjtitle{Journal of Geophysical Research: Space Physics}
\bvolume{98}(\bissue{A10}),
\bfpage{17467}--\blpage{17473}
(\byear{1993}).
doi:\doiurl{10.1029/93JA01630}.
\burl{http://dx.doi.org/10.1029/93JA01630}
\end{barticle}
\endbibitem

\bibitem[\protect\citeauthoryear{{Carbone} et~al.}{2009a}]{carbone09b}
\begin{barticle}
\bauthor{\binits{V.} \bsnm{{Carbone}}},
\bauthor{\binits{L.} \bsnm{{Sorriso-Valvo}}},
\bauthor{\binits{R.} \bsnm{{Marino}}},
\batitle{{On the turbulent energy cascade in anisotropic magnetohydrodynamic
  turbulence}}.
\bjtitle{EPL (Europhysics Letters)}
\bvolume{88},
\bfpage{25001}
(\byear{2009}a).
doi:\doiurl{10.1209/0295-5075/88/25001}
\end{barticle}
\endbibitem

\bibitem[\protect\citeauthoryear{{Carbone} et~al.}{2009b}]{carbone09}
\begin{barticle}
\bauthor{\binits{V.} \bsnm{{Carbone}}},
\bauthor{\binits{R.} \bsnm{{Marino}}},
\bauthor{\binits{L.} \bsnm{{Sorriso-Valvo}}},
\bauthor{\binits{A.} \bsnm{{Noullez}}},
\bauthor{\binits{R.} \bsnm{{Bruno}}},
\batitle{{Scaling Laws of Turbulence and Heating of Fast Solar Wind: The Role
  of Density Fluctuations}}.
\bjtitle{\prl}
\bvolume{103}(\bissue{6}),
\bfpage{061102}
(\byear{2009}b).
doi:\doiurl{10.1103/PhysRevLett.103.061102}
\end{barticle}
\endbibitem

\bibitem[\protect\citeauthoryear{Carbone et~al.}{1995}]{carbone95}
\begin{barticle}
\bauthor{\binits{V.} \bsnm{Carbone}},
\bauthor{\binits{P.} \bsnm{Veltri}},
\bauthor{\binits{R.} \bsnm{Bruno}},
\batitle{Experimental evidence for differences in the extended self-similarity
  scaling laws between fluid and magnetohydrodynamic turbulent flows}.
\bjtitle{Phys. Rev. Lett.}
\bvolume{75},
\bfpage{3110}--\blpage{3113}
(\byear{1995}).
doi:\doiurl{10.1103/PhysRevLett.75.3110}.
\burl{http://link.aps.org/doi/10.1103/PhysRevLett.75.3110}
\end{barticle}
\endbibitem

\bibitem[\protect\citeauthoryear{{Celnikier} et~al.}{1983}]{celnikier83}
\begin{barticle}
\bauthor{\binits{L.M.} \bsnm{{Celnikier}}},
\bauthor{\binits{C.C.} \bsnm{{Harvey}}},
\bauthor{\binits{R.} \bsnm{{Jegou}}},
\bauthor{\binits{P.} \bsnm{{Moricet}}},
\bauthor{\binits{M.} \bsnm{{Kemp}}},
\batitle{{A determination of the electron density fluctuation spectrum in the
  solar wind, using the ISEE propagation experiment}}.
\bjtitle{\aap}
\bvolume{126},
\bfpage{293}--\blpage{298}
(\byear{1983})
\end{barticle}
\endbibitem

\bibitem[\protect\citeauthoryear{{Chandran} et~al.}{2009}]{chandran09c}
\begin{barticle}
\bauthor{\binits{B.D.G.} \bsnm{{Chandran}}},
\bauthor{\binits{E.} \bsnm{{Quataert}}},
\bauthor{\binits{G.G.} \bsnm{{Howes}}},
\bauthor{\binits{Q.} \bsnm{{Xia}}},
\bauthor{\binits{P.} \bsnm{{Pongkitiwanichakul}}},
\batitle{{Constraining Low-Frequency Alfv{\'e}nic Turbulence in the Solar Wind
  Using Density-Fluctuation Measurements}}.
\bjtitle{\apj}
\bvolume{707},
\bfpage{1668}--\blpage{1675}
(\byear{2009}).
doi:\doiurl{10.1088/0004-637X/707/2/1668}
\end{barticle}
\endbibitem

\bibitem[\protect\citeauthoryear{{Chen} et~al.}{2010a}]{chen10b}
\begin{barticle}
\bauthor{\binits{C.H.K.} \bsnm{{Chen}}},
\bauthor{\binits{T.S.} \bsnm{{Horbury}}},
\bauthor{\binits{A.A.} \bsnm{{Schekochihin}}},
\bauthor{\binits{R.T.} \bsnm{{Wicks}}},
\bauthor{\binits{O.} \bsnm{{Alexandrova}}},
\bauthor{\binits{J.} \bsnm{{Mitchell}}},
\batitle{{Anisotropy of Solar Wind Turbulence between Ion and Electron
  Scales}}.
\bjtitle{\prl}
\bvolume{104},
\bfpage{255002}
(\byear{2010}a).
doi:\doiurl{10.1103/PhysRevLett.104.255002}
\end{barticle}
\endbibitem

\bibitem[\protect\citeauthoryear{{Chen} et~al.}{2010b}]{chen10a}
\begin{barticle}
\bauthor{\binits{C.H.K.} \bsnm{{Chen}}},
\bauthor{\binits{R.T.} \bsnm{{Wicks}}},
\bauthor{\binits{T.S.} \bsnm{{Horbury}}},
\bauthor{\binits{A.A.} \bsnm{{Schekochihin}}},
\batitle{{Interpreting Power Anisotropy Measurements in Plasma Turbulence}}.
\bjtitle{\apjl}
\bvolume{711},
\bfpage{79}--\blpage{83}
(\byear{2010}b).
doi:\doiurl{10.1088/2041-8205/711/2/L79}
\end{barticle}
\endbibitem

\bibitem[\protect\citeauthoryear{{Chen} et~al.}{2011a}]{chen11a}
\begin{barticle}
\bauthor{\binits{C.H.K.} \bsnm{{Chen}}},
\bauthor{\binits{A.} \bsnm{{Mallet}}},
\bauthor{\binits{T.A.} \bsnm{{Yousef}}},
\bauthor{\binits{A.A.} \bsnm{{Schekochihin}}},
\bauthor{\binits{T.S.} \bsnm{{Horbury}}},
\batitle{{Anisotropy of Alfv\'enic Turbulence in the Solar Wind and Numerical
  Simulations}}.
\bjtitle{\mnras}
\bvolume{415},
\bfpage{3219}
(\byear{2011}a).
doi:\doiurl{doi:10.1111/j.1365-2966.2011.18933.x}
\end{barticle}
\endbibitem

\bibitem[\protect\citeauthoryear{{Chen} et~al.}{2011b}]{chen11b}
\begin{barticle}
\bauthor{\binits{C.H.K.} \bsnm{{Chen}}},
\bauthor{\binits{S.D.} \bsnm{{Bale}}},
\bauthor{\binits{C.} \bsnm{{Salem}}},
\bauthor{\binits{F.S.} \bsnm{{Mozer}}},
\batitle{{Frame Dependence of the Electric Field Spectrum of Solar Wind
  Turbulence}}.
\bjtitle{\apjl}
\bvolume{737},
\bfpage{41}
(\byear{2011}b).
doi:\doiurl{10.1088/2041-8205/737/2/L41}
\end{barticle}
\endbibitem

\bibitem[\protect\citeauthoryear{{Chen} et~al.}{2012a}]{chen12a}
\begin{barticle}
\bauthor{\binits{C.H.K.} \bsnm{{Chen}}},
\bauthor{\binits{C.S.} \bsnm{{Salem}}},
\bauthor{\binits{J.W.} \bsnm{{Bonnell}}},
\bauthor{\binits{F.S.} \bsnm{{Mozer}}},
\bauthor{\binits{S.D.} \bsnm{{Bale}}},
\batitle{{Density Fluctuation Spectrum of Solar Wind Turbulence between Ion and
  Electron Scales}}.
\bjtitle{\prl}
\bvolume{109}(\bissue{3}),
\bfpage{035001}
(\byear{2012}a).
doi:\doiurl{10.1103/PhysRevLett.109.035001}
\end{barticle}
\endbibitem

\bibitem[\protect\citeauthoryear{{Chen} et~al.}{2012b}]{chen12b}
\begin{barticle}
\bauthor{\binits{C.H.K.} \bsnm{{Chen}}},
\bauthor{\binits{A.} \bsnm{{Mallet}}},
\bauthor{\binits{A.A.} \bsnm{{Schekochihin}}},
\bauthor{\binits{T.S.} \bsnm{{Horbury}}},
\bauthor{\binits{R.T.} \bsnm{{Wicks}}},
\bauthor{\binits{S.D.} \bsnm{{Bale}}},
\batitle{{Three-dimensional Structure of Solar Wind Turbulence}}.
\bjtitle{\apj}
\bvolume{758},
\bfpage{120}
(\byear{2012}b).
doi:\doiurl{10.1088/0004-637X/758/2/120}
\end{barticle}
\endbibitem

\bibitem[\protect\citeauthoryear{{Chen} et~al.}{2013a}]{chen12c}
\begin{barticle}
\bauthor{\binits{C.H.K.} \bsnm{{Chen}}},
\bauthor{\binits{G.G.} \bsnm{{Howes}}},
\bauthor{\binits{J.W.} \bsnm{{Bonnell}}},
\bauthor{\binits{F.S.} \bsnm{{Mozer}}},
\bauthor{\binits{K.G.} \bsnm{{Klein}}},
\bauthor{\binits{S.D.} \bsnm{{Bale}}},
\batitle{Kinetic scale density fluctuations in the solar wind}.
\bjtitle{{Solar Wind 13 Proceedings, arXiv:1210.0127}}
\bvolume{1539},
\bfpage{143}--\blpage{146}
(\byear{2013}a)
\end{barticle}
\endbibitem

\bibitem[\protect\citeauthoryear{{Chen} et~al.}{2013b}]{chen13a}
\begin{barticle}
\bauthor{\binits{C.H.K.} \bsnm{{Chen}}},
\bauthor{\binits{S.D.} \bsnm{{Bale}}},
\bauthor{\binits{C.S.} \bsnm{{Salem}}},
\bauthor{\binits{B.A.} \bsnm{{Maruca}}},
\batitle{Residual energy spectrum of solar wind turbulence}.
\bjtitle{\apj}
\bvolume{770},
\bfpage{125}
(\byear{2013}b).
doi:\doiurl{10.1088/0004-637X/770/2/125}
\end{barticle}
\endbibitem

\bibitem[\protect\citeauthoryear{{Chen} et~al.}{2013c}]{chen13b}
\begin{barticle}
\bauthor{\binits{C.H.K.} \bsnm{{Chen}}},
\bauthor{\binits{S.} \bsnm{{Boldyrev}}},
\bauthor{\binits{Q.} \bsnm{{Xia}}},
\bauthor{\binits{J.C.} \bsnm{{Perez}}},
\batitle{{The Nature of Subproton Scale Turbulence in the Solar Wind}}.
\bjtitle{Phys. Rev. Lett.}
\bvolume{110},
\bfpage{225002}
(\byear{2013}c).
doi:\doiurl{10.1103/PhysRevLett.110.225002}
\end{barticle}
\endbibitem

\bibitem[\protect\citeauthoryear{{Chen} et~al.}{1993}]{chen93}
\begin{barticle}
\bauthor{\binits{S.} \bsnm{{Chen}}},
\bauthor{\binits{G.} \bsnm{{Doolen}}},
\bauthor{\binits{J.R.} \bsnm{{Herring}}},
\bauthor{\binits{R.H.} \bsnm{{Kraichnan}}},
\bauthor{\binits{S.A.} \bsnm{{Orszag}}},
\bauthor{\binits{Z.S.} \bsnm{{She}}},
\batitle{{Far-dissipation range of turbulence}}.
\bjtitle{Physical Review Letters}
\bvolume{70},
\bfpage{3051}--\blpage{3054}
(\byear{1993}).
doi:\doiurl{10.1103/PhysRevLett.70.3051}
\end{barticle}
\endbibitem

\bibitem[\protect\citeauthoryear{{Chew} et~al.}{1956}]{CGL56}
\begin{barticle}
\bauthor{\binits{G.F.} \bsnm{{Chew}}},
\bauthor{\binits{M.L.} \bsnm{{Goldberger}}},
\bauthor{\binits{F.E.} \bsnm{{Low}}},
\batitle{{The Boltzmann Equation and the One-Fluid Hydromagnetic Equations in
  the Absence of Particle Collisions}}.
\bjtitle{Royal Society of London Proceedings Series A}
\bvolume{236},
\bfpage{112}--\blpage{118}
(\byear{1956}).
doi:\doiurl{10.1098/rspa.1956.0116}
\end{barticle}
\endbibitem

\bibitem[\protect\citeauthoryear{{Cho} and {Lazarian}}{2004}]{cho04}
\begin{barticle}
\bauthor{\binits{J.} \bsnm{{Cho}}},
\bauthor{\binits{A.} \bsnm{{Lazarian}}},
\batitle{{The Anisotropy of Electron Magnetohydrodynamic Turbulence}}.
\bjtitle{\apjl}
\bvolume{615},
\bfpage{41}--\blpage{44}
(\byear{2004}).
doi:\doiurl{10.1086/425215}
\end{barticle}
\endbibitem

\bibitem[\protect\citeauthoryear{{Cho} and {Vishniac}}{2000}]{cho00}
\begin{barticle}
\bauthor{\binits{J.} \bsnm{{Cho}}},
\bauthor{\binits{E.T.} \bsnm{{Vishniac}}},
\batitle{{The Anisotropy of Magnetohydrodynamic Alfv{\'e}nic Turbulence}}.
\bjtitle{\apj}
\bvolume{539},
\bfpage{273}--\blpage{282}
(\byear{2000}).
doi:\doiurl{10.1086/309213}
\end{barticle}
\endbibitem

\bibitem[\protect\citeauthoryear{{Coburn} et~al.}{2012}]{coburn12}
\begin{barticle}
\bauthor{\binits{J.T.} \bsnm{{Coburn}}},
\bauthor{\binits{C.W.} \bsnm{{Smith}}},
\bauthor{\binits{B.J.} \bsnm{{Vasquez}}},
\bauthor{\binits{J.E.} \bsnm{{Stawarz}}},
\bauthor{\binits{M.A.} \bsnm{{Forman}}},
\batitle{{The Turbulent Cascade and Proton Heating in the Solar Wind during
  Solar Minimum}}.
\bjtitle{\apj}
\bvolume{754},
\bfpage{93}
(\byear{2012}).
doi:\doiurl{10.1088/0004-637X/754/2/93}
\end{barticle}
\endbibitem

\bibitem[\protect\citeauthoryear{{Danaila} et~al.}{2001}]{danaila01}
\begin{barticle}
\bauthor{\binits{L.} \bsnm{{Danaila}}},
\bauthor{\binits{F.} \bsnm{{Anselmet}}},
\bauthor{\binits{T.} \bsnm{{Zhou}}},
\bauthor{\binits{R.A.} \bsnm{{Antonia}}},
\batitle{{Turbulent energy scale budget equations in a fully developed channel
  flow}}.
\bjtitle{Journal of Fluid Mechanics}
\bvolume{430},
\bfpage{87}--\blpage{109}
(\byear{2001}).
doi:\doiurl{10.1017/S0022112000002767}.
\burl{http://dx.doi.org/10.1017/S0022112000002767}
\end{barticle}
\endbibitem

\bibitem[\protect\citeauthoryear{Davidson}{2004}]{Davidson04book}
\begin{bbook}
\bauthor{\binits{P.A.} \bsnm{Davidson}},
\bbtitle{Turbulence : an Introduction for Scientists and Engineers}
\byear{2004}
\end{bbook}
\endbibitem

\bibitem[\protect\citeauthoryear{{Denskat} et~al.}{1983}]{denskat83}
\begin{barticle}
\bauthor{\binits{K.U.} \bsnm{{Denskat}}},
\bauthor{\binits{H.J.} \bsnm{{Beinroth}}},
\bauthor{\binits{F.M.} \bsnm{{Neubauer}}},
\batitle{{Interplanetary magnetic field power spectra with frequencies from 2.4
  X 10 to the -5th HZ to 470 HZ from HELIOS-observations during solar minimum
  conditions}}.
\bjtitle{\jg}
\bvolume{54},
\bfpage{60}--\blpage{67}
(\byear{1983})
\end{barticle}
\endbibitem

\bibitem[\protect\citeauthoryear{{Dobrowolny} et~al.}{1980}]{dobrowolny80}
\begin{barticle}
\bauthor{\binits{M.} \bsnm{{Dobrowolny}}},
\bauthor{\binits{A.} \bsnm{{Mangeney}}},
\bauthor{\binits{P.} \bsnm{{Veltri}}},
\batitle{{Fully developed anisotropic hydromagnetic turbulence in
  interplanetary space}}.
\bjtitle{\prl}
\bvolume{45},
\bfpage{144}--\blpage{147}
(\byear{1980}).
doi:\doiurl{10.1103/PhysRevLett.45.144}
\end{barticle}
\endbibitem

\bibitem[\protect\citeauthoryear{{Dudok de Wit} et~al.}{2013}]{dudokdewit13}
\begin{barticle}
\bauthor{\binits{T.} \bsnm{{Dudok de Wit}}},
\bauthor{\binits{O.} \bsnm{{Alexandrova}}},
\bauthor{\binits{I.} \bsnm{{Furno}}},
\bauthor{\binits{L.} \bsnm{{Sorriso-Valvo}}},
\bauthor{\binits{G.} \bsnm{{Zimbardo}}},
\batitle{{Methods for Characterising Microphysical Processes in Plasmas}}.
\bjtitle{\ssr}
(\byear{2013}).
doi:\doiurl{10.1007/s11214-013-9974-9}
\end{barticle}
\endbibitem

\bibitem[\protect\citeauthoryear{{Frisch}}{1995}]{frisch95}
\begin{bbook}
\bauthor{\binits{U.} \bsnm{{Frisch}}},
\bbtitle{{Turbulence}}
(\bpublisher{Cambridge University Press}, \blocation{???}, \byear{1995})
\end{bbook}
\endbibitem

\bibitem[\protect\citeauthoryear{{Galtier}}{2006}]{galtier06b}
\begin{barticle}
\bauthor{\binits{S.} \bsnm{{Galtier}}},
\batitle{{Wave turbulence in incompressible Hall magnetohydrodynamics}}.
\bjtitle{\jplp}
\bvolume{72},
\bfpage{721}--\blpage{769}
(\byear{2006}).
doi:\doiurl{10.1017/S0022377806004521}
\end{barticle}
\endbibitem

\bibitem[\protect\citeauthoryear{{Galtier} et~al.}{2005}]{galtier05a}
\begin{barticle}
\bauthor{\binits{S.} \bsnm{{Galtier}}},
\bauthor{\binits{A.} \bsnm{{Pouquet}}},
\bauthor{\binits{A.} \bsnm{{Mangeney}}},
\batitle{{On spectral scaling laws for incompressible anisotropic
  magnetohydrodynamic turbulence}}.
\bjtitle{Physics of Plasmas}
\bvolume{12}(\bissue{9}),
\bfpage{092310}
(\byear{2005}).
doi:\doiurl{10.1063/1.2052507}
\end{barticle}
\endbibitem

\bibitem[\protect\citeauthoryear{{Gary} et~al.}{1996}]{gary96}
\begin{barticle}
\bauthor{\binits{P.C.} \bsnm{{Gary}}},
\bauthor{\binits{C.W.} \bsnm{{Smith}}},
\bauthor{\binits{W.H.} \bsnm{{Matthaeus}}},
\bauthor{\binits{N.F.} \bsnm{{Otani}}},
\batitle{{Heating of the solar wind by pickup ion driven Alfv{\'e}n ion
  cyclotron instability}}.
\bjtitle{\grl}
\bvolume{23},
\bfpage{113}--\blpage{116}
(\byear{1996}).
doi:\doiurl{10.1029/95GL03707}
\end{barticle}
\endbibitem

\bibitem[\protect\citeauthoryear{{Gary}}{1993}]{gary93book}
\begin{bbook}
\bauthor{\binits{S.P.} \bsnm{{Gary}}},
\bbtitle{Theory of Space Plasma Microinstabilities}
\byear{1993}
\end{bbook}
\endbibitem

\bibitem[\protect\citeauthoryear{{Gary} and {Smith}}{2009}]{gary09}
\begin{barticle}
\bauthor{\binits{S.P.} \bsnm{{Gary}}},
\bauthor{\binits{C.W.} \bsnm{{Smith}}},
\batitle{{Short-wavelength turbulence in the solar wind: Linear theory of
  whistler and kinetic Alfv\'en fluctuations}}.
\bjtitle{\jgr}
\bvolume{114},
\bfpage{12105}
(\byear{2009}).
doi:\doiurl{10.1029/2009JA014525}
\end{barticle}
\endbibitem

\bibitem[\protect\citeauthoryear{{Gary} et~al.}{1976}]{gary76}
\begin{barticle}
\bauthor{\binits{S.P.} \bsnm{{Gary}}},
\bauthor{\binits{M.D.} \bsnm{{Montgomery}}},
\bauthor{\binits{W.C.} \bsnm{{Feldman}}},
\bauthor{\binits{D.W.} \bsnm{{Forslund}}},
\batitle{{Proton temperature anisotropy instabilities in the solar wind}}.
\bjtitle{\jgr}
\bvolume{81},
\bfpage{1241}--\blpage{1246}
(\byear{1976}).
doi:\doiurl{10.1029/JA081i007p01241}
\end{barticle}
\endbibitem

\bibitem[\protect\citeauthoryear{{Gary} et~al.}{2001}]{gary01}
\begin{barticle}
\bauthor{\binits{S.P.} \bsnm{{Gary}}},
\bauthor{\binits{R.M.} \bsnm{{Skoug}}},
\bauthor{\binits{J.T.} \bsnm{{Steinberg}}},
\bauthor{\binits{C.W.} \bsnm{{Smith}}},
\batitle{{Proton temperature anisotropy constraint in the solar wind: ACE
  observations}}.
\bjtitle{\grl}
\bvolume{28},
\bfpage{2759}--\blpage{2762}
(\byear{2001}).
doi:\doiurl{10.1029/2001GL013165}
\end{barticle}
\endbibitem

\bibitem[\protect\citeauthoryear{{Ghosh} et~al.}{1996}]{ghosh96}
\begin{barticle}
\bauthor{\binits{S.} \bsnm{{Ghosh}}},
\bauthor{\binits{E.} \bsnm{{Siregar}}},
\bauthor{\binits{D.A.} \bsnm{{Roberts}}},
\bauthor{\binits{M.L.} \bsnm{{Goldstein}}},
\batitle{{Simulation of high-frequency solar wind power spectra using Hall
  magnetohydrodynamics}}.
\bjtitle{\jgr}
\bvolume{101},
\bfpage{2493}--\blpage{2504}
(\byear{1996}).
doi:\doiurl{10.1029/95JA03201}
\end{barticle}
\endbibitem

\bibitem[\protect\citeauthoryear{{Goldreich} and {Sridhar}}{1995}]{goldreich95}
\begin{barticle}
\bauthor{\binits{P.} \bsnm{{Goldreich}}},
\bauthor{\binits{S.} \bsnm{{Sridhar}}},
\batitle{{Toward a theory of interstellar turbulence. II. Strong Alfv\'enic
  turbulence}}.
\bjtitle{\apj}
\bvolume{438},
\bfpage{763}--\blpage{775}
(\byear{1995}).
doi:\doiurl{10.1086/175121}
\end{barticle}
\endbibitem

\bibitem[\protect\citeauthoryear{{Goldreich} and {Sridhar}}{1997}]{goldreich97}
\begin{barticle}
\bauthor{\binits{P.} \bsnm{{Goldreich}}},
\bauthor{\binits{S.} \bsnm{{Sridhar}}},
\batitle{{Magnetohydrodynamic Turbulence Revisited}}.
\bjtitle{\apj}
\bvolume{485},
\bfpage{680}
(\byear{1997}).
doi:\doiurl{10.1086/304442}
\end{barticle}
\endbibitem

\bibitem[\protect\citeauthoryear{{Goldstein} et~al.}{1994}]{goldstein94}
\begin{barticle}
\bauthor{\binits{M.L.} \bsnm{{Goldstein}}},
\bauthor{\binits{D.A.} \bsnm{{Roberts}}},
\bauthor{\binits{C.A.} \bsnm{{Fitch}}},
\batitle{{Properties of the fluctuating magnetic helicity in the inertial and
  dissipation ranges of solar wind turbulence}}.
\bjtitle{\jgr}
\bvolume{99},
\bfpage{11519}--\blpage{11538}
(\byear{1994}).
doi:\doiurl{10.1029/94JA00789}
\end{barticle}
\endbibitem

\bibitem[\protect\citeauthoryear{{Grant} et~al.}{1962}]{grant62}
\begin{barticle}
\bauthor{\binits{H.L.} \bsnm{{Grant}}},
\bauthor{\binits{R.W.} \bsnm{{Stewart}}},
\bauthor{\binits{A.} \bsnm{{Moilliet}}},
\batitle{{Turbulence spectra from a tidal channel}}.
\bjtitle{Journal of Fluid Mechanics}
\bvolume{12},
\bfpage{241}--\blpage{268}
(\byear{1962}).
doi:\doiurl{10.1017/S002211206200018X}
\end{barticle}
\endbibitem

\bibitem[\protect\citeauthoryear{{Grappin} et~al.}{1983}]{grappin83}
\begin{barticle}
\bauthor{\binits{R.} \bsnm{{Grappin}}},
\bauthor{\binits{J.} \bsnm{{Leorat}}},
\bauthor{\binits{A.} \bsnm{{Pouquet}}},
\batitle{{Dependence of MHD turbulence spectra on the velocity field-magnetic
  field correlation}}.
\bjtitle{\aap}
\bvolume{126},
\bfpage{51}--\blpage{58}
(\byear{1983})
\end{barticle}
\endbibitem

\bibitem[\protect\citeauthoryear{{Grappin} et~al.}{1990}]{grappin90}
\begin{barticle}
\bauthor{\binits{R.} \bsnm{{Grappin}}},
\bauthor{\binits{A.} \bsnm{{Mangeney}}},
\bauthor{\binits{E.} \bsnm{{Marsch}}},
\batitle{{On the origin of solar wind MHD turbulence - HELIOS data revisited}}.
\bjtitle{\jgr}
\bvolume{95},
\bfpage{8197}--\blpage{8209}
(\byear{1990}).
doi:\doiurl{10.1029/JA095iA06p08197}
\end{barticle}
\endbibitem

\bibitem[\protect\citeauthoryear{{Grappin} et~al.}{1991}]{grappin91}
\begin{barticle}
\bauthor{\binits{R.} \bsnm{{Grappin}}},
\bauthor{\binits{M.} \bsnm{{Velli}}},
\bauthor{\binits{A.} \bsnm{{Mangeney}}},
\batitle{{``Alfv\'enic'' versus ``standard'' turbulence in the solar wind}}.
\bjtitle{\ang}
\bvolume{9},
\bfpage{416}--\blpage{426}
(\byear{1991})
\end{barticle}
\endbibitem

\bibitem[\protect\citeauthoryear{{Greco} et~al.}{2009}]{greco09}
\begin{barticle}
\bauthor{\binits{A.} \bsnm{{Greco}}},
\bauthor{\binits{W.H.} \bsnm{{Matthaeus}}},
\bauthor{\binits{S.} \bsnm{{Servidio}}},
\bauthor{\binits{P.} \bsnm{{Chuychai}}},
\bauthor{\binits{P.} \bsnm{{Dmitruk}}},
\batitle{{Statistical Analysis of Discontinuities in Solar Wind ACE Data and
  Comparison with Intermittent MHD Turbulence}}.
\bjtitle{\apjl}
\bvolume{691},
\bfpage{111}--\blpage{114}
(\byear{2009}).
doi:\doiurl{10.1088/0004-637X/691/2/L111}
\end{barticle}
\endbibitem

\bibitem[\protect\citeauthoryear{{Greco} et~al.}{2010}]{greco10}
\begin{barticle}
\bauthor{\binits{A.} \bsnm{{Greco}}},
\bauthor{\binits{S.} \bsnm{{Servidio}}},
\bauthor{\binits{W.H.} \bsnm{{Matthaeus}}},
\bauthor{\binits{P.} \bsnm{{Dmitruk}}},
\batitle{{Intermittent structures and magnetic discontinuities on small scales
  in MHD simulations and solar wind}}.
\bjtitle{\pss}
\bvolume{58},
\bfpage{1895}--\blpage{1899}
(\byear{2010}).
doi:\doiurl{10.1016/j.pss.2010.08.019}
\end{barticle}
\endbibitem

\bibitem[\protect\citeauthoryear{{Greco} et~al.}{2012}]{greco12}
\begin{barticle}
\bauthor{\binits{A.} \bsnm{{Greco}}},
\bauthor{\binits{W.H.} \bsnm{{Matthaeus}}},
\bauthor{\binits{R.} \bsnm{{D'Amicis}}},
\bauthor{\binits{S.} \bsnm{{Servidio}}},
\bauthor{\binits{P.} \bsnm{{Dmitruk}}},
\batitle{{Evidence for Nonlinear Development of Magnetohydrodynamic Scale
  Intermittency in the Inner Heliosphere}}.
\bjtitle{\apj}
\bvolume{749},
\bfpage{105}
(\byear{2012}).
doi:\doiurl{10.1088/0004-637X/749/2/105}
\end{barticle}
\endbibitem

\bibitem[\protect\citeauthoryear{{Hamilton} et~al.}{2008}]{hamilton08}
\begin{barticle}
\bauthor{\binits{K.} \bsnm{{Hamilton}}},
\bauthor{\binits{C.W.} \bsnm{{Smith}}},
\bauthor{\binits{B.J.} \bsnm{{Vasquez}}},
\bauthor{\binits{R.J.} \bsnm{{Leamon}}},
\batitle{{Anisotropies and helicities in the solar wind inertial and
  dissipation ranges at 1 AU}}.
\bjtitle{\jgr}
\bvolume{113}(\bissue{A12}),
\bfpage{1106}
(\byear{2008}).
doi:\doiurl{10.1029/2007JA012559}
\end{barticle}
\endbibitem

\bibitem[\protect\citeauthoryear{{Hasegawa}}{1969}]{hasegawa69}
\begin{barticle}
\bauthor{\binits{A.} \bsnm{{Hasegawa}}},
\batitle{{Drift mirror instability of the magnetosphere.}}
\bjtitle{Physics of Fluids}
\bvolume{12},
\bfpage{2642}--\blpage{2650}
(\byear{1969}).
doi:\doiurl{10.1063/1.1692407}
\end{barticle}
\endbibitem

\bibitem[\protect\citeauthoryear{{Haverkorn} and
  {Spangler}}{2013}]{haverkorn13}
\begin{botherref}
\oauthor{\binits{M.} \bsnm{{Haverkorn}}},
\oauthor{\binits{S.R.} \bsnm{{Spangler}}},
{Plasma Diagnostics of the Interstellar Medium with Radio Astronomy}.
submitted to \ssr
(2013)
\end{botherref}
\endbibitem

\bibitem[\protect\citeauthoryear{{He} et~al.}{2011a}]{he11a}
\begin{barticle}
\bauthor{\binits{J.-S.} \bsnm{{He}}},
\bauthor{\binits{E.} \bsnm{{Marsch}}},
\bauthor{\binits{C.-Y.} \bsnm{{Tu}}},
\bauthor{\binits{Q.-G.} \bsnm{{Zong}}},
\bauthor{\binits{S.} \bsnm{{Yao}}},
\bauthor{\binits{H.} \bsnm{{Tian}}},
\batitle{{Two-dimensional correlation functions for density and magnetic field
  fluctuations in magnetosheath turbulence measured by the Cluster
  spacecraft}}.
\bjtitle{\jgr}
\bvolume{116}(\bissue{A15}),
\bfpage{06207}
(\byear{2011}a).
doi:\doiurl{10.1029/2010JA015974}
\end{barticle}
\endbibitem

\bibitem[\protect\citeauthoryear{{He} et~al.}{2011b}]{he11b}
\begin{barticle}
\bauthor{\binits{J.} \bsnm{{He}}},
\bauthor{\binits{E.} \bsnm{{Marsch}}},
\bauthor{\binits{C.} \bsnm{{Tu}}},
\bauthor{\binits{S.} \bsnm{{Yao}}},
\bauthor{\binits{H.} \bsnm{{Tian}}},
\batitle{{Possible Evidence of Alfv{\'e}n-cyclotron Waves in the Angle
  Distribution of Magnetic Helicity of Solar Wind Turbulence}}.
\bjtitle{\apj}
\bvolume{731},
\bfpage{85}
(\byear{2011}b).
doi:\doiurl{10.1088/0004-637X/731/2/85}
\end{barticle}
\endbibitem

\bibitem[\protect\citeauthoryear{{Hellinger} and
  {Matsumoto}}{2000}]{hellinger00}
\begin{barticle}
\bauthor{\binits{P.} \bsnm{{Hellinger}}},
\bauthor{\binits{H.} \bsnm{{Matsumoto}}},
\batitle{{New kinetic instability: Oblique Alfv{\'e}n fire hose}}.
\bjtitle{\jgr}
\bvolume{105},
\bfpage{10519}--\blpage{10526}
(\byear{2000}).
doi:\doiurl{10.1029/1999JA000297}
\end{barticle}
\endbibitem

\bibitem[\protect\citeauthoryear{{Hellinger} and
  {Matsumoto}}{2001}]{hellinger01}
\begin{barticle}
\bauthor{\binits{P.} \bsnm{{Hellinger}}},
\bauthor{\binits{H.} \bsnm{{Matsumoto}}},
\batitle{{Nonlinear competition between the whistler and Alfv{\'e}n fire
  hoses}}.
\bjtitle{\jgr}
\bvolume{106},
\bfpage{13215}--\blpage{13218}
(\byear{2001}).
doi:\doiurl{10.1029/2001JA900026}
\end{barticle}
\endbibitem

\bibitem[\protect\citeauthoryear{{Hellinger} et~al.}{2006}]{hellinger06a}
\begin{barticle}
\bauthor{\binits{P.} \bsnm{{Hellinger}}},
\bauthor{\binits{P.} \bsnm{{Tr{\'a}vn{\'{\i}}{\v c}ek}}},
\bauthor{\binits{J.C.} \bsnm{{Kasper}}},
\bauthor{\binits{A.J.} \bsnm{{Lazarus}}},
\batitle{{Solar wind proton temperature anisotropy: Linear theory and WIND/SWE
  observations}}.
\bjtitle{\grl}
\bvolume{33},
\bfpage{09101}
(\byear{2006}).
doi:\doiurl{10.1029/2006GL025925}
\end{barticle}
\endbibitem

\bibitem[\protect\citeauthoryear{{Hellinger} et~al.}{2011}]{hellinger11}
\begin{barticle}
\bauthor{\binits{P.} \bsnm{{Hellinger}}},
\bauthor{\binits{L.} \bsnm{{Matteini}}},
\bauthor{\binits{{\v S}.} \bsnm{{{\v S}tver{\'a}k}}},
\bauthor{\binits{P.M.} \bsnm{{Tr{\'a}vn{\'{\i}}{\v c}ek}}},
\bauthor{\binits{E.} \bsnm{{Marsch}}},
\batitle{{Heating and cooling of protons in the fast solar wind between 0.3 and
  1 AU: Helios revisited}}.
\bjtitle{Journal of Geophysical Research (Space Physics)}
\bvolume{116},
\bfpage{9105}
(\byear{2011}).
doi:\doiurl{10.1029/2011JA016674}
\end{barticle}
\endbibitem

\bibitem[\protect\citeauthoryear{{Hellinger} et~al.}{2013}]{hellinger13}
\begin{botherref}
\oauthor{\binits{P.} \bsnm{{Hellinger}}},
\oauthor{\binits{P.M.} \bsnm{{Tr{\'a}vn{\'{\i}}{\v c}ek}}},
\oauthor{\binits{{\v S}.} \bsnm{{{\v S}tver{\'a}k}}},
\oauthor{\binits{L.} \bsnm{{Matteini}}},
\oauthor{\binits{M.} \bsnm{{Velli}}},
{Proton thermal energetics in the solar wind: Helios reloaded}.
Journal of Geophysical Research (Space Physics)
\textbf{118}
(2013).
doi:\doiurl{10.1002/jgra.50107}
\end{botherref}
\endbibitem

\bibitem[\protect\citeauthoryear{{Henri} et~al.}{2011}]{henri11}
\begin{barticle}
\bauthor{\binits{P.} \bsnm{{Henri}}},
\bauthor{\binits{F.} \bsnm{{Califano}}},
\bauthor{\binits{C.} \bsnm{{Briand}}},
\bauthor{\binits{A.} \bsnm{{Mangeney}}},
\batitle{{Low-energy Langmuir cavitons: Asymptotic limit of weak turbulence}}.
\bjtitle{EPL (Europhysics Letters)}
\bvolume{96},
\bfpage{55004}
(\byear{2011}).
doi:\doiurl{10.1209/0295-5075/96/55004}
\end{barticle}
\endbibitem

\bibitem[\protect\citeauthoryear{{Higdon}}{1984}]{higdon84}
\begin{barticle}
\bauthor{\binits{J.C.} \bsnm{{Higdon}}},
\batitle{{Density fluctuations in the interstellar medium: Evidence for
  anisotropic magnetogasdynamic turbulence. I. Model and astrophysical sites}}.
\bjtitle{\apj}
\bvolume{285},
\bfpage{109}--\blpage{123}
(\byear{1984}).
doi:\doiurl{10.1086/162481}
\end{barticle}
\endbibitem

\bibitem[\protect\citeauthoryear{{Hnat} et~al.}{2003}]{hnat03}
\begin{barticle}
\bauthor{\binits{B.} \bsnm{{Hnat}}},
\bauthor{\binits{S.C.} \bsnm{{Chapman}}},
\bauthor{\binits{G.} \bsnm{{Rowlands}}},
\batitle{{Intermittency, scaling, and the Fokker-Planck approach to
  fluctuations of the solar wind bulk plasma parameters as seen by the WIND
  spacecraft}}.
\bjtitle{\pre}
\bvolume{67}(\bissue{5}),
\bfpage{056404}
(\byear{2003}).
doi:\doiurl{10.1103/PhysRevE.67.056404}
\end{barticle}
\endbibitem

\bibitem[\protect\citeauthoryear{{Hnat} et~al.}{2005}]{hnat05}
\begin{barticle}
\bauthor{\binits{B.} \bsnm{{Hnat}}},
\bauthor{\binits{S.C.} \bsnm{{Chapman}}},
\bauthor{\binits{G.} \bsnm{{Rowlands}}},
\batitle{{Compressibility in Solar Wind Plasma Turbulence}}.
\bjtitle{\prl}
\bvolume{94}(\bissue{20}),
\bfpage{204502}
(\byear{2005}).
doi:\doiurl{10.1103/PhysRevLett.94.204502}
\end{barticle}
\endbibitem

\bibitem[\protect\citeauthoryear{{Horbury} et~al.}{2008}]{horbury08}
\begin{barticle}
\bauthor{\binits{T.S.} \bsnm{{Horbury}}},
\bauthor{\binits{M.} \bsnm{{Forman}}},
\bauthor{\binits{S.} \bsnm{{Oughton}}},
\batitle{{Anisotropic Scaling of Magnetohydrodynamic Turbulence}}.
\bjtitle{\prl}
\bvolume{101}(\bissue{17}),
\bfpage{175005}
(\byear{2008}).
doi:\doiurl{10.1103/PhysRevLett.101.175005}
\end{barticle}
\endbibitem

\bibitem[\protect\citeauthoryear{{Horbury} et~al.}{2005}]{horbury05}
\begin{barticle}
\bauthor{\binits{T.S.} \bsnm{{Horbury}}},
\bauthor{\binits{M.A.} \bsnm{{Forman}}},
\bauthor{\binits{S.} \bsnm{{Oughton}}},
\batitle{{Spacecraft observations of solar wind turbulence: an overview}}.
\bjtitle{\ppcf}
\bvolume{47},
\bfpage{703}--\blpage{717}
(\byear{2005}).
doi:\doiurl{10.1088/0741-3335/47/12B/S52}
\end{barticle}
\endbibitem

\bibitem[\protect\citeauthoryear{{Howes} and {Quataert}}{2010}]{howes10}
\begin{barticle}
\bauthor{\binits{G.G.} \bsnm{{Howes}}},
\bauthor{\binits{E.} \bsnm{{Quataert}}},
\batitle{{On the Interpretation of Magnetic Helicity Signatures in the
  Dissipation Range Of Solar Wind Turbulence}}.
\bjtitle{\apjl}
\bvolume{709},
\bfpage{49}--\blpage{52}
(\byear{2010}).
doi:\doiurl{10.1088/2041-8205/709/1/L49}
\end{barticle}
\endbibitem

\bibitem[\protect\citeauthoryear{{Howes} et~al.}{2011}]{howes11c}
\begin{barticle}
\bauthor{\binits{G.G.} \bsnm{{Howes}}},
\bauthor{\binits{J.M.} \bsnm{{TenBarge}}},
\bauthor{\binits{W.} \bsnm{{Dorland}}},
\batitle{{A weakened cascade model for turbulence in astrophysical plasmas}}.
\bjtitle{\pop}
\bvolume{18}(\bissue{10}),
\bfpage{102305}
(\byear{2011}).
doi:\doiurl{10.1063/1.3646400}
\end{barticle}
\endbibitem

\bibitem[\protect\citeauthoryear{{Howes} et~al.}{2006}]{howes06}
\begin{barticle}
\bauthor{\binits{G.G.} \bsnm{{Howes}}},
\bauthor{\binits{S.C.} \bsnm{{Cowley}}},
\bauthor{\binits{W.} \bsnm{{Dorland}}},
\bauthor{\binits{G.W.} \bsnm{{Hammett}}},
\bauthor{\binits{E.} \bsnm{{Quataert}}},
\bauthor{\binits{A.A.} \bsnm{{Schekochihin}}},
\batitle{{Astrophysical Gyrokinetics: Basic Equations and Linear Theory}}.
\bjtitle{\apj}
\bvolume{651},
\bfpage{590}--\blpage{614}
(\byear{2006}).
doi:\doiurl{10.1086/506172}
\end{barticle}
\endbibitem

\bibitem[\protect\citeauthoryear{{Howes} et~al.}{2008}]{howes08a}
\begin{barticle}
\bauthor{\binits{G.G.} \bsnm{{Howes}}},
\bauthor{\binits{S.C.} \bsnm{{Cowley}}},
\bauthor{\binits{W.} \bsnm{{Dorland}}},
\bauthor{\binits{G.W.} \bsnm{{Hammett}}},
\bauthor{\binits{E.} \bsnm{{Quataert}}},
\bauthor{\binits{A.A.} \bsnm{{Schekochihin}}},
\batitle{{A model of turbulence in magnetized plasmas: Implications for the
  dissipation range in the solar wind}}.
\bjtitle{\jgr}
\bvolume{113}(\bissue{A12}),
\bfpage{5103}
(\byear{2008}).
doi:\doiurl{10.1029/2007JA012665}
\end{barticle}
\endbibitem

\bibitem[\protect\citeauthoryear{{Howes} et~al.}{2011}]{howes11a}
\begin{barticle}
\bauthor{\binits{G.G.} \bsnm{{Howes}}},
\bauthor{\binits{J.M.} \bsnm{{TenBarge}}},
\bauthor{\binits{W.} \bsnm{{Dorland}}},
\bauthor{\binits{E.} \bsnm{{Quataert}}},
\bauthor{\binits{A.A.} \bsnm{{Schekochihin}}},
\bauthor{\binits{R.} \bsnm{{Numata}}},
\bauthor{\binits{T.} \bsnm{{Tatsuno}}},
\batitle{{Gyrokinetic Simulations of Solar Wind Turbulence from Ion to Electron
  Scales}}.
\bjtitle{\prl}
\bvolume{107}(\bissue{3}),
\bfpage{035004}
(\byear{2011}).
doi:\doiurl{10.1103/PhysRevLett.107.035004}
\end{barticle}
\endbibitem

\bibitem[\protect\citeauthoryear{{Howes} et~al.}{2012a}]{howes12}
\begin{barticle}
\bauthor{\binits{G.G.} \bsnm{{Howes}}},
\bauthor{\binits{S.D.} \bsnm{{Bale}}},
\bauthor{\binits{K.G.} \bsnm{{Klein}}},
\bauthor{\binits{C.H.K.} \bsnm{{Chen}}},
\bauthor{\binits{C.S.} \bsnm{{Salem}}},
\bauthor{\binits{J.M.} \bsnm{{TenBarge}}},
\batitle{{The Slow-mode Nature of Compressible Wave Power in Solar Wind
  Turbulence}}.
\bjtitle{\apjl}
\bvolume{753},
\bfpage{19}
(\byear{2012}a).
doi:\doiurl{10.1088/2041-8205/753/1/L19}
\end{barticle}
\endbibitem

\bibitem[\protect\citeauthoryear{{Howes} et~al.}{2012b}]{howes11b}
\begin{barticle}
\bauthor{\binits{G.G.} \bsnm{{Howes}}},
\bauthor{\binits{S.D.} \bsnm{{Bale}}},
\bauthor{\binits{K.G.} \bsnm{{Klein}}},
\bauthor{\binits{C.H.K.} \bsnm{{Chen}}},
\bauthor{\binits{C.S.} \bsnm{{Salem}}},
\bauthor{\binits{J.M.} \bsnm{{TenBarge}}},
\batitle{{The Slow-mode Nature of Compressible Wave Power in Solar Wind
  Turbulence}}.
\bjtitle{\apjl}
\bvolume{753},
\bfpage{19}
(\byear{2012}b).
doi:\doiurl{10.1088/2041-8205/753/1/L19}
\end{barticle}
\endbibitem

\bibitem[\protect\citeauthoryear{{Iroshnikov}}{1963}]{iroshnikov63}
\begin{barticle}
\bauthor{\binits{P.S.} \bsnm{{Iroshnikov}}},
\batitle{{Turbulence of a Conducting Fluid in a Strong Magnetic Field}}.
\bjtitle{Astron. Zh.}
\bvolume{40},
\bfpage{742}
(\byear{1963})
\end{barticle}
\endbibitem

\bibitem[\protect\citeauthoryear{{Isenberg} et~al.}{2001}]{isenberg01}
\begin{barticle}
\bauthor{\binits{P.A.} \bsnm{{Isenberg}}},
\bauthor{\binits{M.A.} \bsnm{{Lee}}},
\bauthor{\binits{J.V.} \bsnm{{Hollweg}}},
\batitle{{The kinetic shell model of coronal heating and acceleration by ion
  cyclotron waves: 1. Outward propagating waves}}.
\bjtitle{\jgr}
\bvolume{106},
\bfpage{5649}--\blpage{5660}
(\byear{2001}).
doi:\doiurl{10.1029/2000JA000099}
\end{barticle}
\endbibitem

\bibitem[\protect\citeauthoryear{{Issautier} et~al.}{2010}]{issautier10}
\begin{barticle}
\bauthor{\binits{K.} \bsnm{{Issautier}}},
\bauthor{\binits{A.} \bsnm{{Mangeney}}},
\bauthor{\binits{O.} \bsnm{{Alexandrova}}},
\batitle{{Spectrum of the electron density fluctuations: preliminary results
  from Ulysses observations}}.
\bjtitle{\aipcp}
\bvolume{1216},
\bfpage{148}--\blpage{151}
(\byear{2010}).
doi:\doiurl{10.1063/1.3395822}
\end{barticle}
\endbibitem

\bibitem[\protect\citeauthoryear{Jankovicova et~al.}{2008}]{jankovicova08}
\begin{barticle}
\bauthor{\binits{D.} \bsnm{Jankovicova}},
\bauthor{\binits{Z.} \bsnm{Voros}},
\bauthor{\binits{J.} \bsnm{Simkanin}},
\batitle{The influence of solar wind turbulence on geomagnetic activity}.
\bjtitle{NONLINEAR PROCESSES IN GEOPHYSICS}
\bvolume{15}(\bissue{1}),
\bfpage{53}--\blpage{59}
(\byear{2008})
\end{barticle}
\endbibitem

\bibitem[\protect\citeauthoryear{{Karimabadi} et~al.}{2013}]{karimabadi13}
\begin{barticle}
\bauthor{\binits{H.} \bsnm{{Karimabadi}}},
\bauthor{\binits{V.} \bsnm{{Roytershteyn}}},
\bauthor{\binits{M.} \bsnm{{Wan}}},
\bauthor{\binits{W.H.} \bsnm{{Matthaeus}}},
\bauthor{\binits{W.} \bsnm{{Daughton}}},
\bauthor{\binits{P.} \bsnm{{Wu}}},
\bauthor{\binits{M.} \bsnm{{Shay}}},
\bauthor{\binits{B.} \bsnm{{Loring}}},
\bauthor{\binits{J.} \bsnm{{Borovsky}}},
\bauthor{\binits{E.} \bsnm{{Leonardis}}},
\bauthor{\binits{S.C.} \bsnm{{Chapman}}},
\bauthor{\binits{T.K.M.} \bsnm{{Nakamura}}},
\batitle{{Coherent structures, intermittent turbulence, and dissipation in
  high-temperature plasmas}}.
\bjtitle{Physics of Plasmas}
\bvolume{20}(\bissue{1}),
\bfpage{012303}
(\byear{2013}).
doi:\doiurl{10.1063/1.4773205}
\end{barticle}
\endbibitem

\bibitem[\protect\citeauthoryear{{Kasper}}{2002}]{kasper02phd}
\begin{botherref}
\oauthor{\binits{J.C.} \bsnm{{Kasper}}},
Solar Wind Plasma: Kinetic Properties and Micro- Instabilities,
PhD thesis,
MASSACHUSETTS INSTITUTE OF TECHNOLOGY,
2002
\end{botherref}
\endbibitem

\bibitem[\protect\citeauthoryear{{Kasper} et~al.}{2008}]{kasper08}
\begin{barticle}
\bauthor{\binits{J.C.} \bsnm{{Kasper}}},
\bauthor{\binits{A.J.} \bsnm{{Lazarus}}},
\bauthor{\binits{S.P.} \bsnm{{Gary}}},
\batitle{{Hot Solar-Wind Helium: Direct Evidence for Local Heating by
  Alfv{\'e}n-Cyclotron Dissipation}}.
\bjtitle{Physical Review Letters}
\bvolume{101}(\bissue{26}),
\bfpage{261103}
(\byear{2008}).
doi:\doiurl{10.1103/PhysRevLett.101.261103}
\end{barticle}
\endbibitem

\bibitem[\protect\citeauthoryear{{Kasper} et~al.}{2013}]{kasper13}
\begin{barticle}
\bauthor{\binits{J.C.} \bsnm{{Kasper}}},
\bauthor{\binits{B.A.} \bsnm{{Maruca}}},
\bauthor{\binits{M.L.} \bsnm{{Stevens}}},
\bauthor{\binits{A.} \bsnm{{Zaslavsky}}},
\batitle{{Sensitive Test for Ion-Cyclotron Resonant Heating in the Solar
  Wind}}.
\bjtitle{Physical Review Letters}
\bvolume{110}(\bissue{9}),
\bfpage{091102}
(\byear{2013}).
doi:\doiurl{10.1103/PhysRevLett.110.091102}
\end{barticle}
\endbibitem

\bibitem[\protect\citeauthoryear{{Kellogg} and {Horbury}}{2005}]{kellogg05}
\begin{barticle}
\bauthor{\binits{P.J.} \bsnm{{Kellogg}}},
\bauthor{\binits{T.S.} \bsnm{{Horbury}}},
\batitle{{Rapid density fluctuations in the solar wind}}.
\bjtitle{\ang}
\bvolume{23},
\bfpage{3765}--\blpage{3773}
(\byear{2005})
\end{barticle}
\endbibitem

\bibitem[\protect\citeauthoryear{{Kiyani} et~al.}{2009}]{kiyani09a}
\begin{barticle}
\bauthor{\binits{K.H.} \bsnm{{Kiyani}}},
\bauthor{\binits{S.C.} \bsnm{{Chapman}}},
\bauthor{\binits{Y.V.} \bsnm{{Khotyaintsev}}},
\bauthor{\binits{M.W.} \bsnm{{Dunlop}}},
\bauthor{\binits{F.} \bsnm{{Sahraoui}}},
\batitle{{Global Scale-Invariant Dissipation in Collisionless Plasma
  Turbulence}}.
\bjtitle{\prl}
\bvolume{103}(\bissue{7}),
\bfpage{075006}
(\byear{2009}).
doi:\doiurl{10.1103/PhysRevLett.103.075006}
\end{barticle}
\endbibitem

\bibitem[\protect\citeauthoryear{{Kiyani} et~al.}{2013}]{kiyani13}
\begin{barticle}
\bauthor{\binits{K.H.} \bsnm{{Kiyani}}},
\bauthor{\binits{S.C.} \bsnm{{Chapman}}},
\bauthor{\binits{F.} \bsnm{{Sahraoui}}},
\bauthor{\binits{B.} \bsnm{{Hnat}}},
\bauthor{\binits{O.} \bsnm{{Fauvarque}}},
\bauthor{\binits{Y.V.} \bsnm{{Khotyaintsev}}},
\batitle{{Enhanced Magnetic Compressibility and Isotropic Scale Invariance at
  Sub-ion Larmor Scales in Solar Wind Turbulence}}.
\bjtitle{\apj}
\bvolume{763},
\bfpage{10}
(\byear{2013}).
doi:\doiurl{10.1088/0004-637X/763/1/10}
\end{barticle}
\endbibitem

\bibitem[\protect\citeauthoryear{{Klein} et~al.}{2012}]{klein12}
\begin{barticle}
\bauthor{\binits{K.G.} \bsnm{{Klein}}},
\bauthor{\binits{G.G.} \bsnm{{Howes}}},
\bauthor{\binits{J.M.} \bsnm{{TenBarge}}},
\bauthor{\binits{S.D.} \bsnm{{Bale}}},
\bauthor{\binits{C.H.K.} \bsnm{{Chen}}},
\bauthor{\binits{C.S.} \bsnm{{Salem}}},
\batitle{{Using Synthetic Spacecraft Data to Interpret Compressible
  Fluctuations in Solar Wind Turbulence}}.
\bjtitle{\apj}
\bvolume{755},
\bfpage{159}
(\byear{2012}).
doi:\doiurl{10.1088/0004-637X/755/2/159}
\end{barticle}
\endbibitem

\bibitem[\protect\citeauthoryear{{Kolmogorov}}{1941a}]{k41}
\begin{barticle}
\bauthor{\binits{A.} \bsnm{{Kolmogorov}}},
\batitle{{The Local Structure of Turbulence in Incompressible Viscous Fluid for
  Very Large Reynolds' Numbers}}.
\bjtitle{Akademiia Nauk SSSR Doklady}
\bvolume{30},
\bfpage{301}--\blpage{305}
(\byear{1941}a)
\end{barticle}
\endbibitem

\bibitem[\protect\citeauthoryear{{Kolmogorov}}{1941b}]{kolmogorov41a}
\begin{barticle}
\bauthor{\binits{A.N.} \bsnm{{Kolmogorov}}},
\batitle{{The Local Structure of Turbulence in Incompressible Viscous Fluid for
  Very Large Reynolds' Numbers}}.
\bjtitle{\ansd}
\bvolume{30},
\bfpage{299}--\blpage{303}
(\byear{1941}b)
\end{barticle}
\endbibitem

\bibitem[\protect\citeauthoryear{{Kraichnan}}{1965}]{kraichnan65}
\begin{barticle}
\bauthor{\binits{R.H.} \bsnm{{Kraichnan}}},
\batitle{{Inertial-Range Spectrum of Hydromagnetic Turbulence}}.
\bjtitle{\pof}
\bvolume{8},
\bfpage{1385}--\blpage{1387}
(\byear{1965})
\end{barticle}
\endbibitem

\bibitem[\protect\citeauthoryear{{Leamon} et~al.}{1998}]{leamon98a}
\begin{barticle}
\bauthor{\binits{R.J.} \bsnm{{Leamon}}},
\bauthor{\binits{C.W.} \bsnm{{Smith}}},
\bauthor{\binits{N.F.} \bsnm{{Ness}}},
\bauthor{\binits{W.H.} \bsnm{{Matthaeus}}},
\bauthor{\binits{H.K.} \bsnm{{Wong}}},
\batitle{{Observational constraints on the dynamics of the interplanetary
  magnetic field dissipation range}}.
\bjtitle{\jgr}
\bvolume{103},
\bfpage{4775}
(\byear{1998})
\end{barticle}
\endbibitem

\bibitem[\protect\citeauthoryear{{Leamon} et~al.}{1999}]{leamon99}
\begin{barticle}
\bauthor{\binits{R.J.} \bsnm{{Leamon}}},
\bauthor{\binits{C.W.} \bsnm{{Smith}}},
\bauthor{\binits{N.F.} \bsnm{{Ness}}},
\bauthor{\binits{H.K.} \bsnm{{Wong}}},
\batitle{{Dissipation range dynamics: Kinetic Alfv{\'e}n waves and the
  importance of $\beta_e$}}.
\bjtitle{\jgr}
\bvolume{104},
\bfpage{22331}--\blpage{22344}
(\byear{1999}).
doi:\doiurl{10.1029/1999JA900158}
\end{barticle}
\endbibitem

\bibitem[\protect\citeauthoryear{{Leamon} et~al.}{2000}]{leamon00}
\begin{barticle}
\bauthor{\binits{R.J.} \bsnm{{Leamon}}},
\bauthor{\binits{W.H.} \bsnm{{Matthaeus}}},
\bauthor{\binits{C.W.} \bsnm{{Smith}}},
\bauthor{\binits{G.P.} \bsnm{{Zank}}},
\bauthor{\binits{D.J.} \bsnm{{Mullan}}},
\bauthor{\binits{S.} \bsnm{{Oughton}}},
\batitle{{MHD-driven Kinetic Dissipation in the Solar Wind and Corona}}.
\bjtitle{\apj}
\bvolume{537},
\bfpage{1054}--\blpage{1062}
(\byear{2000}).
doi:\doiurl{10.1086/309059}
\end{barticle}
\endbibitem

\bibitem[\protect\citeauthoryear{{Lepping} et~al.}{1995}]{lepping95_wind}
\begin{barticle}
\bauthor{\binits{R.P.} \bsnm{{Lepping}}},
\bauthor{\binits{M.H.} \bsnm{{Ac{\~u}na}}},
\bauthor{\binits{L.F.} \bsnm{{Burlaga}}},
\bauthor{\binits{W.M.} \bsnm{{Farrell}}},
\bauthor{\binits{J.A.} \bsnm{{Slavin}}},
\bauthor{\binits{K.H.} \bsnm{{Schatten}}},
\bauthor{\binits{F.} \bsnm{{Mariani}}},
\bauthor{\binits{N.F.} \bsnm{{Ness}}},
\bauthor{\binits{F.M.} \bsnm{{Neubauer}}},
\bauthor{\binits{Y.C.} \bsnm{{Whang}}},
\bauthor{\binits{J.B.} \bsnm{{Byrnes}}},
\bauthor{\binits{R.S.} \bsnm{{Kennon}}},
\bauthor{\binits{P.V.} \bsnm{{Panetta}}},
\bauthor{\binits{J.} \bsnm{{Scheifele}}},
\bauthor{\binits{E.M.} \bsnm{{Worley}}},
\batitle{{The Wind Magnetic Field Investigation}}.
\bjtitle{\ssr}
\bvolume{71},
\bfpage{207}--\blpage{229}
(\byear{1995}).
doi:\doiurl{10.1007/BF00751330}
\end{barticle}
\endbibitem

\bibitem[\protect\citeauthoryear{Leubner and Voros}{2005}]{leubner05}
\begin{barticle}
\bauthor{\binits{M.P.} \bsnm{Leubner}},
\bauthor{\binits{Z.} \bsnm{Voros}},
\batitle{A nonextensive entropy approach to solar wind intermittency}.
\bjtitle{The Astrophysical Journal}
\bvolume{618}(\bissue{1}),
\bfpage{547}
(\byear{2005}).
\burl{http://stacks.iop.org/0004-637X/618/i=1/a=547}
\end{barticle}
\endbibitem

\bibitem[\protect\citeauthoryear{{Li} et~al.}{2001}]{li01}
\begin{barticle}
\bauthor{\binits{H.} \bsnm{{Li}}},
\bauthor{\binits{S.P.} \bsnm{{Gary}}},
\bauthor{\binits{O.} \bsnm{{Stawicki}}},
\batitle{{On the dissipation of magnetic fluctuations in the solar wind}}.
\bjtitle{\grl}
\bvolume{28},
\bfpage{1347}--\blpage{1350}
(\byear{2001}).
doi:\doiurl{10.1029/2000GL012501}
\end{barticle}
\endbibitem

\bibitem[\protect\citeauthoryear{{Lithwick} and {Goldreich}}{2001}]{lithwick01}
\begin{barticle}
\bauthor{\binits{Y.} \bsnm{{Lithwick}}},
\bauthor{\binits{P.} \bsnm{{Goldreich}}},
\batitle{{Compressible Magnetohydrodynamic Turbulence in Interstellar
  Plasmas}}.
\bjtitle{\apj}
\bvolume{562},
\bfpage{279}--\blpage{296}
(\byear{2001}).
doi:\doiurl{10.1086/323470}
\end{barticle}
\endbibitem

\bibitem[\protect\citeauthoryear{{Luo} and {Wu}}{2010}]{luo10}
\begin{barticle}
\bauthor{\binits{Q.Y.} \bsnm{{Luo}}},
\bauthor{\binits{D.J.} \bsnm{{Wu}}},
\batitle{{Observations of Anisotropic Scaling of Solar Wind Turbulence}}.
\bjtitle{\apjl}
\bvolume{714},
\bfpage{138}--\blpage{141}
(\byear{2010}).
doi:\doiurl{10.1088/2041-8205/714/1/L138}
\end{barticle}
\endbibitem

\bibitem[\protect\citeauthoryear{{MacBride} et~al.}{2005}]{macbride05}
\begin{bchapter}
\bauthor{\binits{B.T.} \bsnm{{MacBride}}},
\bauthor{\binits{M.A.} \bsnm{{Forman}}},
\bauthor{\binits{C.W.} \bsnm{{Smith}}},
\bctitle{Turbulence and Third Moment of Fluctuations: Kolmogorov's 4/5 Law and
  Its Mhd Analogues in the Solar Wind},
in \bbtitle{Solar Wind 11/SOHO 16, Connecting Sun and Heliosphere},
ed. by \beditor{\binits{B.} \bsnm{{Fleck}}},
\beditor{\binits{T.H.} \bsnm{{Zurbuchen}}},
\beditor{\binits{H.} \bsnm{{Lacoste}}}
\bsertitle{ESA Special Publication},
vol. \bseriesno{592},
\byear{2005},
p. \bfpage{613}
\end{bchapter}
\endbibitem

\bibitem[\protect\citeauthoryear{{MacBride} et~al.}{2008}]{macbride08}
\begin{barticle}
\bauthor{\binits{B.T.} \bsnm{{MacBride}}},
\bauthor{\binits{C.W.} \bsnm{{Smith}}},
\bauthor{\binits{M.A.} \bsnm{{Forman}}},
\batitle{{The Turbulent Cascade at 1 AU: Energy Transfer and the Third-Order
  Scaling for MHD}}.
\bjtitle{\apj}
\bvolume{679},
\bfpage{1644}--\blpage{1660}
(\byear{2008}).
doi:\doiurl{10.1086/529575}
\end{barticle}
\endbibitem

\bibitem[\protect\citeauthoryear{{MacBride} et~al.}{2010}]{macbride10}
\begin{barticle}
\bauthor{\binits{B.T.} \bsnm{{MacBride}}},
\bauthor{\binits{C.W.} \bsnm{{Smith}}},
\bauthor{\binits{B.J.} \bsnm{{Vasquez}}},
\batitle{{Inertial-range anisotropies in the solar wind from 0.3 to 1 AU:
  Helios 1 observations}}.
\bjtitle{Journal of Geophysical Research (Space Physics)}
\bvolume{115}(\bissue{A14}),
\bfpage{7105}
(\byear{2010}).
doi:\doiurl{10.1029/2009JA014939}
\end{barticle}
\endbibitem

\bibitem[\protect\citeauthoryear{{Malara} et~al.}{2000}]{malara00}
\begin{barticle}
\bauthor{\binits{F.} \bsnm{{Malara}}},
\bauthor{\binits{L.} \bsnm{{Primavera}}},
\bauthor{\binits{P.} \bsnm{{Veltri}}},
\batitle{{Nonlinear evolution of parametric instability of a large-amplitude
  nonmonochromatic Alfv{\'e}n wave}}.
\bjtitle{Physics of Plasmas}
\bvolume{7},
\bfpage{2866}--\blpage{2877}
(\byear{2000}).
doi:\doiurl{10.1063/1.874136}
\end{barticle}
\endbibitem

\bibitem[\protect\citeauthoryear{{Malara} et~al.}{2001}]{malara01}
\begin{barticle}
\bauthor{\binits{F.} \bsnm{{Malara}}},
\bauthor{\binits{L.} \bsnm{{Primavera}}},
\bauthor{\binits{P.} \bsnm{{Veltri}}},
\batitle{{Nonlinear evolution of the parametric instability: numerical
  predictions versus observations in the heliosphere}}.
\bjtitle{Nonlinear Processes in Geophysics}
\bvolume{8},
\bfpage{159}--\blpage{166}
(\byear{2001})
\end{barticle}
\endbibitem

\bibitem[\protect\citeauthoryear{{Mangeney}}{2012}]{mangeney12}
\begin{bchapter}
\bauthor{\binits{A.} \bsnm{{Mangeney}}},
\bctitle{Intermittency and Regularity in the Alfvenic Range of Solar Wind
  Turbulence},
in \bbtitle{American Institute of Physics Conference Series},
ed. by \beditor{\binits{P.-L.} \bsnm{{Sulem}}},
\beditor{\binits{M.} \bsnm{{Mond}}}
\bsertitle{American Institute of Physics Conference Series},
vol. \bseriesno{1439},
\byear{2012},
pp. \bfpage{26}--\blpage{41}.
doi:\doiurl{10.1063/1.3701349}
\end{bchapter}
\endbibitem

\bibitem[\protect\citeauthoryear{{Mangeney} et~al.}{1991}]{mangeney91}
\begin{bchapter}
\bauthor{\binits{A.} \bsnm{{Mangeney}}},
\bauthor{\binits{R.} \bsnm{{Grappin}}},
\bauthor{\binits{M.} \bsnm{{Velli}}},
\bctitle{Magnetohydrodynamic Turbulence in the Solar Wind},
in \bbtitle{Advances in Solar System Magnetohydrodynamics},
ed. by \beditor{\binits{E.R.} \bsnm{{Priest}}},
\beditor{\binits{A.W.} \bsnm{{Hood}}},
\byear{1991},
p. \bfpage{327}
\end{bchapter}
\endbibitem

\bibitem[\protect\citeauthoryear{{Mangeney} et~al.}{2001}]{mangeney01}
\begin{bchapter}
\bauthor{\binits{A.} \bsnm{{Mangeney}}},
\bauthor{\binits{C.} \bsnm{{Salem}}},
\bauthor{\binits{P.L.} \bsnm{{Veltri}}},
\bauthor{\binits{B.} \bsnm{{Cecconi}}},
\bctitle{Intermittency in the Solar Wind Turbulence and the Haar Wavelet
  Transform},
in \bbtitle{Sheffield Space Plasma Meeting: Multipoint Measurements versus
  Theory},
ed. by \beditor{\binits{B.} \bsnm{Warmbein}}
\bsertitle{ESA Special Pub.},
vol. \bseriesno{492},
\byear{2001},
p. \bfpage{53}
\end{bchapter}
\endbibitem

\bibitem[\protect\citeauthoryear{{Mangeney} et~al.}{2006}]{mangeney06}
\begin{barticle}
\bauthor{\binits{A.} \bsnm{{Mangeney}}},
\bauthor{\binits{C.} \bsnm{{Lacombe}}},
\bauthor{\binits{M.} \bsnm{{Maksimovic}}},
\bauthor{\binits{A.A.} \bsnm{{Samsonov}}},
\bauthor{\binits{N.} \bsnm{{Cornilleau-Wehrlin}}},
\bauthor{\binits{C.C.} \bsnm{{Harvey}}},
\bauthor{\binits{J.-M.} \bsnm{{Bosqued}}},
\bauthor{\binits{P.} \bsnm{{Tr{\'a}vn{\'{\i}}{\v c}ek}}},
\batitle{{Cluster observations in the magnetosheath - Part 1: Anisotropies of
  the wave vector distribution of the turbulence at electron scales}}.
\bjtitle{\ang}
\bvolume{24},
\bfpage{3507}--\blpage{3521}
(\byear{2006}).
doi:\doiurl{10.5194/angeo-24-3507-2006}
\end{barticle}
\endbibitem

\bibitem[\protect\citeauthoryear{{Manoharan} et~al.}{1994}]{manoharan94}
\begin{barticle}
\bauthor{\binits{P.K.} \bsnm{{Manoharan}}},
\bauthor{\binits{M.} \bsnm{{Kojima}}},
\bauthor{\binits{H.} \bsnm{{Misawa}}},
\batitle{{The spectrum of electron density fluctuations in the solar wind and
  its variations with solar wind speed}}.
\bjtitle{\jgr}
\bvolume{99},
\bfpage{23411}
(\byear{1994}).
doi:\doiurl{10.1029/94JA01955}
\end{barticle}
\endbibitem

\bibitem[\protect\citeauthoryear{{Marino} et~al.}{2008}]{marino08}
\begin{barticle}
\bauthor{\binits{R.} \bsnm{{Marino}}},
\bauthor{\binits{L.} \bsnm{{Sorriso-Valvo}}},
\bauthor{\binits{V.} \bsnm{{Carbone}}},
\bauthor{\binits{A.} \bsnm{{Noullez}}},
\bauthor{\binits{R.} \bsnm{{Bruno}}},
\bauthor{\binits{B.} \bsnm{{Bavassano}}},
\batitle{{Heating the Solar Wind by a Magnetohydrodynamic Turbulent Energy
  Cascade}}.
\bjtitle{\apjl}
\bvolume{677},
\bfpage{71}--\blpage{74}
(\byear{2008}).
doi:\doiurl{10.1086/587957}
\end{barticle}
\endbibitem

\bibitem[\protect\citeauthoryear{{Marino} et~al.}{2011}]{marino11}
\begin{barticle}
\bauthor{\binits{R.} \bsnm{{Marino}}},
\bauthor{\binits{L.} \bsnm{{Sorriso-Valvo}}},
\bauthor{\binits{V.} \bsnm{{Carbone}}},
\bauthor{\binits{P.} \bsnm{{Veltri}}},
\bauthor{\binits{A.} \bsnm{{Noullez}}},
\bauthor{\binits{R.} \bsnm{{Bruno}}},
\batitle{{The magnetohydrodynamic turbulent cascade in the ecliptic solar wind:
  Study of Ulysses data}}.
\bjtitle{\pss}
\bvolume{59},
\bfpage{592}--\blpage{597}
(\byear{2011}).
doi:\doiurl{10.1016/j.pss.2010.06.005}
\end{barticle}
\endbibitem

\bibitem[\protect\citeauthoryear{{Marino} et~al.}{2012}]{marino12}
\begin{barticle}
\bauthor{\binits{R.} \bsnm{{Marino}}},
\bauthor{\binits{L.} \bsnm{{Sorriso-Valvo}}},
\bauthor{\binits{R.} \bsnm{{D'Amicis}}},
\bauthor{\binits{V.} \bsnm{{Carbone}}},
\bauthor{\binits{R.} \bsnm{{Bruno}}},
\bauthor{\binits{P.} \bsnm{{Veltri}}},
\batitle{{On the Occurrence of the Third-order Scaling in High Latitude Solar
  Wind}}.
\bjtitle{\apj}
\bvolume{750},
\bfpage{41}
(\byear{2012}).
doi:\doiurl{10.1088/0004-637X/750/1/41}
\end{barticle}
\endbibitem

\bibitem[\protect\citeauthoryear{{Markovskii} et~al.}{2008}]{markovskii08}
\begin{barticle}
\bauthor{\binits{S.A.} \bsnm{{Markovskii}}},
\bauthor{\binits{B.J.} \bsnm{{Vasquez}}},
\bauthor{\binits{C.W.} \bsnm{{Smith}}},
\batitle{{Statistical Analysis of the High-Frequency Spectral Break of the
  Solar Wind Turbulence at 1 AU}}.
\bjtitle{\apj}
\bvolume{675},
\bfpage{1576}--\blpage{1583}
(\byear{2008}).
doi:\doiurl{10.1086/527431}
\end{barticle}
\endbibitem

\bibitem[\protect\citeauthoryear{{Maron} and {Goldreich}}{2001}]{maron01}
\begin{barticle}
\bauthor{\binits{J.} \bsnm{{Maron}}},
\bauthor{\binits{P.} \bsnm{{Goldreich}}},
\batitle{{Simulations of Incompressible Magnetohydrodynamic Turbulence}}.
\bjtitle{\apj}
\bvolume{554},
\bfpage{1175}--\blpage{1196}
(\byear{2001}).
doi:\doiurl{10.1086/321413}
\end{barticle}
\endbibitem

\bibitem[\protect\citeauthoryear{{Marsch}}{2006}]{marsch06}
\begin{barticle}
\bauthor{\binits{E.} \bsnm{{Marsch}}},
\batitle{{Kinetic Physics of the Solar Corona and Solar Wind}}.
\bjtitle{\lrsp}
\bvolume{3},
\bfpage{1}
(\byear{2006})
\end{barticle}
\endbibitem

\bibitem[\protect\citeauthoryear{{Marsch} and {Bourouaine}}{2011}]{marsch11}
\begin{barticle}
\bauthor{\binits{E.} \bsnm{{Marsch}}},
\bauthor{\binits{S.} \bsnm{{Bourouaine}}},
\batitle{{Velocity-space diffusion of solar wind protons in oblique waves and
  weak turbulence}}.
\bjtitle{Annales Geophysicae}
\bvolume{29},
\bfpage{2089}--\blpage{2099}
(\byear{2011}).
doi:\doiurl{10.5194/angeo-29-2089-2011}
\end{barticle}
\endbibitem

\bibitem[\protect\citeauthoryear{{Marsch} and {Mangeney}}{1987}]{marsch87}
\begin{barticle}
\bauthor{\binits{E.} \bsnm{{Marsch}}},
\bauthor{\binits{A.} \bsnm{{Mangeney}}},
\batitle{{Ideal MHD equations in terms of compressive Elsaesser variables}}.
\bjtitle{\jgr}
\bvolume{92},
\bfpage{7363}--\blpage{7367}
(\byear{1987}).
doi:\doiurl{10.1029/JA092iA07p07363}
\end{barticle}
\endbibitem

\bibitem[\protect\citeauthoryear{{Marsch} and {Tu}}{1990}]{marsch90b}
\begin{barticle}
\bauthor{\binits{E.} \bsnm{{Marsch}}},
\bauthor{\binits{C.-Y.} \bsnm{{Tu}}},
\batitle{{Spectral and spatial evolution of compressible turbulence in the
  inner solar wind}}.
\bjtitle{\jgr}
\bvolume{95},
\bfpage{11945}--\blpage{11956}
(\byear{1990})
\end{barticle}
\endbibitem

\bibitem[\protect\citeauthoryear{{Marsch} and {Tu}}{2001}]{marsch01}
\begin{barticle}
\bauthor{\binits{E.} \bsnm{{Marsch}}},
\bauthor{\binits{C.-Y.} \bsnm{{Tu}}},
\batitle{{Evidence for pitch angle diffusion of solar wind protons in resonance
  with cyclotron waves}}.
\bjtitle{\jgr}
\bvolume{106},
\bfpage{8357}--\blpage{8362}
(\byear{2001}).
doi:\doiurl{10.1029/2000JA000414}
\end{barticle}
\endbibitem

\bibitem[\protect\citeauthoryear{{Marsch} et~al.}{1982}]{marsch82}
\begin{barticle}
\bauthor{\binits{E.} \bsnm{{Marsch}}},
\bauthor{\binits{R.} \bsnm{{Schwenn}}},
\bauthor{\binits{H.} \bsnm{{Rosenbauer}}},
\bauthor{\binits{K.-H.} \bsnm{{Muehlhaeuser}}},
\bauthor{\binits{W.} \bsnm{{Pilipp}}},
\bauthor{\binits{F.M.} \bsnm{{Neubauer}}},
\batitle{{Solar wind protons - Three-dimensional velocity distributions and
  derived plasma parameters measured between 0.3 and 1 AU}}.
\bjtitle{\jgr}
\bvolume{87},
\bfpage{52}--\blpage{72}
(\byear{1982}).
doi:\doiurl{10.1029/JA087iA01p00052}
\end{barticle}
\endbibitem

\bibitem[\protect\citeauthoryear{{Matteini} et~al.}{2007}]{matteini07}
\begin{barticle}
\bauthor{\binits{L.} \bsnm{{Matteini}}},
\bauthor{\binits{S.} \bsnm{{Landi}}},
\bauthor{\binits{P.} \bsnm{{Hellinger}}},
\bauthor{\binits{F.} \bsnm{{Pantellini}}},
\bauthor{\binits{M.} \bsnm{{Maksimovic}}},
\bauthor{\binits{M.} \bsnm{{Velli}}},
\bauthor{\binits{B.E.} \bsnm{{Goldstein}}},
\bauthor{\binits{E.} \bsnm{{Marsch}}},
\batitle{{Evolution of the solar wind proton temperature anisotropy from 0.3 to
  2.5 AU}}.
\bjtitle{\grl}
\bvolume{34},
\bfpage{20105}
(\byear{2007}).
doi:\doiurl{10.1029/2007GL030920}
\end{barticle}
\endbibitem

\bibitem[\protect\citeauthoryear{{Matteini} et~al.}{2011}]{matteini11}
\begin{botherref}
\oauthor{\binits{L.} \bsnm{{Matteini}}},
\oauthor{\binits{P.} \bsnm{{Hellinger}}},
\oauthor{\binits{S.} \bsnm{{Landi}}},
\oauthor{\binits{P.M.} \bsnm{{Tr{\'a}vn{\'{\i}}{\v c}ek}}},
\oauthor{\binits{M.} \bsnm{{Velli}}},
{Ion Kinetics in the Solar Wind: Coupling Global Expansion to Local
  Microphysics}.
\ssr,
128
(2011).
doi:\doiurl{10.1007/s11214-011-9774-z}
\end{botherref}
\endbibitem

\bibitem[\protect\citeauthoryear{{Matthaeus} and
  {Goldstein}}{1986}]{matthaeus86}
\begin{barticle}
\bauthor{\binits{W.H.} \bsnm{{Matthaeus}}},
\bauthor{\binits{M.L.} \bsnm{{Goldstein}}},
\batitle{{Low-frequency 1/f noise in the interplanetary magnetic field}}.
\bjtitle{\prl}
\bvolume{57},
\bfpage{495}--\blpage{498}
(\byear{1986})
\end{barticle}
\endbibitem

\bibitem[\protect\citeauthoryear{{Matthaeus} and {Velli}}{2011}]{matthaeus11}
\begin{barticle}
\bauthor{\binits{W.H.} \bsnm{{Matthaeus}}},
\bauthor{\binits{M.} \bsnm{{Velli}}},
\batitle{{Who Needs Turbulence?. A Review of Turbulence Effects in the
  Heliosphere and on the Fundamental Process of Reconnection}}.
\bjtitle{\ssr}
\bvolume{160},
\bfpage{145}--\blpage{168}
(\byear{2011}).
doi:\doiurl{10.1007/s11214-011-9793-9}
\end{barticle}
\endbibitem

\bibitem[\protect\citeauthoryear{{Matthaeus} et~al.}{1990}]{matthaeus90}
\begin{barticle}
\bauthor{\binits{W.H.} \bsnm{{Matthaeus}}},
\bauthor{\binits{M.L.} \bsnm{{Goldstein}}},
\bauthor{\binits{D.A.} \bsnm{{Roberts}}},
\batitle{{Evidence for the presence of quasi-two-dimensional nearly
  incompressible fluctuations in the solar wind}}.
\bjtitle{\jgr}
\bvolume{95},
\bfpage{20673}--\blpage{20683}
(\byear{1990})
\end{barticle}
\endbibitem

\bibitem[\protect\citeauthoryear{{Matthaeus} et~al.}{1982}]{matthaeus82c}
\begin{barticle}
\bauthor{\binits{W.H.} \bsnm{{Matthaeus}}},
\bauthor{\binits{M.L.} \bsnm{{Goldstein}}},
\bauthor{\binits{C.} \bsnm{{Smith}}},
\batitle{{Evaluation of magnetic helicity in homogeneous turbulence}}.
\bjtitle{\prl}
\bvolume{48},
\bfpage{1256}--\blpage{1259}
(\byear{1982}).
doi:\doiurl{10.1103/PhysRevLett.48.1256}
\end{barticle}
\endbibitem

\bibitem[\protect\citeauthoryear{{Matthaeus} et~al.}{2008}]{matthaeus08b}
\begin{barticle}
\bauthor{\binits{W.H.} \bsnm{{Matthaeus}}},
\bauthor{\binits{S.} \bsnm{{Servidio}}},
\bauthor{\binits{P.} \bsnm{{Dmitruk}}},
\batitle{{Comment on ``Kinetic Simulations of Magnetized Turbulence in
  Astrophysical Plasmas''}}.
\bjtitle{\prl}
\bvolume{101}(\bissue{14}),
\bfpage{149501}
(\byear{2008}).
doi:\doiurl{10.1103/PhysRevLett.101.149501}
\end{barticle}
\endbibitem

\bibitem[\protect\citeauthoryear{{Matthaeus} et~al.}{2010}]{matthaeus10a}
\begin{barticle}
\bauthor{\binits{W.H.} \bsnm{{Matthaeus}}},
\bauthor{\binits{S.} \bsnm{{Servidio}}},
\bauthor{\binits{P.} \bsnm{{Dmitruk}}},
\batitle{{Dispersive Effects of Hall Electric Field in Turbulence}}.
\bjtitle{\aipcp}
\bvolume{1216},
\bfpage{184}--\blpage{187}
(\byear{2010}).
doi:\doiurl{10.1063/1.3395832}
\end{barticle}
\endbibitem

\bibitem[\protect\citeauthoryear{{Matthaeus} et~al.}{2012}]{matthaeus12}
\begin{barticle}
\bauthor{\binits{W.H.} \bsnm{{Matthaeus}}},
\bauthor{\binits{S.} \bsnm{{Servidio}}},
\bauthor{\binits{P.} \bsnm{{Dmitruk}}},
\bauthor{\binits{V.} \bsnm{{Carbone}}},
\bauthor{\binits{S.} \bsnm{{Oughton}}},
\bauthor{\binits{M.} \bsnm{{Wan}}},
\bauthor{\binits{K.T.} \bsnm{{Osman}}},
\batitle{{Local Anisotropy, Higher Order Statistics, and Turbulence Spectra}}.
\bjtitle{\apj}
\bvolume{750},
\bfpage{103}
(\byear{2012}).
doi:\doiurl{10.1088/0004-637X/750/2/103}
\end{barticle}
\endbibitem

\bibitem[\protect\citeauthoryear{{Meyer-Vernet}}{2007}]{meyer07}
\begin{bbook}
\bauthor{\binits{N.} \bsnm{{Meyer-Vernet}}},
\bbtitle{Basics of the Solar Wind}
(\bpublisher{Cambridge University Press}, \blocation{???}, \byear{2007})
\end{bbook}
\endbibitem

\bibitem[\protect\citeauthoryear{{Milano} et~al.}{2001}]{milano01}
\begin{barticle}
\bauthor{\binits{L.J.} \bsnm{{Milano}}},
\bauthor{\binits{W.H.} \bsnm{{Matthaeus}}},
\bauthor{\binits{P.} \bsnm{{Dmitruk}}},
\bauthor{\binits{D.C.} \bsnm{{Montgomery}}},
\batitle{{Local anisotropy in incompressible magnetohydrodynamic turbulence}}.
\bjtitle{Physics of Plasmas}
\bvolume{8},
\bfpage{2673}--\blpage{2681}
(\byear{2001}).
doi:\doiurl{10.1063/1.1369658}
\end{barticle}
\endbibitem

\bibitem[\protect\citeauthoryear{{M{\"u}ller} and {Grappin}}{2005}]{muller05}
\begin{barticle}
\bauthor{\binits{W.-C.} \bsnm{{M{\"u}ller}}},
\bauthor{\binits{R.} \bsnm{{Grappin}}},
\batitle{{Spectral Energy Dynamics in Magnetohydrodynamic Turbulence}}.
\bjtitle{\prl}
\bvolume{95}(\bissue{11}),
\bfpage{114502}
(\byear{2005}).
doi:\doiurl{10.1103/PhysRevLett.95.114502}
\end{barticle}
\endbibitem

\bibitem[\protect\citeauthoryear{{Narita} et~al.}{2011}]{narita11}
\begin{barticle}
\bauthor{\binits{Y.} \bsnm{{Narita}}},
\bauthor{\binits{S.P.} \bsnm{{Gary}}},
\bauthor{\binits{S.} \bsnm{{Saito}}},
\bauthor{\binits{K.-H.} \bsnm{{Glassmeier}}},
\bauthor{\binits{U.} \bsnm{{Motschmann}}},
\batitle{{Dispersion relation analysis of solar wind turbulence}}.
\bjtitle{\grl}
\bvolume{38},
\bfpage{5101}
(\byear{2011}).
doi:\doiurl{10.1029/2010GL046588}
\end{barticle}
\endbibitem

\bibitem[\protect\citeauthoryear{{Osman} et~al.}{2011}]{osman11a}
\begin{barticle}
\bauthor{\binits{K.T.} \bsnm{{Osman}}},
\bauthor{\binits{W.H.} \bsnm{{Matthaeus}}},
\bauthor{\binits{A.} \bsnm{{Greco}}},
\bauthor{\binits{S.} \bsnm{{Servidio}}},
\batitle{{Evidence for Inhomogeneous Heating in the Solar Wind}}.
\bjtitle{\apjl}
\bvolume{727},
\bfpage{11}
(\byear{2011}).
doi:\doiurl{10.1088/2041-8205/727/1/L11}
\end{barticle}
\endbibitem

\bibitem[\protect\citeauthoryear{{Osman} et~al.}{2012}]{osman12_kin}
\begin{barticle}
\bauthor{\binits{K.T.} \bsnm{{Osman}}},
\bauthor{\binits{W.H.} \bsnm{{Matthaeus}}},
\bauthor{\binits{B.} \bsnm{{Hnat}}},
\bauthor{\binits{S.C.} \bsnm{{Chapman}}},
\batitle{{Kinetic Signatures and Intermittent Turbulence in the Solar Wind
  Plasma}}.
\bjtitle{Physical Review Letters}
\bvolume{108}(\bissue{26}),
\bfpage{261103}
(\byear{2012}).
doi:\doiurl{10.1103/PhysRevLett.108.261103}
\end{barticle}
\endbibitem

\bibitem[\protect\citeauthoryear{Owens et~al.}{{2011}}]{owens11}
\begin{barticle}
\bauthor{\binits{M.J.} \bsnm{Owens}},
\bauthor{\binits{R.T.} \bsnm{Wicks}},
\bauthor{\binits{T.S.} \bsnm{Horbury}},
\batitle{{Magnetic Discontinuities in the Near-Earth Solar Wind: Evidence of
  In-Transit Turbulence or Remnants of Coronal Structure?}}
\bjtitle{{SOLAR PHYSICS}}
\bvolume{{269}}(\bissue{{2}}),
\bfpage{411}--\blpage{420}
(\byear{{2011}}).
doi:\doiurl{{10.1007/s11207-010-9695-0}}
\end{barticle}
\endbibitem

\bibitem[\protect\citeauthoryear{{Perri} and {Balogh}}{2010}]{perri10a}
\begin{barticle}
\bauthor{\binits{S.} \bsnm{{Perri}}},
\bauthor{\binits{A.} \bsnm{{Balogh}}},
\batitle{{Differences in solar wind cross-helicity and residual energy during
  the last two solar minima}}.
\bjtitle{\grl}
\bvolume{37},
\bfpage{17102}
(\byear{2010}).
doi:\doiurl{10.1029/2010GL044570}
\end{barticle}
\endbibitem

\bibitem[\protect\citeauthoryear{{Perri} et~al.}{2010}]{perri10b}
\begin{barticle}
\bauthor{\binits{S.} \bsnm{{Perri}}},
\bauthor{\binits{V.} \bsnm{{Carbone}}},
\bauthor{\binits{P.} \bsnm{{Veltri}}},
\batitle{{Where Does Fluid-like Turbulence Break Down in the Solar Wind?}}
\bjtitle{Astrophys. J.l}
\bvolume{725},
\bfpage{52}--\blpage{55}
(\byear{2010}).
doi:\doiurl{10.1088/2041-8205/725/1/L52}
\end{barticle}
\endbibitem

\bibitem[\protect\citeauthoryear{{Perri} et~al.}{2012}]{perri12a}
\begin{barticle}
\bauthor{\binits{S.} \bsnm{{Perri}}},
\bauthor{\binits{M.L.} \bsnm{{Goldstein}}},
\bauthor{\binits{J.C.} \bsnm{{Dorelli}}},
\bauthor{\binits{F.} \bsnm{{Sahraoui}}},
\batitle{{Detection of Small-Scale Structures in the Dissipation Regime of
  Solar-Wind Turbulence}}.
\bjtitle{Physical Review Letters}
\bvolume{109}(\bissue{19}),
\bfpage{191101}
(\byear{2012}).
doi:\doiurl{10.1103/PhysRevLett.109.191101}
\end{barticle}
\endbibitem

\bibitem[\protect\citeauthoryear{{Perrone} et~al.}{2013}]{perrone13}
\begin{barticle}
\bauthor{\binits{D.} \bsnm{{Perrone}}},
\bauthor{\binits{F.} \bsnm{{Valentini}}},
\bauthor{\binits{S.} \bsnm{{Servidio}}},
\bauthor{\binits{S.} \bsnm{{Dalena}}},
\bauthor{\binits{P.} \bsnm{{Veltri}}},
\batitle{{Vlasov Simulations of Multi-ion Plasma Turbulence in the Solar
  Wind}}.
\bjtitle{\apj}
\bvolume{762},
\bfpage{99}
(\byear{2013}).
doi:\doiurl{10.1088/0004-637X/762/2/99}
\end{barticle}
\endbibitem

\bibitem[\protect\citeauthoryear{Petviashvili and Pokhotelov}{1992}]{pp92}
\begin{bbook}
\bauthor{\binits{V.I.} \bsnm{Petviashvili}},
\bauthor{\binits{O.A.} \bsnm{Pokhotelov}},
\bbtitle{Solitary Waves in Plasmas and in the Atmosphere}
(\bpublisher{Gordon \& Breach Science Pub}, \blocation{???}, \byear{1992}).
\bisbn{2881247873}
\end{bbook}
\endbibitem

\bibitem[\protect\citeauthoryear{{Pietarila Graham} et~al.}{2006}]{pietarila06}
\begin{barticle}
\bauthor{\binits{J.} \bsnm{{Pietarila Graham}}},
\bauthor{\binits{D.D.} \bsnm{{Holm}}},
\bauthor{\binits{P.} \bsnm{{Mininni}}},
\bauthor{\binits{A.} \bsnm{{Pouquet}}},
\batitle{{Inertial range scaling, K{\'a}rm{\'a}n-Howarth theorem, and
  intermittency for forced and decaying Lagrangian averaged magnetohydrodynamic
  equations in two dimensions}}.
\bjtitle{Physics of Fluids}
\bvolume{18}(\bissue{4}),
\bfpage{045106}
(\byear{2006}).
doi:\doiurl{10.1063/1.2194966}
\end{barticle}
\endbibitem

\bibitem[\protect\citeauthoryear{{Podesta}}{2009}]{podesta09a}
\begin{barticle}
\bauthor{\binits{J.J.} \bsnm{{Podesta}}},
\batitle{{Dependence of Solar-Wind Power Spectra on the Direction of the Local
  Mean Magnetic Field}}.
\bjtitle{\apj}
\bvolume{698},
\bfpage{986}--\blpage{999}
(\byear{2009}).
doi:\doiurl{10.1088/0004-637X/698/2/986}
\end{barticle}
\endbibitem

\bibitem[\protect\citeauthoryear{{Podesta}}{2011}]{podesta11c}
\begin{barticle}
\bauthor{\binits{J.J.} \bsnm{{Podesta}}},
\batitle{{On the energy cascade rate of solar wind turbulence in high cross
  helicity flows}}.
\bjtitle{\jgr}
\bvolume{116}(\bissue{A15}),
\bfpage{05101}
(\byear{2011}).
doi:\doiurl{10.1029/2010JA016306}
\end{barticle}
\endbibitem

\bibitem[\protect\citeauthoryear{{Podesta} and {Gary}}{2011}]{podesta11d}
\begin{barticle}
\bauthor{\binits{J.J.} \bsnm{{Podesta}}},
\bauthor{\binits{S.P.} \bsnm{{Gary}}},
\batitle{{Magnetic Helicity Spectrum of Solar Wind Fluctuations as a Function
  of the Angle with Respect to the Local Mean Magnetic Field}}.
\bjtitle{\apj}
\bvolume{734},
\bfpage{15}
(\byear{2011}).
doi:\doiurl{10.1088/0004-637X/734/1/15}
\end{barticle}
\endbibitem

\bibitem[\protect\citeauthoryear{{Podesta} et~al.}{2007}]{podesta07a}
\begin{barticle}
\bauthor{\binits{J.J.} \bsnm{{Podesta}}},
\bauthor{\binits{D.A.} \bsnm{{Roberts}}},
\bauthor{\binits{M.L.} \bsnm{{Goldstein}}},
\batitle{{Spectral Exponents of Kinetic and Magnetic Energy Spectra in Solar
  Wind Turbulence}}.
\bjtitle{\apj}
\bvolume{664},
\bfpage{543}--\blpage{548}
(\byear{2007}).
doi:\doiurl{10.1086/519211}
\end{barticle}
\endbibitem

\bibitem[\protect\citeauthoryear{{Podesta} et~al.}{2009a}]{podesta09c}
\begin{barticle}
\bauthor{\binits{J.J.} \bsnm{{Podesta}}},
\bauthor{\binits{M.A.} \bsnm{{Forman}}},
\bauthor{\binits{C.W.} \bsnm{{Smith}}},
\bauthor{\binits{D.C.} \bsnm{{Elton}}},
\bauthor{\binits{Y.} \bsnm{{Mal{\'e}cot}}},
\bauthor{\binits{Y.} \bsnm{{Gagne}}},
\batitle{{Accurate estimation of third-order moments from turbulence
  measurements}}.
\bjtitle{\npg}
\bvolume{16},
\bfpage{99}--\blpage{110}
(\byear{2009}a)
\end{barticle}
\endbibitem

\bibitem[\protect\citeauthoryear{{Podesta} et~al.}{2009b}]{podesta09e}
\begin{barticle}
\bauthor{\binits{J.J.} \bsnm{{Podesta}}},
\bauthor{\binits{B.D.G.} \bsnm{{Chandran}}},
\bauthor{\binits{A.} \bsnm{{Bhattacharjee}}},
\bauthor{\binits{D.A.} \bsnm{{Roberts}}},
\bauthor{\binits{M.L.} \bsnm{{Goldstein}}},
\batitle{{Scale-dependent angle of alignment between velocity and magnetic
  field fluctuations in solar wind turbulence}}.
\bjtitle{\jgr}
\bvolume{114}(\bissue{A13}),
\bfpage{1107}
(\byear{2009}b).
doi:\doiurl{10.1029/2008JA013504}
\end{barticle}
\endbibitem

\bibitem[\protect\citeauthoryear{{Politano} and {Pouquet}}{1998}]{politano98}
\begin{barticle}
\bauthor{\binits{H.} \bsnm{{Politano}}},
\bauthor{\binits{A.} \bsnm{{Pouquet}}},
\batitle{{von K{\'a}rm{\'a}n-Howarth equation for magnetohydrodynamics and its
  consequences on third-order longitudinal structure and correlation
  functions}}.
\bjtitle{\pre}
\bvolume{57},
\bfpage{21}
(\byear{1998}).
doi:\doiurl{10.1103/PhysRevE.57.R21}
\end{barticle}
\endbibitem

\bibitem[\protect\citeauthoryear{{Rezeau} et~al.}{1993}]{rezeau93}
\begin{barticle}
\bauthor{\binits{L.} \bsnm{{Rezeau}}},
\bauthor{\binits{A.} \bsnm{{Roux}}},
\bauthor{\binits{C.T.} \bsnm{{Russell}}},
\batitle{{Characterization of small-scale structures at the magnetopause from
  ISEE measurements}}.
\bjtitle{J. Geophys. Res.}
\bvolume{98}(\bissue{.17}),
\bfpage{179}--\blpage{186}
(\byear{1993})
\end{barticle}
\endbibitem

\bibitem[\protect\citeauthoryear{{Roberts} et~al.}{2013}]{roberts13}
\begin{barticle}
\bauthor{\binits{O.W.} \bsnm{{Roberts}}},
\bauthor{\binits{X.} \bsnm{{Li}}},
\bauthor{\binits{B.} \bsnm{{Li}}},
\batitle{{Kinetic Plasma Turbulence in the Fast Solar Wind Measured by
  Cluster}}.
\bjtitle{\apj}
\bvolume{769},
\bfpage{58}
(\byear{2013}).
doi:\doiurl{10.1088/0004-637X/769/1/58}
\end{barticle}
\endbibitem

\bibitem[\protect\citeauthoryear{{Rudakov} et~al.}{2011}]{rudakov11}
\begin{barticle}
\bauthor{\binits{L.} \bsnm{{Rudakov}}},
\bauthor{\binits{M.} \bsnm{{Mithaiwala}}},
\bauthor{\binits{G.} \bsnm{{Ganguli}}},
\bauthor{\binits{C.} \bsnm{{Crabtree}}},
\batitle{{Linear and nonlinear Landau resonance of kinetic Alfv{\'e}n waves:
  Consequences for electron distribution and wave spectrum in the solar wind}}.
\bjtitle{Physics of Plasmas}
\bvolume{18}(\bissue{1}),
\bfpage{012307}
(\byear{2011}).
doi:\doiurl{10.1063/1.3532819}
\end{barticle}
\endbibitem

\bibitem[\protect\citeauthoryear{{Sahraoui} et~al.}{2009}]{sahraoui09}
\begin{barticle}
\bauthor{\binits{F.} \bsnm{{Sahraoui}}},
\bauthor{\binits{M.L.} \bsnm{{Goldstein}}},
\bauthor{\binits{P.} \bsnm{{Robert}}},
\bauthor{\binits{Y.V.} \bsnm{{Khotyaintsev}}},
\batitle{Evidence of a cascade and dissipation of solar-wind turbulence at the
  electron gyroscale}.
\bjtitle{\prl}
\bvolume{102}(\bissue{23}),
\bfpage{231102}
(\byear{2009}).
doi:\doiurl{10.1103/PhysRevLett.102.231102}
\end{barticle}
\endbibitem

\bibitem[\protect\citeauthoryear{{Sahraoui} et~al.}{2010}]{sahraoui10}
\begin{barticle}
\bauthor{\binits{F.} \bsnm{{Sahraoui}}},
\bauthor{\binits{M.L.} \bsnm{{Goldstein}}},
\bauthor{\binits{G.} \bsnm{{Belmont}}},
\bauthor{\binits{P.} \bsnm{{Canu}}},
\bauthor{\binits{L.} \bsnm{{Rezeau}}},
\batitle{{Three Dimensional Anisotropic k Spectra of Turbulence at Subproton
  Scales in the Solar Wind}}.
\bjtitle{\prl}
\bvolume{105},
\bfpage{131101}
(\byear{2010})
\end{barticle}
\endbibitem

\bibitem[\protect\citeauthoryear{{Salem}}{2000}]{salem00}
\begin{botherref}
\oauthor{\binits{C.} \bsnm{{Salem}}},
Ondes, turbulence et ph{\'e}nom{\'e}nes dissipatifs dans le vent solaire {\`a}
  partir des observations de la sonde wind.
PhD thesis, Univ. Paris VII
(2000)
\end{botherref}
\endbibitem

\bibitem[\protect\citeauthoryear{{Salem} et~al.}{2012}]{salem12}
\begin{barticle}
\bauthor{\binits{C.S.} \bsnm{{Salem}}},
\bauthor{\binits{G.G.} \bsnm{{Howes}}},
\bauthor{\binits{D.} \bsnm{{Sundkvist}}},
\bauthor{\binits{S.D.} \bsnm{{Bale}}},
\bauthor{\binits{C.C.} \bsnm{{Chaston}}},
\bauthor{\binits{C.H.K.} \bsnm{{Chen}}},
\bauthor{\binits{F.S.} \bsnm{{Mozer}}},
\batitle{{Identification of Kinetic Alfv{\'e}n Wave Turbulence in the Solar
  Wind}}.
\bjtitle{\apjl}
\bvolume{745},
\bfpage{9}
(\byear{2012}).
doi:\doiurl{10.1088/2041-8205/745/1/L9}
\end{barticle}
\endbibitem

\bibitem[\protect\citeauthoryear{{Salem} et~al.}{2009}]{salem09}
\begin{barticle}
\bauthor{\binits{C.} \bsnm{{Salem}}},
\bauthor{\binits{A.} \bsnm{{Mangeney}}},
\bauthor{\binits{S.D.} \bsnm{{Bale}}},
\bauthor{\binits{P.} \bsnm{{Veltri}}},
\batitle{{Solar Wind Magnetohydrodynamics Turbulence: Anomalous Scaling and
  Role of Intermittency}}.
\bjtitle{\apj}
\bvolume{702},
\bfpage{537}--\blpage{553}
(\byear{2009}).
doi:\doiurl{10.1088/0004-637X/702/1/537}
\end{barticle}
\endbibitem

\bibitem[\protect\citeauthoryear{{Schekochihin} et~al.}{2009}]{schekochihin09}
\begin{barticle}
\bauthor{\binits{A.A.} \bsnm{{Schekochihin}}},
\bauthor{\binits{S.C.} \bsnm{{Cowley}}},
\bauthor{\binits{W.} \bsnm{{Dorland}}},
\bauthor{\binits{G.W.} \bsnm{{Hammett}}},
\bauthor{\binits{G.G.} \bsnm{{Howes}}},
\bauthor{\binits{E.} \bsnm{{Quataert}}},
\bauthor{\binits{T.} \bsnm{{Tatsuno}}},
\batitle{{Astrophysical Gyrokinetics: Kinetic and Fluid Turbulent Cascades in
  Magnetized Weakly Collisional Plasmas}}.
\bjtitle{\apjs}
\bvolume{182},
\bfpage{310}--\blpage{377}
(\byear{2009}).
doi:\doiurl{10.1088/0067-0049/182/1/310}
\end{barticle}
\endbibitem

\bibitem[\protect\citeauthoryear{{Servidio} et~al.}{2007}]{servidio07}
\begin{barticle}
\bauthor{\binits{S.} \bsnm{{Servidio}}},
\bauthor{\binits{V.} \bsnm{{Carbone}}},
\bauthor{\binits{L.} \bsnm{{Primavera}}},
\bauthor{\binits{P.} \bsnm{{Veltri}}},
\bauthor{\binits{K.} \bsnm{{Stasiewicz}}},
\batitle{{Compressible turbulence in Hall Magnetohydrodynamics}}.
\bjtitle{\pss}
\bvolume{55},
\bfpage{2239}--\blpage{2243}
(\byear{2007}).
doi:\doiurl{10.1016/j.pss.2007.05.023}
\end{barticle}
\endbibitem

\bibitem[\protect\citeauthoryear{{Servidio} et~al.}{2011}]{servidio11_npg}
\begin{barticle}
\bauthor{\binits{S.} \bsnm{{Servidio}}},
\bauthor{\binits{P.} \bsnm{{Dmitruk}}},
\bauthor{\binits{A.} \bsnm{{Greco}}},
\bauthor{\binits{M.} \bsnm{{Wan}}},
\bauthor{\binits{S.} \bsnm{{Donato}}},
\bauthor{\binits{P.A.} \bsnm{{Cassak}}},
\bauthor{\binits{M.A.} \bsnm{{Shay}}},
\bauthor{\binits{V.} \bsnm{{Carbone}}},
\bauthor{\binits{W.H.} \bsnm{{Matthaeus}}},
\batitle{{Magnetic reconnection as an element of turbulence}}.
\bjtitle{Nonlinear Processes in Geophysics}
\bvolume{18},
\bfpage{675}--\blpage{695}
(\byear{2011}).
doi:\doiurl{10.5194/npg-18-675-2011}
\end{barticle}
\endbibitem

\bibitem[\protect\citeauthoryear{{Servidio} et~al.}{2012}]{servidio12}
\begin{barticle}
\bauthor{\binits{S.} \bsnm{{Servidio}}},
\bauthor{\binits{F.} \bsnm{{Valentini}}},
\bauthor{\binits{F.} \bsnm{{Califano}}},
\bauthor{\binits{P.} \bsnm{{Veltri}}},
\batitle{{Local Kinetic Effects in Two-Dimensional Plasma Turbulence}}.
\bjtitle{Physical Review Letters}
\bvolume{108}(\bissue{4}),
\bfpage{045001}
(\byear{2012}).
doi:\doiurl{10.1103/PhysRevLett.108.045001}
\end{barticle}
\endbibitem

\bibitem[\protect\citeauthoryear{{Shebalin} et~al.}{1983}]{shebalin83}
\begin{barticle}
\bauthor{\binits{J.V.} \bsnm{{Shebalin}}},
\bauthor{\binits{W.H.} \bsnm{{Matthaeus}}},
\bauthor{\binits{D.} \bsnm{{Montgomery}}},
\batitle{{Anisotropy in MHD turbulence due to a mean magnetic field}}.
\bjtitle{\jopp}
\bvolume{29},
\bfpage{525}--\blpage{547}
(\byear{1983})
\end{barticle}
\endbibitem

\bibitem[\protect\citeauthoryear{{Smith} et~al.}{2012}]{smith12}
\begin{barticle}
\bauthor{\binits{C.W.} \bsnm{{Smith}}},
\bauthor{\binits{B.J.} \bsnm{{Vasquez}}},
\bauthor{\binits{J.V.} \bsnm{{Hollweg}}},
\batitle{{Observational Constraints on the Role of Cyclotron Damping and
  Kinetic Alfv{\'e}n Waves in the Solar Wind}}.
\bjtitle{\apj}
\bvolume{745},
\bfpage{8}
(\byear{2012}).
doi:\doiurl{10.1088/0004-637X/745/1/8}
\end{barticle}
\endbibitem

\bibitem[\protect\citeauthoryear{{Smith} et~al.}{1998}]{smith98_ace}
\begin{barticle}
\bauthor{\binits{C.W.} \bsnm{{Smith}}},
\bauthor{\binits{J.} \bsnm{{L'Heureux}}},
\bauthor{\binits{N.F.} \bsnm{{Ness}}},
\bauthor{\binits{M.H.} \bsnm{{Acu{\~n}a}}},
\bauthor{\binits{L.F.} \bsnm{{Burlaga}}},
\bauthor{\binits{J.} \bsnm{{Scheifele}}},
\batitle{{The ACE Magnetic Fields Experiment}}.
\bjtitle{\ssr}
\bvolume{86},
\bfpage{613}--\blpage{632}
(\byear{1998}).
doi:\doiurl{10.1023/A:1005092216668}
\end{barticle}
\endbibitem

\bibitem[\protect\citeauthoryear{{Smith} et~al.}{2006}]{smith06a}
\begin{barticle}
\bauthor{\binits{C.W.} \bsnm{{Smith}}},
\bauthor{\binits{K.} \bsnm{{Hamilton}}},
\bauthor{\binits{B.J.} \bsnm{{Vasquez}}},
\bauthor{\binits{R.J.} \bsnm{{Leamon}}},
\batitle{{Dependence of the Dissipation Range Spectrum of Interplanetary
  Magnetic Fluctuationson the Rate of Energy Cascade}}.
\bjtitle{\apjl}
\bvolume{645},
\bfpage{85}--\blpage{88}
(\byear{2006}).
doi:\doiurl{10.1086/506151}
\end{barticle}
\endbibitem

\bibitem[\protect\citeauthoryear{{Smith} et~al.}{2009}]{smith09}
\begin{barticle}
\bauthor{\binits{C.W.} \bsnm{{Smith}}},
\bauthor{\binits{J.E.} \bsnm{{Stawarz}}},
\bauthor{\binits{B.J.} \bsnm{{Vasquez}}},
\bauthor{\binits{M.A.} \bsnm{{Forman}}},
\bauthor{\binits{B.T.} \bsnm{{MacBride}}},
\batitle{{Turbulent Cascade at 1 AU in High Cross-Helicity Flows}}.
\bjtitle{Physical Review Letters}
\bvolume{103}(\bissue{20}),
\bfpage{201101}
(\byear{2009}).
doi:\doiurl{10.1103/PhysRevLett.103.201101}
\end{barticle}
\endbibitem

\bibitem[\protect\citeauthoryear{Sorriso-Valvo et~al.}{2010}]{sorriso-valvo10}
\begin{barticle}
\bauthor{\binits{L.} \bsnm{Sorriso-Valvo}},
\bauthor{\binits{E.} \bsnm{Yordanova}},
\bauthor{\binits{V.} \bsnm{Carbone}},
\batitle{On the scaling properties of anisotropy of interplanetary magnetic
  turbulent fluctuations}.
\bjtitle{EPL (Europhysics Letters)}
\bvolume{90}(\bissue{5}),
\bfpage{59001}
(\byear{2010}).
\burl{http://stacks.iop.org/0295-5075/90/i=5/a=59001}
\end{barticle}
\endbibitem

\bibitem[\protect\citeauthoryear{{Sorriso-Valvo}
  et~al.}{1999}]{sorriso-valvo99}
\begin{barticle}
\bauthor{\binits{L.} \bsnm{{Sorriso-Valvo}}},
\bauthor{\binits{V.} \bsnm{{Carbone}}},
\bauthor{\binits{P.} \bsnm{{Veltri}}},
\bauthor{\binits{G.} \bsnm{{Consolini}}},
\bauthor{\binits{R.} \bsnm{{Bruno}}},
\batitle{{Intermittency in the solar wind turbulence through probability
  distribution functions of fluctuations}}.
\bjtitle{Geophys. Res. Lett.}
\bvolume{26},
\bfpage{1801}--\blpage{1804}
(\byear{1999}).
doi:\doiurl{10.1029/1999GL900270}
\end{barticle}
\endbibitem

\bibitem[\protect\citeauthoryear{{Sorriso-Valvo}
  et~al.}{2001}]{sorriso-valvo01}
\begin{barticle}
\bauthor{\binits{L.} \bsnm{{Sorriso-Valvo}}},
\bauthor{\binits{V.} \bsnm{{Carbone}}},
\bauthor{\binits{P.} \bsnm{{Giuliani}}},
\bauthor{\binits{P.} \bsnm{{Veltri}}},
\bauthor{\binits{R.} \bsnm{{Bruno}}},
\bauthor{\binits{V.} \bsnm{{Antoni}}},
\bauthor{\binits{E.} \bsnm{{Martines}}},
\batitle{{Intermittency in plasma turbulence}}.
\bjtitle{\pss}
\bvolume{49},
\bfpage{1193}--\blpage{1200}
(\byear{2001})
\end{barticle}
\endbibitem

\bibitem[\protect\citeauthoryear{{Sorriso-Valvo}
  et~al.}{2002}]{sorriso-valvo02}
\begin{barticle}
\bauthor{\binits{L.} \bsnm{{Sorriso-Valvo}}},
\bauthor{\binits{V.} \bsnm{{Carbone}}},
\bauthor{\binits{A.} \bsnm{{Noullez}}},
\bauthor{\binits{H.} \bsnm{{Politano}}},
\bauthor{\binits{A.} \bsnm{{Pouquet}}},
\bauthor{\binits{P.} \bsnm{{Veltri}}},
\batitle{{Analysis of cancellation in two-dimensional magnetohydrodynamic
  turbulence}}.
\bjtitle{Physics of Plasmas}
\bvolume{9},
\bfpage{89}--\blpage{95}
(\byear{2002}).
doi:\doiurl{10.1063/1.1420738}
\end{barticle}
\endbibitem

\bibitem[\protect\citeauthoryear{{Sorriso-Valvo}
  et~al.}{2007}]{sorriso-valvo07}
\begin{barticle}
\bauthor{\binits{L.} \bsnm{{Sorriso-Valvo}}},
\bauthor{\binits{R.} \bsnm{{Marino}}},
\bauthor{\binits{V.} \bsnm{{Carbone}}},
\bauthor{\binits{A.} \bsnm{{Noullez}}},
\bauthor{\binits{F.} \bsnm{{Lepreti}}},
\bauthor{\binits{P.} \bsnm{{Veltri}}},
\bauthor{\binits{R.} \bsnm{{Bruno}}},
\bauthor{\binits{B.} \bsnm{{Bavassano}}},
\bauthor{\binits{E.} \bsnm{{Pietropaolo}}},
\batitle{{Observation of Inertial Energy Cascade in Interplanetary Space
  Plasma}}.
\bjtitle{Physical Review Letters}
\bvolume{99}(\bissue{11}),
\bfpage{115001}
(\byear{2007}).
doi:\doiurl{10.1103/PhysRevLett.99.115001}
\end{barticle}
\endbibitem

\bibitem[\protect\citeauthoryear{{Spangler} and {Gwinn}}{1990}]{spangler90}
\begin{barticle}
\bauthor{\binits{S.R.} \bsnm{{Spangler}}},
\bauthor{\binits{C.R.} \bsnm{{Gwinn}}},
\batitle{{Evidence for an inner scale to the density turbulence in the
  interstellar medium}}.
\bjtitle{\apjl}
\bvolume{353},
\bfpage{29}--\blpage{32}
(\byear{1990}).
doi:\doiurl{10.1086/185700}
\end{barticle}
\endbibitem

\bibitem[\protect\citeauthoryear{{Stawarz} et~al.}{2009}]{stawarz09}
\begin{barticle}
\bauthor{\binits{J.E.} \bsnm{{Stawarz}}},
\bauthor{\binits{C.W.} \bsnm{{Smith}}},
\bauthor{\binits{B.J.} \bsnm{{Vasquez}}},
\bauthor{\binits{M.A.} \bsnm{{Forman}}},
\bauthor{\binits{B.T.} \bsnm{{MacBride}}},
\batitle{{The Turbulent Cascade and Proton Heating in the Solar Wind at 1 AU}}.
\bjtitle{\apj}
\bvolume{697},
\bfpage{1119}--\blpage{1127}
(\byear{2009}).
doi:\doiurl{10.1088/0004-637X/697/2/1119}
\end{barticle}
\endbibitem

\bibitem[\protect\citeauthoryear{{Stawarz} et~al.}{2010}]{stawarz10}
\begin{barticle}
\bauthor{\binits{J.E.} \bsnm{{Stawarz}}},
\bauthor{\binits{C.W.} \bsnm{{Smith}}},
\bauthor{\binits{B.J.} \bsnm{{Vasquez}}},
\bauthor{\binits{M.A.} \bsnm{{Forman}}},
\bauthor{\binits{B.T.} \bsnm{{MacBride}}},
\batitle{{The Turbulent Cascade for High Cross-Helicity States at 1 AU}}.
\bjtitle{\apj}
\bvolume{713},
\bfpage{920}--\blpage{934}
(\byear{2010}).
doi:\doiurl{10.1088/0004-637X/713/2/920}
\end{barticle}
\endbibitem

\bibitem[\protect\citeauthoryear{{Stawarz} et~al.}{2011}]{stawarz11}
\begin{barticle}
\bauthor{\binits{J.E.} \bsnm{{Stawarz}}},
\bauthor{\binits{B.J.} \bsnm{{Vasquez}}},
\bauthor{\binits{C.W.} \bsnm{{Smith}}},
\bauthor{\binits{M.A.} \bsnm{{Forman}}},
\bauthor{\binits{J.} \bsnm{{Klewicki}}},
\batitle{{Third Moments and the Role of Anisotropy from Velocity Shear in the
  Solar Wind}}.
\bjtitle{\apj}
\bvolume{736},
\bfpage{44}
(\byear{2011}).
doi:\doiurl{10.1088/0004-637X/736/1/44}
\end{barticle}
\endbibitem

\bibitem[\protect\citeauthoryear{{Stawicki} et~al.}{2001}]{stawicki01}
\begin{barticle}
\bauthor{\binits{O.} \bsnm{{Stawicki}}},
\bauthor{\binits{S.P.} \bsnm{{Gary}}},
\bauthor{\binits{H.} \bsnm{{Li}}},
\batitle{{Solar wind magnetic fluctuation spectra: Dispersion versus damping}}.
\bjtitle{\jgr}
\bvolume{106},
\bfpage{8273}--\blpage{8282}
(\byear{2001}).
doi:\doiurl{10.1029/2000JA000446}
\end{barticle}
\endbibitem

\bibitem[\protect\citeauthoryear{{Taylor}}{1938}]{taylor38}
\begin{barticle}
\bauthor{\binits{G.I.} \bsnm{{Taylor}}},
\batitle{{The Spectrum of Turbulence}}.
\bjtitle{\rslpsa}
\bvolume{164},
\bfpage{476}--\blpage{490}
(\byear{1938})
\end{barticle}
\endbibitem

\bibitem[\protect\citeauthoryear{{TenBarge} et~al.}{2012}]{tenbarge12b}
\begin{barticle}
\bauthor{\binits{J.M.} \bsnm{{TenBarge}}},
\bauthor{\binits{J.J.} \bsnm{{Podesta}}},
\bauthor{\binits{K.G.} \bsnm{{Klein}}},
\bauthor{\binits{G.G.} \bsnm{{Howes}}},
\batitle{{Interpreting Magnetic Variance Anisotropy Measurements in the Solar
  Wind}}.
\bjtitle{\apj}
\bvolume{753},
\bfpage{107}
(\byear{2012}).
doi:\doiurl{10.1088/0004-637X/753/2/107}
\end{barticle}
\endbibitem

\bibitem[\protect\citeauthoryear{{Tu} and {Marsch}}{1995}]{tu95}
\begin{barticle}
\bauthor{\binits{C.-Y.} \bsnm{{Tu}}},
\bauthor{\binits{E.} \bsnm{{Marsch}}},
\batitle{{MHD structures, waves and turbulence in the solar wind: Observations
  and theories}}.
\bjtitle{\ssr}
\bvolume{73},
\bfpage{1}--\blpage{2}
(\byear{1995})
\end{barticle}
\endbibitem

\bibitem[\protect\citeauthoryear{{Turner} et~al.}{2011}]{turner11}
\begin{barticle}
\bauthor{\binits{A.J.} \bsnm{{Turner}}},
\bauthor{\binits{G.} \bsnm{{Gogoberidze}}},
\bauthor{\binits{S.C.} \bsnm{{Chapman}}},
\bauthor{\binits{B.} \bsnm{{Hnat}}},
\bauthor{\binits{W.-C.} \bsnm{{M{\"u}ller}}},
\batitle{{Nonaxisymmetric Anisotropy of Solar Wind Turbulence}}.
\bjtitle{\prl}
\bvolume{107}(\bissue{9}),
\bfpage{095002}
(\byear{2011}).
doi:\doiurl{10.1103/PhysRevLett.107.095002}
\end{barticle}
\endbibitem

\bibitem[\protect\citeauthoryear{{{\v S}afr{\'a}nkov{\'a}}
  et~al.}{2013}]{safrankova13}
\begin{barticle}
\bauthor{\binits{J.} \bsnm{{{\v S}afr{\'a}nkov{\'a}}}},
\bauthor{\binits{Z.} \bsnm{{N{\v e}me{\v c}ek}}},
\bauthor{\binits{L.} \bsnm{{P{\v r}ech}}},
\bauthor{\binits{G.N.} \bsnm{{Zastenker}}},
\batitle{{Ion Kinetic Scale in the Solar Wind Observed}}.
\bjtitle{\prl}
\bvolume{110}(\bissue{2}),
\bfpage{025004}
(\byear{2013}).
doi:\doiurl{10.1103/PhysRevLett.110.025004}
\end{barticle}
\endbibitem

\bibitem[\protect\citeauthoryear{{Vasquez} et~al.}{2007}]{vasquez07}
\begin{barticle}
\bauthor{\binits{B.J.} \bsnm{{Vasquez}}},
\bauthor{\binits{V.I.} \bsnm{{Abramenko}}},
\bauthor{\binits{D.K.} \bsnm{{Haggerty}}},
\bauthor{\binits{C.W.} \bsnm{{Smith}}},
\batitle{{Numerous small magnetic field discontinuities of Bartels rotation
  2286 and the potential role of Alfv{\'e}nic turbulence}}.
\bjtitle{Journal of Geophysical Research (Space Physics)}
\bvolume{112}(\bissue{A11}),
\bfpage{11102}
(\byear{2007}).
doi:\doiurl{10.1029/2007JA012504}
\end{barticle}
\endbibitem

\bibitem[\protect\citeauthoryear{{Veltri}}{1999}]{veltri99b}
\begin{barticle}
\bauthor{\binits{P.} \bsnm{{Veltri}}},
\batitle{{MHD turbulence in the solar wind: self-similarity, intermittency and
  coherent structures}}.
\bjtitle{Plasma Physics and Controlled Fusion}
\bvolume{41},
\bfpage{787}--\blpage{795}
(\byear{1999})
\end{barticle}
\endbibitem

\bibitem[\protect\citeauthoryear{{Veltri} and {Mangeney}}{1999}]{veltri99a}
\begin{bchapter}
\bauthor{\binits{P.} \bsnm{{Veltri}}},
\bauthor{\binits{A.} \bsnm{{Mangeney}}},
\bctitle{Scaling Laws and Intermittent Structures in Solar Wind Mhd
  Turbulence},
in \bbtitle{Solar Wind Nine},
ed. by \beditor{\binits{S.R.} \bsnm{{Habbal}}},
\beditor{\binits{R.} \bsnm{{Esser}}},
\beditor{\binits{J.V.} \bsnm{{Hollweg}}},
\beditor{\binits{P.A.} \bsnm{{Isenberg}}}
\bsertitle{American Institute of Physics Conference Series},
vol. \bseriesno{471},
\byear{1999},
p. \bfpage{543}
\end{bchapter}
\endbibitem

\bibitem[\protect\citeauthoryear{{Veltri} et~al.}{2005}]{veltri05}
\begin{barticle}
\bauthor{\binits{P.} \bsnm{{Veltri}}},
\bauthor{\binits{G.} \bsnm{{Nigro}}},
\bauthor{\binits{F.} \bsnm{{Malara}}},
\bauthor{\binits{V.} \bsnm{{Carbone}}},
\bauthor{\binits{A.} \bsnm{{Mangeney}}},
\batitle{{Intermittency in MHD turbulence and coronal nanoflares modelling}}.
\bjtitle{Nonlinear Processes in Geophysics}
\bvolume{12},
\bfpage{245}--\blpage{255}
(\byear{2005})
\end{barticle}
\endbibitem

\bibitem[\protect\citeauthoryear{{Verdini} et~al.}{2012}]{verdini12}
\begin{barticle}
\bauthor{\binits{A.} \bsnm{{Verdini}}},
\bauthor{\binits{R.} \bsnm{{Grappin}}},
\bauthor{\binits{R.} \bsnm{{Pinto}}},
\bauthor{\binits{M.} \bsnm{{Velli}}},
\batitle{{On the Origin of the 1/f Spectrum in the Solar Wind Magnetic Field}}.
\bjtitle{\apjl}
\bvolume{750},
\bfpage{33}
(\byear{2012}).
doi:\doiurl{10.1088/2041-8205/750/2/L33}
\end{barticle}
\endbibitem

\bibitem[\protect\citeauthoryear{Wan et~al.}{2009}]{wan09}
\begin{botherref}
\oauthor{\binits{M.} \bsnm{Wan}},
\oauthor{\binits{S.} \bsnm{Servidio}},
\oauthor{\binits{S.} \bsnm{Oughton}},
\oauthor{\binits{W.H.} \bsnm{Matthaeus}},
The third-order law for increments in magnetohydrodynamic turbulence with
  constant shear.
Phys. Plasmas
\textbf{16}
(2009).
doi:\doiurl{10.1063/1.3240333}
\end{botherref}
\endbibitem

\bibitem[\protect\citeauthoryear{{Wan} et~al.}{2012}]{wan12prl}
\begin{barticle}
\bauthor{\binits{M.} \bsnm{{Wan}}},
\bauthor{\binits{W.H.} \bsnm{{Matthaeus}}},
\bauthor{\binits{H.} \bsnm{{Karimabadi}}},
\bauthor{\binits{V.} \bsnm{{Roytershteyn}}},
\bauthor{\binits{M.} \bsnm{{Shay}}},
\bauthor{\binits{P.} \bsnm{{Wu}}},
\bauthor{\binits{W.} \bsnm{{Daughton}}},
\bauthor{\binits{B.} \bsnm{{Loring}}},
\bauthor{\binits{S.C.} \bsnm{{Chapman}}},
\batitle{{Intermittent Dissipation at Kinetic Scales in Collisionless Plasma
  Turbulence}}.
\bjtitle{Physical Review Letters}
\bvolume{109}(\bissue{19}),
\bfpage{195001}
(\byear{2012}).
doi:\doiurl{10.1103/PhysRevLett.109.195001}
\end{barticle}
\endbibitem

\bibitem[\protect\citeauthoryear{{Wicks} et~al.}{2010}]{wicks10}
\begin{barticle}
\bauthor{\binits{R.T.} \bsnm{{Wicks}}},
\bauthor{\binits{T.S.} \bsnm{{Horbury}}},
\bauthor{\binits{C.H.K.} \bsnm{{Chen}}},
\bauthor{\binits{A.A.} \bsnm{{Schekochihin}}},
\batitle{{Power and spectral index anisotropy of the entire inertial range of
  turbulence in the fast solar wind}}.
\bjtitle{\mnras}
\bvolume{407},
\bfpage{31}--\blpage{35}
(\byear{2010}).
doi:\doiurl{10.1111/j.1745-3933.2010.00898.x}
\end{barticle}
\endbibitem

\bibitem[\protect\citeauthoryear{{Wicks} et~al.}{2011}]{wicks11a}
\begin{barticle}
\bauthor{\binits{R.T.} \bsnm{{Wicks}}},
\bauthor{\binits{T.S.} \bsnm{{Horbury}}},
\bauthor{\binits{C.H.K.} \bsnm{{Chen}}},
\bauthor{\binits{A.A.} \bsnm{{Schekochihin}}},
\batitle{{Anisotropy of Imbalanced Alfv\'enic Turbulence in Fast Solar Wind}}.
\bjtitle{\prl}
\bvolume{106},
\bfpage{045001}
(\byear{2011})
\end{barticle}
\endbibitem

\bibitem[\protect\citeauthoryear{{Wicks} et~al.}{2013}]{wicks13}
\begin{barticle}
\bauthor{\binits{R.T.} \bsnm{{Wicks}}},
\bauthor{\binits{A.} \bsnm{{Mallet}}},
\bauthor{\binits{T.S.} \bsnm{{Horbury}}},
\bauthor{\binits{C.H.K.} \bsnm{{Chen}}},
\bauthor{\binits{A.A.} \bsnm{{Schekochihin}}},
\bauthor{\binits{J.J.} \bsnm{{Mitchell}}},
\batitle{{Alignment and Scaling of Large-Scale Fluctuations in the Solar
  Wind}}.
\bjtitle{Physical Review Letters}
\bvolume{110}(\bissue{2}),
\bfpage{025003}
(\byear{2013}).
doi:\doiurl{10.1103/PhysRevLett.110.025003}
\end{barticle}
\endbibitem

\bibitem[\protect\citeauthoryear{{Wu} et~al.}{2013}]{wu13}
\begin{barticle}
\bauthor{\binits{P.} \bsnm{{Wu}}},
\bauthor{\binits{S.} \bsnm{{Perri}}},
\bauthor{\binits{K.} \bsnm{{Osman}}},
\bauthor{\binits{M.} \bsnm{{Wan}}},
\bauthor{\binits{W.H.} \bsnm{{Matthaeus}}},
\bauthor{\binits{M.A.} \bsnm{{Shay}}},
\bauthor{\binits{M.L.} \bsnm{{Goldstein}}},
\bauthor{\binits{H.} \bsnm{{Karimabadi}}},
\bauthor{\binits{S.} \bsnm{{Chapman}}},
\batitle{{Intermittent Heating in Solar Wind and Kinetic Simulations}}.
\bjtitle{\apjl}
\bvolume{763},
\bfpage{30}
(\byear{2013}).
doi:\doiurl{10.1088/2041-8205/763/2/L30}
\end{barticle}
\endbibitem

\bibitem[\protect\citeauthoryear{Yaglom}{1949}]{yaglom49}
\begin{barticle}
\bauthor{\binits{A.M.} \bsnm{Yaglom}},
\batitle{O lokalnoi strukture polya temperatur v turbulentnom potoke}.
\bjtitle{Dokl. Akad. Nauk. SSSR}
\bvolume{69},
\bfpage{743}--\blpage{746}
(\byear{1949})
\end{barticle}
\endbibitem

\bibitem[\protect\citeauthoryear{{Yao} et~al.}{2011}]{yao11}
\begin{barticle}
\bauthor{\binits{S.} \bsnm{{Yao}}},
\bauthor{\binits{J.-S.} \bsnm{{He}}},
\bauthor{\binits{E.} \bsnm{{Marsch}}},
\bauthor{\binits{C.-Y.} \bsnm{{Tu}}},
\bauthor{\binits{A.} \bsnm{{Pedersen}}},
\bauthor{\binits{H.} \bsnm{{R{\`e}me}}},
\bauthor{\binits{J.G.} \bsnm{{Trotignon}}},
\batitle{{Multi-scale Anti-correlation Between Electron Density and Magnetic
  Field Strength in the Solar Wind}}.
\bjtitle{\apj}
\bvolume{728},
\bfpage{146}
(\byear{2011}).
doi:\doiurl{10.1088/0004-637X/728/2/146}
\end{barticle}
\endbibitem

\bibitem[\protect\citeauthoryear{{Zhdankin} et~al.}{2012}]{zhdankin12_prl}
\begin{barticle}
\bauthor{\binits{V.} \bsnm{{Zhdankin}}},
\bauthor{\binits{S.} \bsnm{{Boldyrev}}},
\bauthor{\binits{J.} \bsnm{{Mason}}},
\bauthor{\binits{J.C.} \bsnm{{Perez}}},
\batitle{{Magnetic Discontinuities in Magnetohydrodynamic Turbulence and in the
  Solar Wind}}.
\bjtitle{Physical Review Letters}
\bvolume{108}(\bissue{17}),
\bfpage{175004}
(\byear{2012}).
doi:\doiurl{10.1103/PhysRevLett.108.175004}
\end{barticle}
\endbibitem

\end{thebibliography}

\end{document}